\documentclass{aa}  
\usepackage{graphicx}
\usepackage{txfonts}
\begin{document}

\title{
Tomography of the environment of the COSMOS/AzTEC-3 submillimeter galaxy at $z\sim5.3$ revealed by Ly$\alpha$ and MUSE observations\thanks{The reduced mosaic of the MUSE observations is available in electronic form at the CDS via anonymous ftp to cdsarc.u-strasbg.fr (130.79.128.5) or via http://cdsweb.u-strasbg.fr/cgi-bin/qcat?J/A+A/}}

\author{L. Guaita \inst{1,2}
\and M. Aravena \inst{2} 
\and S. Gurung-Lopez \inst{3,4}
 \and S. Cantalupo \inst{5,6}
\and R. Marino \inst{6}
 \and D. Riechers \inst{7}
\and E. da Cunha \inst{8}
\and J. Wagg \inst{9}
\and H. S. B. Algera \inst{10,11}
\and H. Dannerbauer \inst{12,13}
\and P. Cox \inst{14}
\fnmsep \thanks{Based on observations made with ESO Telescopes at the Paranal Observatory, under programme ID 094.A-0487.}
}
\offprints{Lucia Guaita, \email{lucia.guaita@unab.cl}}

\institute{Departamento de Ciencias F\'isicas, Universidad Andr\'es Bello, Fernandez Concha 700, Las Condes, Santiago, Chile
\and N\'ucleo de Astronom\'ia, Facultad de Ingenier\'ia y Ciencia, Universidad Diego Portales, Av. Ej\'ercito 441, Santiago, Chile
\and Observatori Astron\`omic, Universitat de Val\`encia, C/ Catedr\'atico Jos\'e Beltran, 2, 46980 Paterna (Val\`encia), Spain
\and Departament d'Astronomia i Astrof\'isica, Universitat de Val\`encia, 46100-Burjassot, Val\`encia, Spain
\and Dipartimento di Fisica, Universit\'a di Milano Bicocca, Piazza della Scienza 3, 20126 Milano, Italy
\and Institute for Astronomy, ETH Zurich, CH-8093 Zurich, Switzerland
\and I. Physikalisches Institut, Universit\"at zu K\"oln, Z\"ulpicher Strasse 77, D-50937 K\"oln, Germany
\and International Centre for Radio Astronomy Research, University of Western Australia, 35 Stirling Hwy, Crawley, WA 6009, Australia
\and SKA Organization, Lower Withington Macclesfield, Cheshire SK11 9DL, UK
\and Hiroshima Astrophysical Science Center, Hiroshima University, 1-3-1 Kagamiyama, Higashi-Hiroshima, Hiroshima 739-8526, Japan
\and National Astronomical Observatory of Japan, 2-21-1, Osawa, Mitaka, Tokyo, Japan
\and Instituto de Astrof\'isica de Canarias (IAC), E-38205 La Laguna, Tenerife, Spain
\and Universidad de La Laguna, Dpto. Astrof\'isica, E-38206 La Laguna, Tenerife, Spain
\and Sorbonne Universit\'e, UPMC Universit\'e Paris 6 \& CNRS, UMR 7095, Institut d'Astrophysique de Paris, 98b Boulevard Arago, 75014 Paris, France
}

\date{Acceptance date Feb. 4th 2022}

 
\abstract
   {Submillimeter galaxies (SMGs) have been proposed as the progenitors of massive ellipticals in the local Universe. Mapping the neutral gas distribution and investigating the gas accretion toward the SMGs at high redshift can provide information on the way SMG environments can evolve into clusters at $z=0$.}
   {In this work, we study the 
members of the protocluster around AzTEC-3, a submillimeter galaxy at $z=5.3$. We use Ly$\alpha$ emission and its synergy with previous CO and [C${\sc II}$]158 $\mu$m observations.} 
   {We analyzed the data from the Multi Unit Spectroscopic Explorer (MUSE) instrument in an area of $1.4x1.4$ arcmin$^2$ around AzTEC-3 and derived information on the Ly$\alpha$ line in emission. 
We compared the Ly$\alpha$ profile of various regions of the environment with the zELDA radiative transfer model, revealing the neutral gas distribution and kinematics.}
   {We identified ten Ly$\alpha$ emitting sources, including two regions with extended emission: one embedding AzTEC-3 and LBG-3, which is a star-forming galaxy located 2$"$ (12 kpc) north of the SMG and another toward LBG-1, which is a star-forming galaxy located 15$''$ (90 kpc) to the southeast. 
The two regions extend for $\sim 27\times38$ kpc$^2$ ($\sim170\times240$ ckpc$^2$) and $\sim 20\times20$ kpc$^2$  ($\sim 125\times125$ ckpc$^2$), respectively. 
The sources appear distributed in an elongated configuration of about 70$''$ (430 kpc) in extent. The number of sources confirms the overdensity around AzTEC-3.

We study the MUSE spectra of the AzTEC-3+LBG-3 system and LBG-1 in detail. 
For the AzTEC-3+LBG-3 system, the Ly$\alpha$ emission appears redshifted and more spatially extended than the [C${\sc II}$] line emission. Similarly, the Ly$\alpha$ line spectrum is broader in
velocity than [C${\sc II}$] for LBG-1. In the former spectrum, the Ly$\alpha$ emission is elongated to the north of LBG-3 and to the south of AzTEC-3, where a faint Ly$\alpha$ emitting galaxy is also located. The elongated structures could resemble tidal features due to the interaction of the two galaxies with AzTEC-3. Also, we find a bridge of gas, revealed by the Ly$\alpha$ emission between AzTEC-3 and LBG-3. 
The Ly$\alpha$ emission toward LBG-1 embeds its three components. The HI kinematics 
support the idea of a merger 
of the three components. 
}
   {Given the availability of CO and [C${\sc II}$] observations from previous campaigns, and the Ly$\alpha$ information from our MUSE dataset, we find evidence of starburst-driven phenomena and interactions around AzTEC-3. The stellar mass of the galaxies of the overdensity and the Ly$\alpha$ luminosity of the HI nebula associated with AzTEC-3
imply a dark matter halo of $\sim10^{12}$ M$_{\odot}$ at $z=5.3$. By comparing this with semi-analytical models, the dark matter halo mass indicates that the region could evolve into a cluster of $2\times10^{13}$ M$_{\odot}$ by $z=2$ and into a Fornax-type cluster at $z=0$ with a typical mass of $2\times10^{14}$ M$_{\odot}$.}  


\titlerunning{COSMOS/AzTEC-3 environment at $z\sim5.3$}
\authorrunning{Guaita L.}

\maketitle
%

\section{Introduction}


In the hierarchical theory of structure formation, initial small density fluctuations give rise to the formation of the first stars and galaxies. These structures subsequently grow larger and more massive via mergers and accretion \citep[e.g.,][]{White1978}. This produces denser and denser regions with a strong gravitational potential that can influence the distribution and the evolution of galaxies.

The densest regions at high redshift can be pinpointed by searching for massive dusty galaxies, such as submillimeter galaxies (SMGs).  Submillimeter galaxies \citep[][]{Blain2002, Casey2014} are characterized by strong star-formation events, typically associated with large amounts of gas and dust which often obscures rest-frame ultraviolet to optical wavelengths \citep{Dudzeviciute2020}. They typically present rapid gas consumption through high star-formation efficiencies that are associated with major mergers. Overdense regions around SMGs are interesting because SMGs are thought to be the progenitors of the most massive galaxies in the local Universe \citep{Chapman2005,Swinbank2008,Stach2021}. 

By studying the distribution of the neutral gas around SMGs and within their overdense environment, we can investigate accretion events and obtain information on their evolution. An efficient method to reveal the distribution and study the kinematics of the neutral Hydrogen (HI) gas at high redshift is the detection of Ly$\alpha$ emission \citep[e.g.,][]{Cantalupo2014,Matthee2019,Daddi2020} 

It is important to note that Ly$\alpha$ is the strongest recombination line of neutral Hydrogen. It is mainly produced in star-forming regions where the ionizing radiation emitted by young, hot O and B stars ionize their surrounding gas which recombines in relatively short timescales, depending on its density ($>1$ atoms cm$^{-3}$) and temperature (10$^4$ K) conditions \citep{Dijkstra2016, Cantalupo2017}. 
Due to their short wavelength, Ly$\alpha$ photons are easily absorbed by dust and, due to the resonant nature of the Ly$\alpha$ transition, they are scattered by HI atoms. Also, the escape of Ly$\alpha$ photons out of the dense interstellar medium (ISM) gas can be favored by HI kinematics, which can produce an asymmetric profile where the red peak is the dominant one in case of an outflowing gas \citep{Verhamme2006,SV:2008, Laursen2009, Orsi2008, Gurung-Lopez2019}. These phenomena produce an emission line with a red and a blue peak and some absorption at the resonant wavelength of Ly$\alpha$. The relative intensity and the separation between the two peaks is related to the amount and distribution of dust and HI atoms, and the Ly$\alpha$ emission is typically spatially more extended than the UV continuum \citep[e.g.,][]{Steidel2011,Wisotzki2016,Leclercq2017}. 
 Another process that can produce Ly$\alpha$ photons and that has been proposed to explain the very extended Ly$\alpha$ nebulae observed at high redshift is collisional excitation \citep{Haiman2000, Fardal2001} with electrons. This process converts the thermal energy of the gas into radiation, and therefore cools the gas. It is typically less energetic than recombination, except in the case of a strong gravitational potential. 

Together with Ly$\alpha$, also [C${\sc II}$]158$\mu$m has been detected in galaxies at $z>4$ and used to investigate the distribution and the properties of their gas \citep[e.g.,][]{Capak2015, Maiolino2015, Pentericci2016, Carniani2018}.
[C${\sc II}$]158$\mu$m is the dominant cooling line of the ISM 
 in star-forming galaxies, where it can carry up to 1\% of the far-infrared luminosity \citep[e.g.,][]{Israel1996}. 
%
It can originate from different phases of the ISM \citep{Hollenbach1999, Stacey1991, Goldsmith2012, Pineda2013}; mainly in the photo-dissociation regions 
where the UV radiation from hot stars can dissociate molecules and ionize atoms \citep{Stacey1991}, also in the cold neutral atomic medium, but sometimes either in the warm or in the ionized medium \citep[e.g.,][]{Pineda2013, Maiolino2015}.
The association with the molecular CO lines can help in understanding its origin from molecular clouds. 
Therefore, [C${\sc II}$]158$\mu$m is an ideal tracer of star formation, but also of the distribution, dynamics, and enrichment of the ISM in star-forming galaxies and also of their circum-galactic medium (CGM).

We expect that the HI gas in dense environments 
affect the stage of evolution of the galaxies inside the dense regions and the shape of the emission of Ly$\alpha$ and possibly [C${\sc II}$]158$\mu$m. For instance, the observations of protoclusters around radio galaxies had shown an inside-out picture in which dusty starburst galaxies are located in the cores and young Ly$\alpha$ emitters are distributed in the outskirts \citep{Kuiper2011}. However, this picture is not so straightforward because there are also dense regions at high redshift composed of more than one dense clump \citep[e.g.,][]{Cortese2006, Zheng2016, Guaita2020} and also dense clumps could be traced by different galaxy populations. \citet[][]{Shi2019}, for example, showed an overdensity of Ly$\alpha$ emitters separated from a peak traced by more massive Lyman Break galaxies, maybe indicating the presence of a stream of gas falling into one side of the structure. 
The SSA22 protocluster \citep{Steidel2000} at $z\sim3$ is an example in which the overdensity is traced by both Ly$\alpha$ emitters and Lyman Break galaxies. The filamentary structure traced by the Ly$\alpha$ emitting sources shows the effect of the gravitational potential that seems to result in the formation of giant Ly$\alpha$ nebulae in the intersection of the filaments.

Furthermore, in dense regions, the major, gas-rich merger rate can be higher than in the field, given the possible galaxy encounters \citep[e.g.,][]{Hine2016,Tacconi2013}. In a merging process, the gas can be compressed either in the central and in the external regions of the progenitor galaxies, favoring strong episodes of star formation \citep[e.g.,][]{Hernquist1989}. The properties of the merging gas and of the star-formation phenomenon could shape the Ly$\alpha$ emission of the merging system. \citet{Yajima2013} studied the formation of extended Ly$\alpha$
emission from interacting galaxies at high redshift using a combination of hydrodynamic simulations with three-dimensional radiative transfer calculations. 
They showed that at the first passage the triggered star-formation event produces two Ly$\alpha$ peaks coinciding with the two progenitor nuclei. The Ly$\alpha$ peaks decrease in intensity after the first passage and increase again in the second passage. 
The Ly$\alpha$ emission continues to decline and its distribution becomes more compact in the final stage of the merger.
Recently, \citet{Romano2021} studied 75 main-sequence star-forming galaxies at $4.4 < z < 5.9$ through their [C${\sc II}$]158$\mu$m emission and reported about the presence of significant merger activity at $z\sim5$. 


%
%
In this paper, we study the environment surrounding the submillimeter galaxy, AzTEC-3, at $z=5.3$ \citep{Capak2011}. AzTEC-3 was discovered \citep{Younger2007} in the AzTEC survey \citep{Austermann2009} and it  
was the first SMG found at $z>5$ with a large overdensity associated with it.
\citet{Capak2011} discovered at least ten star-forming galaxies in a circular area of 2 comoving Mpc (cMpc) around AzTEC-3 and estimated that the SMG makes at least one-fourth of the mass of the entire system. \citet{Riechers2010} detected CO molecular gas emission from a compact ($<2.3$ kpc scale) region on top of the SMG, estimating a CO mass on the order of $5\times10^{10}$ M$_{\odot}$ that could be depleted in 50 Myr at a star-formation rate (SFR) of 1100 M$_{\odot}$ yr$^{-1}$. 
The detection of luminous CO emission implies enrichment with heavy elements in the material that fuels the star formation in AzTEC-3.
The compactness of the CO-emitting region and the consequent high star-formation rate density make AzTEC-3 a special object to study, probably affected by its special environment \citep[see also][]{Riechers2020}. In fact, a Lyman Break galaxy (LBG-3, nomenclature from Riechers et al. 2014) is observed at 2$''$ (about 75 comoving kpc = ckpc) and another one (LBG-1, nomenclature from Riechers et al. 2014) is observed at 15$''$ (about 580 comoving kpc) from AzTEC-3. 
LBG-1  is composed of three main clumps as seen in the rest-frame UV images from the $Hubble$ Space telescope ($HST$) and it makes a significant fraction of the mass of the protocluster (about half of the mass of AzTEC-3). From the spectral energy distribution point of view, LBG-1 can be considered as a typical star-forming galaxy at $z\sim5$, with little dust obscuration and a young starburst in addition to an old underlying stellar population 
\citep{Capak2011}.  No CO emission was detected in LBG-1 \citep{Riechers2014, Pavesi2019} and this is consistent with its low metallicity.
\citet{Riechers2014} used the Atacama Large Millimeter/submillimeter Array (ALMA) 
and detected [C${\sc II}$]158$\mu$m emission from the same compact region of AzTEC-3 as CO. 
Also, they observed a hint of a tidal feature toward LBG-3. [C${\sc II}$]158$\mu$m emission was also detected at the position of LBG-1 with a hint of velocity gradient, which could be related to a merging event within the three clumps. 

Keeping the previous observations in mind, we study here deep VLT/MUSE (Multi Unit Spectroscopic Explorer) observations of the region 
around AzTEC-3 and LBG-1. The MUSE data cover the 800-1500 {\AA} rest-frame wavelengths at $z\simeq5.3$, including Ly$\alpha$. We take advantage of the synergy between the previous [C${\sc II}$] information and the MUSE Ly$\alpha$ detections to study the properties of the galaxies in the protocluster and to investigate whether the environment could play a role in the stage of evolution of AzTEC-3. It is worth noting that many $z>5$ SMGs are not detected at UV wavelengths in deep $HST$ data \citep[e.g., HDF850.1 and GN10 as explained in][]{Riechers2020}, so AzTEC-3 is a relatively rare case where MUSE UV studies are possible.  

The paper is organized as follows. In Sect. \ref{dataMUSE} and \ref{reductionMUSE}, we explain the MUSE observation and the reduction of the data. In Sect. \ref{analysisMUSE}, we describe the method adopted to analyze the MUSE datacube and to detect Ly$\alpha$ emitting sources. In Sect. \ref{results}, we present the sample of Ly$\alpha$ emitting galaxies and star-forming galaxies in the field. In Sect. \ref{AzTEC3MUSE} and \ref{LBG1MUSE}, we discuss in detail the spectroscopic properties of the Ly$\alpha$ emission around AzTEC-3 and LBG-1, the two galaxies with known systemic redshift from previous [C${\sc II}$] observations. 
In Sect. \ref{model}, we present the radiative transfer model used to fit the Ly$\alpha$ emission profiles. In Sect. \ref{discussion}, we provide a discussion about the AzTEC-3 environment coming from our MUSE Ly$\alpha$, the CO and [C${\sc II}$] observations. In Sect. \ref{summary}, we summarize our work. Throughout the paper, we adopt a standard cosmology (H$_0$ = 70 km sec$^{-1}$ Mpc$^{-1}$, $\Omega_{0}=0.3$).

\section{Observations}
\label{dataMUSE}

The field around the AzTEC-3 (RA=150.0863, dec=2.589) submillimeter galaxy was selected to match the central region of the CO Luminosity Density at High-z (COLDz) survey \citep{Pavesi2018, Riechers2019}, which corresponds to the deepest field with CO(2-1) coverage at $z\simeq5.3$. It also 
  matches the COSMOS-XS survey field \citep{VanderVlugt2021,Algera2020}
   which has the deepest radio continuum data in the COSMOS field \citep[see also][]{Algera2021}. 

Data from the $HST$ are available in the F606W, F814W, F105W, F125W, and F160W filters.  
The F814W image was obtained in Cycle 12 and 13 \citep{Scoville2007}, the F606W, F125W, and F160W image s were obtained in 2014 \citep{Riechers2013prop}, and 
the F105W image was obtained in Cycle 22 (PI: P. L. Capak) \citep[see also][]{Barisic2017}.
The environment around AzTEC-3 is quite rich. As we can see in Fig. \ref{NBs}, within a radius of 2$''$ there are at least three sources at similar redshift. At 15$''$ from AzTEC-3, there is also LBG-1 \citep{Riechers2014}.

During December 2014 and February 2015, we observed the AzTEC-3 field with the Multi Unit Spectroscopic Explorer \citep[MUSE,][]{Bacon2010,Bacon2014,Bacon2015}. MUSE is a second generation instrument installed on the Nasmyth focus of UT4 at the Very Large Telescope (Chile). The spectral resolution of MUSE at 7000 {\AA} is about 2700 and each resolution element is sampled by 2.5 pixels along the spectral direction. The spectral sampling is 1.25 {\AA} per pixel.
The pixel scale of MUSE is 0.2$''$ per pixel (about 8 ckpc per pixel) and the point spread function (PSF) of our MUSE data is $0.7''$, corresponding to about 4 kpc or 27 ckpc at $z=5.3$.
The field was observed in four MUSE pointings, two of which contain AzTEC-3 at the edges, and the total area 
 is $1.4\times1.4$ arcmin$^2$.

Each pointing was observed with 20 frames of 900 sec, making a total exposure time of 20 hours in the very center of the overlapping regions. 
For every observation night, we were able to use bias, dark, dome flat, sky flat frames from our run to perform the basic reduction steps, wavelength-calibration files, standard stars, and illumination-correction files close to each observing night were also used for proper calibration.


\subsection{Reduction of the MUSE datacubes}
\label{reductionMUSE}

We used the standard ESO MUSE pipeline \citep[][v2.2]{Weilbacher2014} to perform the initial reduction of the MUSE datacubes. The pipeline carries out bias, dark and flat field corrections, calibrates the data in wavelength and astrometry, and applies a basic illumination correction. Afterwards, we improved the quality of the illumination correction and sky subtraction by applying the CubExtractor (CubEx) software following \citet{Cantalupo2019} \citep[see also][for details]{Borisova2016, Fumagalli2016, Fumagalli2017a, Mackenzie2019}.

We reduced the MUSE dataset in two ways. First, we reduced the datacubes corresponding to each pointing separately. In the final combinations, we generated a mean and median stack of the four pointings. 
Then, we produced a mosaic by combining the information contained in all the 80 MUSE OBJECT\_PIXTABLEs and by generating individual cubes, all of the size of the final mosaic. Basically, NaN values are inserted in the areas of the individual cubes outside the observation field of view. We calculated the frame offset before combination, by cross-matching the coordinates of the sources identified with the SourceExtractor software \citep{bertin1996}. 
We made a mean combination of all the frames ("mosaic"), two mean combinations of the even ("even mosaic") and odd ("odd mosaic") numbered subexposures to have independent sets of data to be used as sanity checks of the detections, and a median combination of all the frames. 

\begin{figure*}
 \centering
\includegraphics[width=7.5cm]{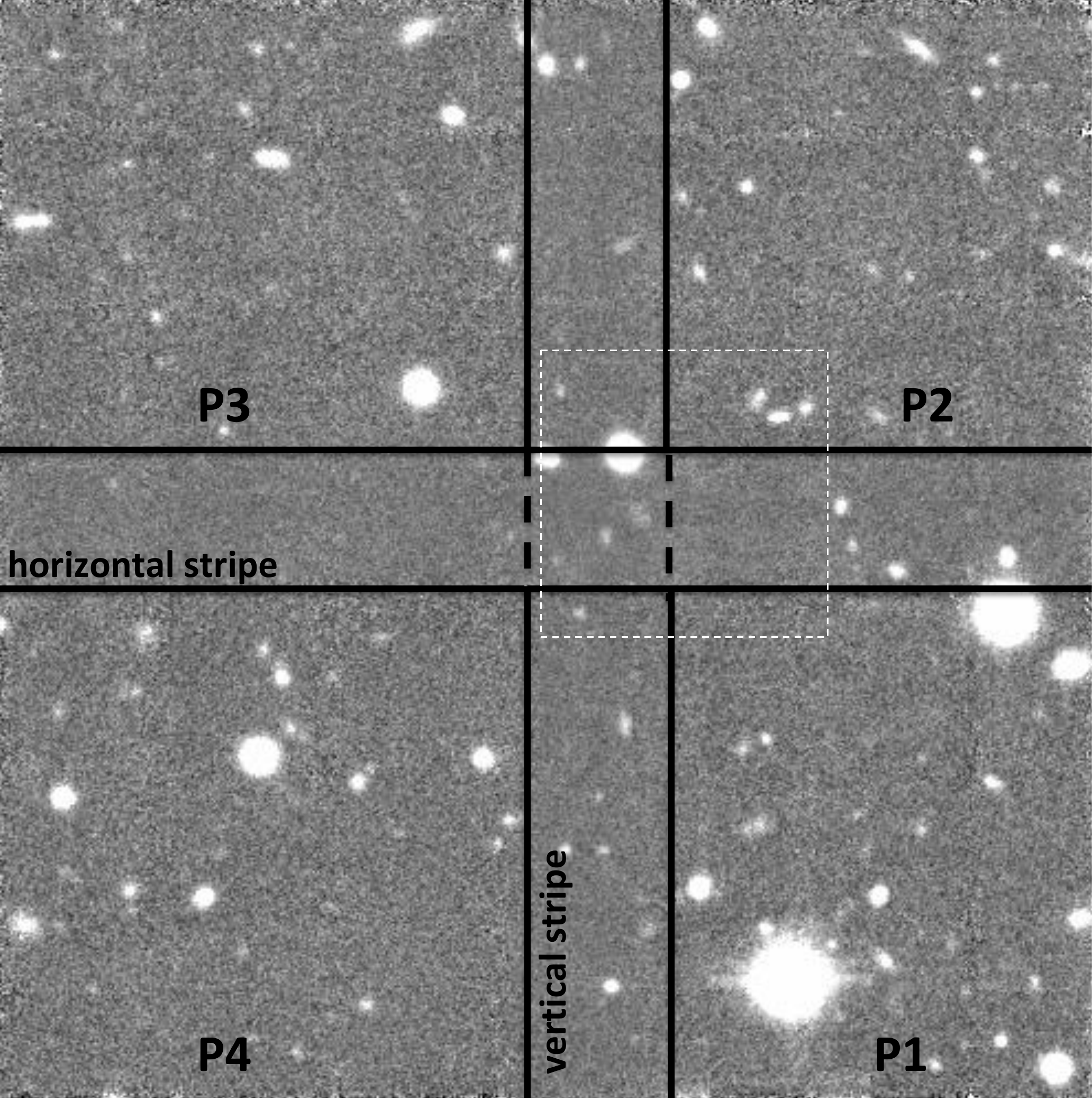}
\includegraphics[width=10.5cm]{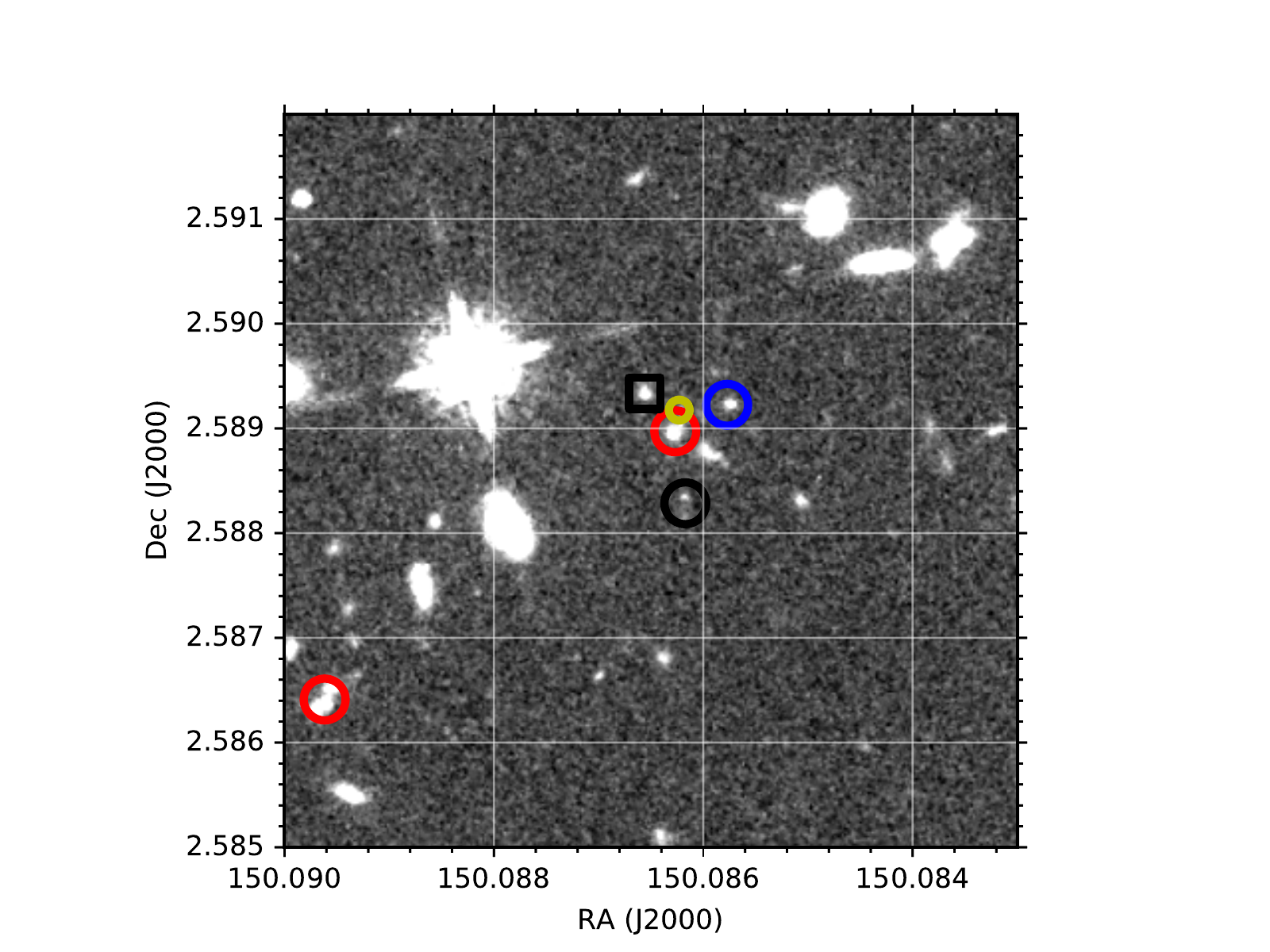} 
\caption{$Left ~panel:$ Mosaic of the entire area observed with MUSE, obtained collapsing the cube information in the 7600--7700 {\AA} wavelength range ("mosaic"), which corresponds to the redshifted Ly$\alpha$ at $z\simeq5.3$. We identified six regions characterized by similar background level, the center of the four pointings (P1, P2, P3, P4), the central vertical and horizontal (black rectangles) stripes of overlapping regions. The white square corresponds to the area zoomed in the right panel. $Right ~Panel:$ Zoom of the region containing AzTEC-3 (upper red circle) and LBG-1 (lower red circle) in the $HST$ F160W image. We show the position of LBG-3 
as black square, the position of LBG-2 (1447523 in Capak et al. 2011 and COSMOS2015\_849887 in Laigle et al. 2016) as blue circle, and the position of  COSMOS2015\_848724 as black circle. 
These sources are all at $z\sim5.3$.
The yellow circle indicates the position of a knot of star formation which can also be part of AzTEC-3. It is worth noting that LBG-1 is located in the center of the mosaic, in the lower corner of the 20 hour overlapping region.
}
\label{NBs}%
\end{figure*}

We performed the two reductions because the mosaic is more difficult to handle in the stage of the analysis due to its size, but it is fundamental to improve the signal-to-noise ration (S/N) of the overlapping areas among the pointings, in particular the region just to the south of AzTEC-3. 
The difficulty comes from the fact that the background level of the mosaic is different at the center of the pointings and in the overlapping areas and this could affect the measured S/N in the source detection. To overcome this difficulty 
we divided the mosaic into six regions, the center of each of the four pointings, the vertical overlapping area (vertical stripe), and the horizontal overlapping area (horizontal stripe), as shown in Fig. \ref{NBs}. However, these regions have sharp separations and the reduction of the four pointings separately are used to identify sources otherwise missed due to border effects. As expected, the detection limit increases in the overlapping regions (Table \ref{table:depths}). For comparison, the 5$\sigma$ detection limit of the white light image of the entire area is 26.2 when the central wavelength of the MUSE coverage 
is taken as the reference wavelength.
We matched the astrometry of MUSE to that of the images of the $Hubble$ Space Telescope ($HST$) and we reached an accuracy better than 0.1$''$ on average.
In Fig. \ref{NBs}, we show the narrow bands obtained collapsing the MUSE datacube in the 7600-7700 {\AA} wavelength range, that contains the observed wavelength of the Ly$\alpha$ emission line at $z=5.3$ ($\lambda\_air \sim 7654$ {\AA}).

\begin{table}
\caption{5$\sigma$ detection limits of the mean-combination narrow bands}  
\label{table:depths}
\centering
\scalebox{0.78}{
\begin{tabular}{c|c}
\hline  
NB & mag\_{7660} \\
\hline
P1 &  26.2   \\
P2 &  26.0   \\
P3 &  26.1   \\
P4 &  26.1   \\
horizontal stripe & 26.7\\
vertical stripe &26.7 \\
entire mosaic & 26.3\\
\hline
\hline
\end{tabular}
}
\tablefoot{The narrow-band (NB) images correspond to the ones showed in Fig. \ref{NBs}, the center of the four pointings, the vertical and horizontal stripes of overlapping regions, and the entire mosaic. The 5$\sigma$ detection limits in the second column are the magnitudes corresponding to a flux density at 7660 {\AA} (mag\_{7660} ) within an aperture of 1$''$. 
The Point Spread Function of the entire mosaic is 0.7$''$.
}
\end{table}


\subsection{Detection and identification of the line emitters}
\label{analysisMUSE}

The inspection of the combined MUSE datacubes was performed with the CubEx software \citep{Cantalupo2019}. 
To maximize the S/N of the detections, we first ran a continuum-subtraction procedure, which is performed by the CubeBKGSub function of the CubEx package. The function subtracts the continuum sources in the field, likely to be low-z objects. The continuum subtraction is performed through a median filtering along the spectral dimension, spaxel by spaxel.
Following \citet{Marino2018}, we chose the size of one cell of continuum in the wavelength direction equal to 50 
and the continuum filter radius in the wavelength direction equal to 3. 

To avoid the subtraction of emission lines in the wavelength range of our interest, some wavelengths were masked in the CubeBKGSub procedure. 
For the specific case of the detections of the members of the AzTEC-3 protocluster, we masked the wavelength range corresponding to the Ly$\alpha$ emission line at the redshift of the AzTEC-3 submillimeter galaxy (2324-2336 pixels in the spectral direction, corresponding to [-40, +550] km sec$^{-1}$ from the AzTEC-3 systemic redshift at the Ly$\alpha$ wavelength). 

Then, CubEx was used to perform extraction, detection, and photometry of sources with arbitrary spatial and spectral shapes directly in the datacubes. 
We rescaled the data variance \citep[see][]{Mackenzie2019} and we applied smoothing before extraction. The smoothing function is a Gaussian with a two-pixel 
radius 
in the spatial direction. The datacube elements, called `voxels', were selected as detections if they matched the following criteria \citep[see also][]{Marino2018, Mackenzie2019, Cantalupo2019}:
signal-to-noise ratio of the individual voxels larger than 3; integrated S/N ratio in the 3D space equal to 7; integrated spectral S/N (using 1ds optimally-extracted spectra) equal to 3; S/N of an individual voxel equal to 3 if the voxel is connected to a previously detected voxel; minimum number of connected voxels for detection equal to 30; minimum number of spectral pixels for detection equal to 3.
The main sources of fake detections were sky line residuals. 

We ran CubEx in the 1000-1100 pixel wavelength range (arbitrary much lower wavelength than 7660 {\AA}) to immediately remove bright stars from the catalog.    
With the CubEx package we also generated optimally extracted images (OEimages) of the detections showing the extension of the detection emission line in the spatial direction \citep[details in][]{Borisova2016}, weighted by signal to noise, and the signal-to-noise maps. We chose OEimages that are smoothed with a box car of a 2-pixel radius in the spatial direction for inspections. 
 
To confirm a line detection at $z\sim5.3$, we therefore, made sure that it did not show any counterpart in the $HST$ F606W image; it had reasonable segmentation and S/N datacubes; 
 it was not at the wavelength of a strong sky line residual; it had a redshift not more than $2000$ km sec$^{-1}$ from the AzTEC-3 systemic redshift; it was not associated to a detection at a different wavelength implying the emission line was not Ly$\alpha$; it had a reasonable S/N also in the even and odd mosaics (integrated S/N ratio larger than 5). This was also useful to exclude detections from cosmic rays \citep[e.g.,][]{Lofthouse2020}. Finally, we checked that
 its 1d spectrum showed a reasonable Ly$\alpha$ shape at the detection wavelength, for example not too narrow like a spike of a few bad pixels. 
%
The criteria were applied in the form of a visual inspection of the OEimages and of the spectra.

\begin{figure}
 \centering
\includegraphics[width=10cm]{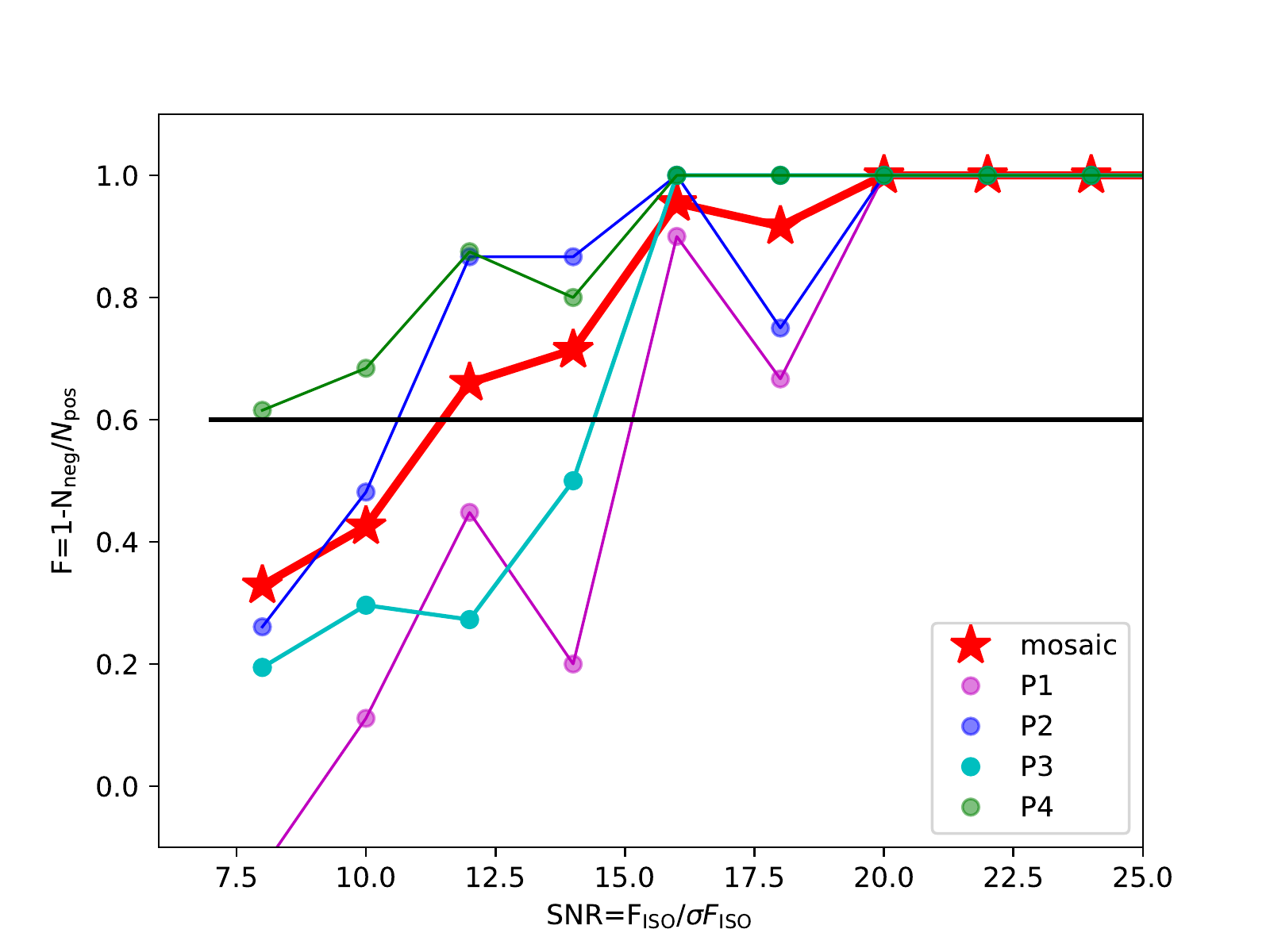}
\caption{Detection fidelity as a function of signal-to-noise ratio. The fidelity is expressed as 1-N$_{\rm neg}/N_{\rm pos}$, where N$_{\rm neg}$is the numbers of the sources detected in the datacubes multiplied by -1 and N$_{\rm pos}$ in the number of the detections in the original datacubes, when we apply the list of detection criteria described in the text. The red stars and line correspond to the detections in the "mosaic", while the other lines correspond to the individual pointings, magenta for P1, blue for P2, cyan for P3, and green for P4. The signal-to-noise ratio in the x-axis correspond to the ratio between the isophotal flux and its error as provided by CubEx. The shapes of the lines reflect the uncertainty in the fidelity measurement.}
\label{negpos}%
\end{figure}

To estimate the fidelity of our detections (i.e., down to which signal to noise the line detections start to be spurious), we applied the same set of criteria for the detection datacubes and for the datacubes multiplied by -1, that is the negative cubes. 
In Fig. \ref{negpos}, we show the ratio between the number of negative and positive detections as a function of signal to noise for the entire mosaic and for the four individual pointings.  
The signal to noise is calculated as the ratio between the isophotal flux integrated over the line detection by CubEx and its error. This signal-to-noise definition is different from the one based on pixel-to-pixel noise and that can produce lower values, but directly uses the output of CubEx.
A S/N equal to 11 allowed that more than 60\% of the detections were likely to be real detections in the mosaic. Therefore, we fixed to 11 the S/N of the line detections that made our "Main" sample. 
We propagated the 40\% chance to have a fake detection in the estimation of the density of the protocluster (see the following sections), even if the visual inspection described above already removed 20\% of the sources in the fidelity-cut catalog.

The choice of the detection parameters was guided by the extensive tests of the code performance made in the mentioned papers \citep{Marino2018, Cantalupo2019, Mackenzie2019, Lofthouse2020}, with the scope of exploiting CubEx for the detection of the faintest Ly$\alpha$ nebulae. However, we also tested the code with a variety of combinations of parameters before choosing the best set that allows a fidelity of 60\% at S/N larger than 11.

We tested the detection of sources in the mosaic as well the detection of fake sources in the negative of the mosaic cube, by changing the spatial and spectral S/N of the individual voxels (values of 2, 3, 7, and 10), the integrated S/N (values of 5, 7, and 10), and the minimum number of connected voxels (values of 5, 10, 30, and 50). We found that with a spatial S/N larger than 6, we only detect a few sources with very high isophotal S/N. For a S/N of connected voxels lower than 3, we detect more fake sources than real sources.  In the case the integrated S/N is larger than 8, we miss 80\% of the real sources for an isophotal S/N of 11.  
We found that the minimum number of voxels does not change the fidelity fraction more than 5\%, however, for a value larger than 30 and for a spatial and spectral S/N of 3, we miss 15\% of sources. Therefore, we fixed this parameter to 30.

As described in the following section, we kept a few lower S/N sources, with a counterpart in the $HST$ F160W image and which Ly$\alpha$ emission line seemed convincing in the 1D spectrum. These sources make the "Supplemental" sample (Table  \ref{finallistMUSEdetectionsat53}).
For one of the supplemental sources, the counterpart in the F160W image is also present in the COSMOS2015 catalog \citep{Laigle2016} with a redshift consistent with that of the Ly$\alpha$ detection (see Table \ref{finallistMUSEdetectionsat53}).  However, given the high density of sources in the $HST$ image, we calculated the probability of chance alignment for all the supplemental sources. We first ran Source Extractor \citep{bertin1996} with the F160W image as detection and a detection threshold of 2$\sigma$. In this run, we identified the counterparts of all the three sources in the Supplemental sample. Then, we calculated the surface density, n(<mag), of the sources detected in the F160W catalog, brighter than those counterparts, and obtained n(m<m\_446) = 0.09 arcsec$^{-2}$, n(m<m\_414) = 0.07 arcsec$^{-2}$,  and n(m<m\_199) = 0.04 arcsec$^{-2}$. By following the discussion in \citet{DP1986} (and references therein), we estimated the probability of chance alignment as $P=1-exp (-n \pi R^2)$, where n is the surface density (sources per arcsec$^2$), $\pi$ R$^2$ is the searching area of the counterparts, and R is the searching radius in arcsec given by the MUSE PSF. The probabilities resulted P\_446 = 0.09, P\_414 = 0.07, and P\_199 = 0.04. A chance probability of less than 10\% supports the idea that the low signal-to-noise sources are unlikely to be chance detections. However, we treat them as candidates in the estimation of the density of the protocluster in Sect. 3.4. 


\section{Results}
\label{results}

We present here the list of Ly$\alpha$ emitting sources found in our MUSE datacube with the method described in Sect. \ref{analysisMUSE}. 
In addition, we also discuss sources from the literature with photometric redshift consistent with $z\simeq5.3$ that could show a Ly$\alpha$ emission in the MUSE datacube, but that do not match our detection criteria. 

\subsection{Line detections compatible with Ly$\alpha$ at $z\simeq5.3$}
\label{detectionsMUSE}

By using the method described in Sect. \ref{analysisMUSE}, we identified ten sources with Ly$\alpha$ emission line compatible to be at $z\simeq5.3$. Eight of them are detected in the mosaic, one is detected in P2, and one is detected in P3. 
In Table \ref{finallistMUSEdetectionsat53}, we report the properties of the ten sources, naming them based on the detection cube. The first seven sources make our Main list. mosaic\_1513 corresponds to the system composed by AzTEC-3 and the Lyman break galaxy (LBG-3) located at 2$''$ north of it. mosaic\_1496 corresponds to LBG-1. 
mosaic\_1513 and mosaic\_1496 are the most extended we have in our list of Ly$\alpha$ emitting sources and are characterized by previous ALMA [C${\sc II}$] detections. They deserve a special discussion (see the following sections). 
In addition, mosaic\_1548,  mosaic\_1520, and mosaic\_199 overlap with sources  
in the COSMOS2015 catalog, where their photometric redshifts agree with the Ly$\alpha$ redshift.


\begin{table*}
\caption{Properties of the sources detected in the MUSE datacube and compatible with $z\simeq5.3$}  
\label{finallistMUSEdetectionsat53}
\centering
\scalebox{0.65}
{
\begin{tabular}{c|c|c|c|c|c|c|c|c|c|c|c|c|c}
\hline  
ID & ra & dec& $\lambda_{\rm{Ly}\alpha}$ & F(Ly$\alpha$)& S/N & zLy$\alpha$-z[C${\sc II}$]$^c$ & M$_{*}$ & M$_{CO}$ & Mdyn([C${\sc II}$]) & M$_{dust}$ & SFR$^d$ & $A_{V}^d$ & IAU name \\ 
    &      &        & {\AA}  &  $10^{-18}$erg sec$^{-1}$cm$^{-2}$ & & km sec$^{-1}$ & $10^{9}$ M${\odot}$ & $10^{9}$ M${\odot}$ & $10^{9}$ M${\odot}$ & $10^{8}$ M${\odot}$ &  M${\odot}$ yr$^{-1}$ & &\\
(1) & (2)  & (3)  & (4)  & (5)  & (6) & (7)  & (8)   & (9)  & (10) & (11) & (12)& (13)& (14)\\
\hline
Main &  &  &  &  & &  &   &  & &  & &\\
mosaic\_1513$^a$ & 150.0864 & 2.5891 & 7663.2 &  16.92$\pm$0.37 &   46   & $380\pm5$ & 20, 0.95 & $57\pm5$, -  & $97\pm16$, - & $2.66\pm0.76$, - & 1100$^e$, 18 & 0.8, 0.0 & LAE J100020.74+023520.8\\ 
mosaic\_1496$^b$ & 150.0897 & 2.5865 & 7662.7 & 5.63 $\pm$0.26 & 22 & $540\pm10$ & $14.8\pm2.0^f$ &  & $19.0\pm5.0$ & $<0.9$ & $<54^e$ & 0.2$^f$ & LAE J100021.53+023511.4\\ 
mosaic\_1548$^g$ & 150.0909 & 2.5770 & 7666.5 & 5.42$\pm$0.34 & 16 & & $12.2 \pm2.2$&   &  & & 15 & 0.0 & LAE J100021.82+023437.2\\
mosaic\_1520$^h$ & 150.0874 & 2.5938 & 7657.9 & 1.89 $\pm$0.17 & 11 & & $2.8\pm3.8$ &   &  & & 130 & 0.7 & LAE J100020.98+023537.7\\
mosaic\_770 & 150.1012 & 2.5915 & 7634.5 & 2.60$\pm$0.24 & 11 & &  &   &  & &  & &   LAE J100024.29+023529.4  \\
mosaic\_1035 & 150.0823 & 2.5959 & 7640.7 & 2.53$\pm$0.23 & 11 & &  &   &  & &  & & LAE J100019.75+023545.2\\
P3\_547 & 150.0895 & 2.5842 & 7643.9 & 1.60$\pm$0.14 & 11 & &  &   &  & &  & & LAE J100021.48+023503.1\\
\hline
Supplemental &  &  &  &  & &  &   &  & &  & \\
P2\_446 & 150.0852 & 2.5877 & 7643.7 & 1.68$\pm$0.19 & 9 & &  &   &  & &  & & LAE J100020.45+023515.7\\
mosaic\_414 & 150.0924 & 2.5752 & 7667.8 & 0.96$\pm$0.10 & 10 & &  &   &  & &  & & LAE J100022.18+023430.7\\
mosaic\_199$^i$ & 150.0860 & 2.5881 & 7662.7 & 5.14$\pm$1.15 & 5 & & $0.5\pm0.4$ &   &  & & 23 & 0.2 &  LAE J100020.64+023517.2 \\
\hline
\hline
\end{tabular}
}
\tablefoot{
Ly$\alpha$ and physical properties of the galaxies detected in the MUSE datacube. In the first part of the table, we list the sources in the `Main' sample, 
in the second part the "Supplemental" sample. 
The columns correspond to 1) ID of the sources, 2) their RA and 3) declination, 4) wavelength of the detected emission line which is interpreted as Ly$\alpha$, 5) isophotal integrated flux of the detected Ly$\alpha$ emission line, 6) signal-to-noise ratio of the detected emission line, 7) velocity difference between the redshift implied by the detected Ly$\alpha$ central wavelength and the systemic redshift inferred by the [C${\sc II}$] emission line, 8) stellar, 9) molecular-gas, 10) dynamical, and 11) dust mass, 12) star-formation rate, 13) dust extinction in magnitude, 14) name of the sources according to the IAU standard. 
The mass estimates are reported only for the sources with CO and/or [C${\sc II}$] detections \citep{Capak2011,Riechers2010,Riechers2014}, and for the sources with a counterpart in the COSMOS2015 catalog \citep{Laigle2016}. Star-formation rate and dust extinction are reported for the sources with a counterpart in the COSMOS2015 catalog. 
$^a$system composed by AzTEC-3 and LBG-3. LBG-3 is named 1447526 in Capak et al. (2011), COLDz.COS.0 in \citet{Riechers2020}, and COSMOS2015\_849732 in the COSMOS2015 catalog.
The physical parameters corresponding to AzTEC-3 (first number) and LBG-3 (second number) are taken from \citet{Riechers2014,Riechers2020}. 
$^b$LBG-1 as defined in \citet{Riechers2014}, 1447524 in Capak et al. (2011), HZ6 in \citet{Capak2015}, and COSMOS2015\_848185 in the COSMOS2015 catalog. The parameter values are taken from \citet{Riechers2014,Pavesi2019, Faisst2020} \citep[see also][for $HST$ and ALMA measurements]{Barisic2017}. 
$^c$velocity reported only for the sources with [C${\sc II}$] detections.
$^d$SFR and $A_{V}$ from the best fit SED fitting that provides the photometric redshift in the COSMOS2015 catalog, except for AzTEC-3 and LBG-1.
$^e$SFR from FIR.
$^f$Capak et al. (2011) reported M$_{*}=3.2 \times 10^9$ M$_{\odot}$ and $A_{V}=0.5$.  $^g$it overlaps with COSMOS2015\_841844 ($z_{phot}\simeq5.4\pm0.1$).  $^h$it overlaps with COSMOS2015\_852474 ($z_{phot}\simeq5.4\pm0.9$).  $^i$it overlaps with COSMOS2015\_848724 ($z_{phot}\simeq5.1\pm1.1$).}
\end{table*}

\begin{figure*}
 \centering
\includegraphics[width=20cm]{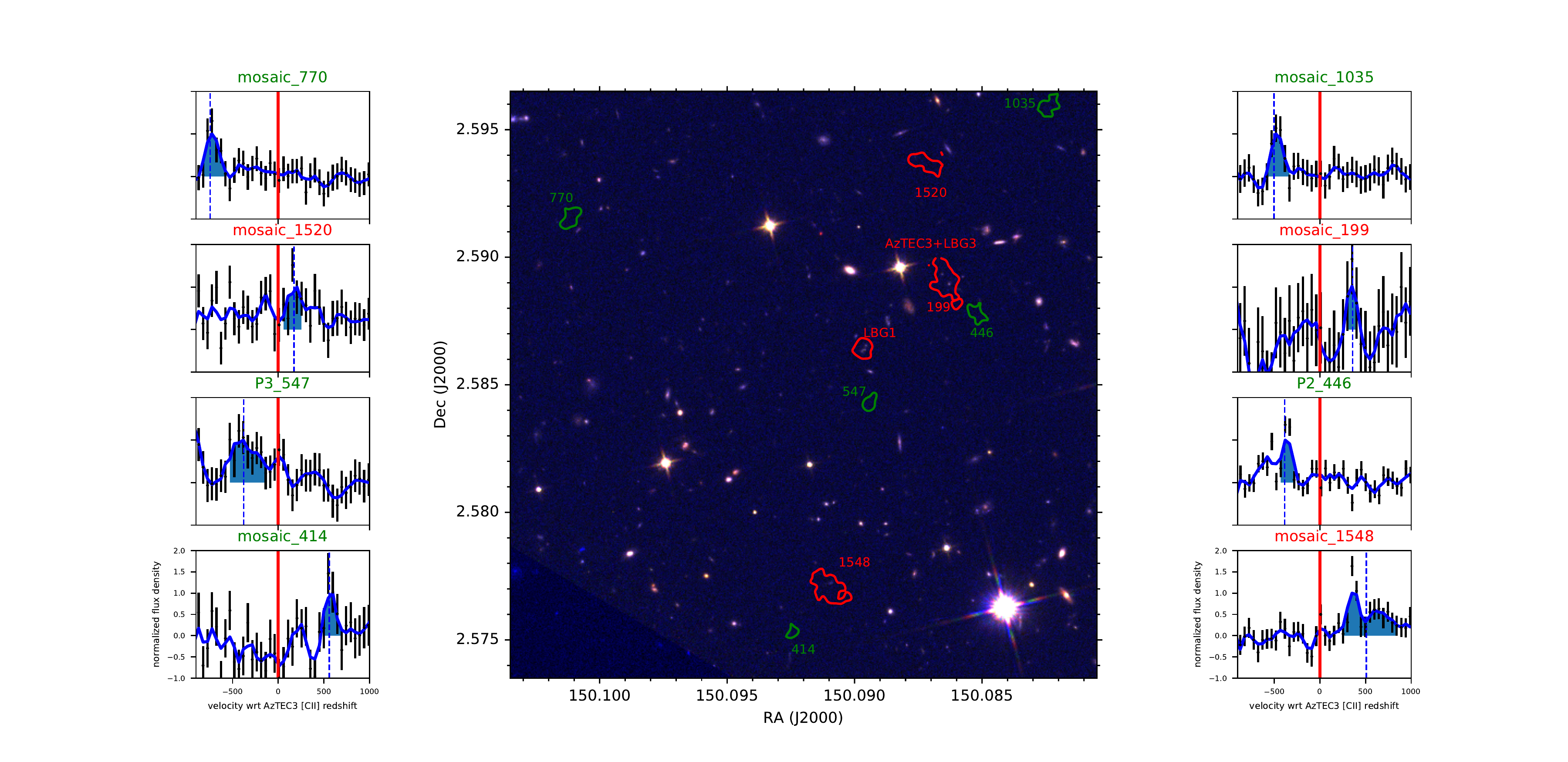} 
\caption{RGB image composed of the $HST$ images in F814W, F125W, and F160W filters covering the area of our MUSE observations. The overlaid contours represent the 3$\sigma$ 
levels of the Ly$\alpha$ emission. Red contours correspond to the sources with counterparts in the COSMOS2015 catalog. Green contours correspond to the Ly$\alpha$ detections without counterparts. 
The inserts contain the normalized Ly$\alpha$ profiles (normalized flux density versus velocity) of the Ly$\alpha$ detections indicated in the title of the inserts. The wavelengths of the maximum of the line detected by CubEx is indicated with a vertical dashed blue line. The blue shaded areas correspond to the spectral pixel ranges of the lines detected by CubEx. The zero velocity corresponds to the systemic redshift of the [C${\sc II}$] detection of AzTEC-3 and it is indicated as a vertical red line in each insert. 
}
\label{NBcandidates}%
\end{figure*}

In Fig. \ref{NBcandidates}, we show the spatial distribution of the ten Ly$\alpha$ detections on the RGB image corresponding to the MUSE mosaic. The inserts show their Ly$\alpha$ emission line in velocity space with respect to the systemic redshift of AzTEC-3. 
As we can see in the inserts, the ranges in spectral pixels of the CubEx detections are large for a few sources. In particular, for the mosaic\_1548 source, CubEx identified 12 spectral pixels which correspond to a line with a tail.  

\begin{figure*}
 \centering
\includegraphics[width=9cm]{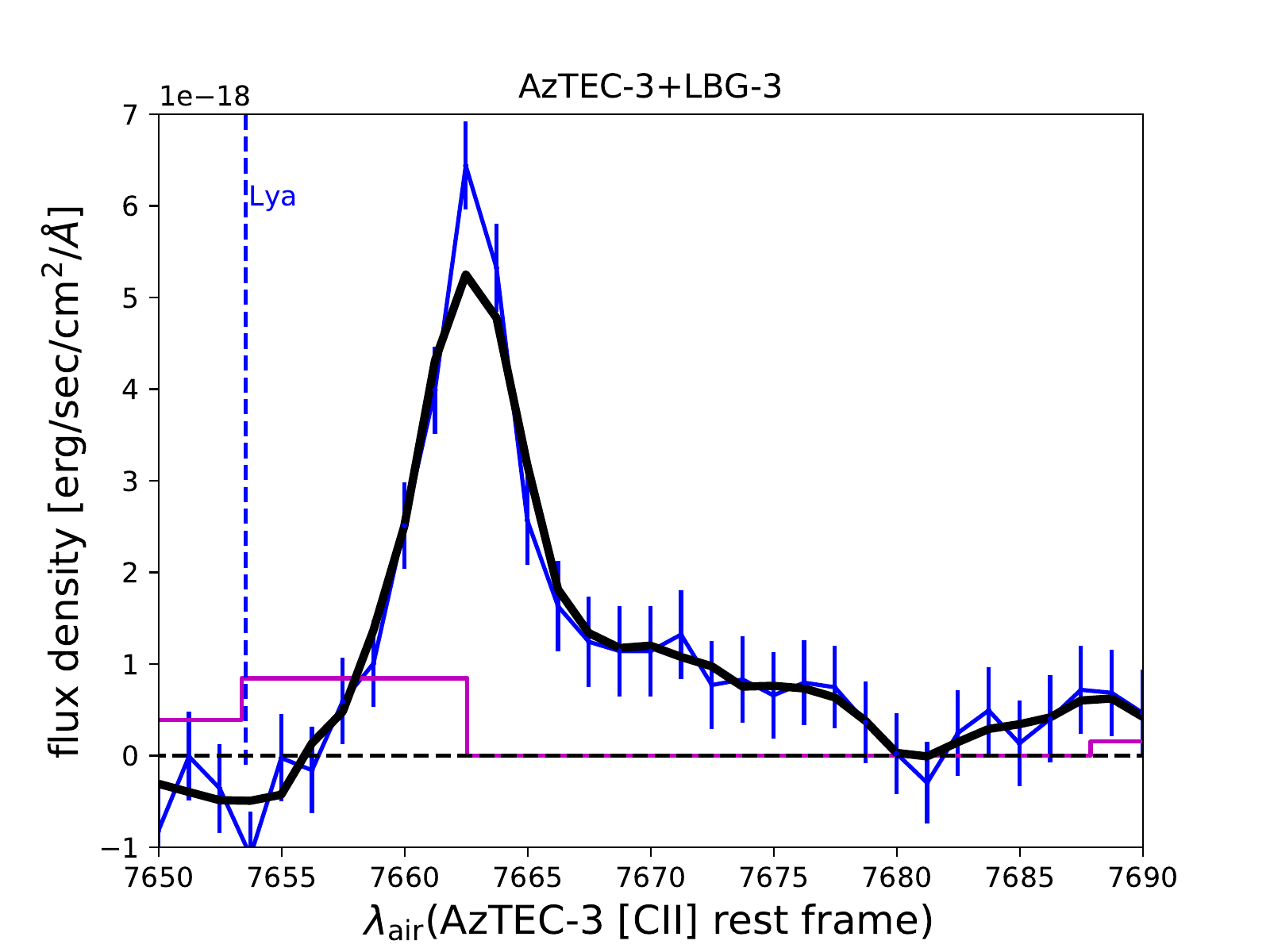} 
\includegraphics[width=9cm]{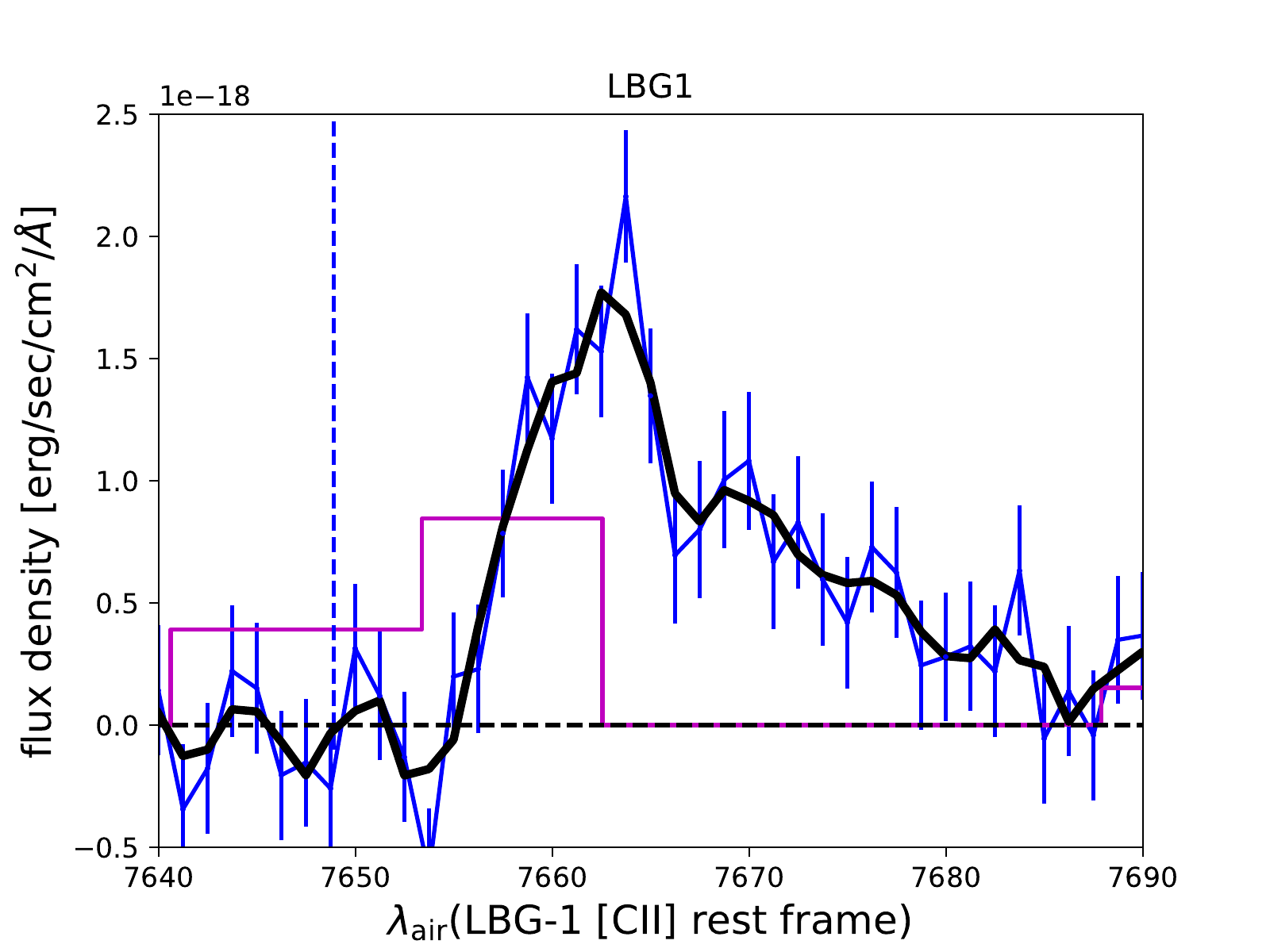} 
\caption{$Left ~panel:$
1D spectrum of the AzTEC-3+LBG-3 system in the Ly$\alpha$ wavelength range. The original-sampling spectrum is shown in blue with error bars, the one smoothed by $3\times3$ wavelength channels in black. 
The error spectrum comes from the rescaled variance cube.
The magenta spectrum is the theoretical sky spectrum from \citet{rou2000} in arbitrary units. 
$Right ~panels:$ same as the left panel for LBG-1.
}
\label{integratedspectrumzoom}%
\end{figure*}

In Fig. \ref{integratedspectrumzoom}, we show the 1D MUSE spectra extracted with CubEx from an area corresponding to the Ly$\alpha$ detection toward AzTEC-3 and LBG-1. These are the only sources in the field with a known systemic redshift from [C${\sc II}$] \citep{Riechers2014}. 
In the next section and in the appendix, we show the optimally extracted narrow-band images of all the ten Ly$\alpha$ detections. 


\subsection{Sources at $z\simeq5.3$ with $HST$ counterparts}
\label{methodHST}

To find additional Ly$\alpha$ emitting sources that might have escaped our detection threshold, we also selected sources with photometric redshifts consistent with that of AzTEC-3 ($4.9<z_{phot}<5.6$ which is $z=5.3\pm 3 \sigma$ given the typical $z_{phot}$ uncertainty at $3<z<6$) in the COSMOS2015 catalog 
and extracted their MUSE spectra 
 %
within a fixed aperture of 0.7$''$ radius (about twice the mosaic PSF size). Five sources in addition to the ones mentioned in the previous section did not show any Ly$\alpha$ emission in the MUSE spectra.

\begin{figure*}
 \centering
\includegraphics[width=10cm]{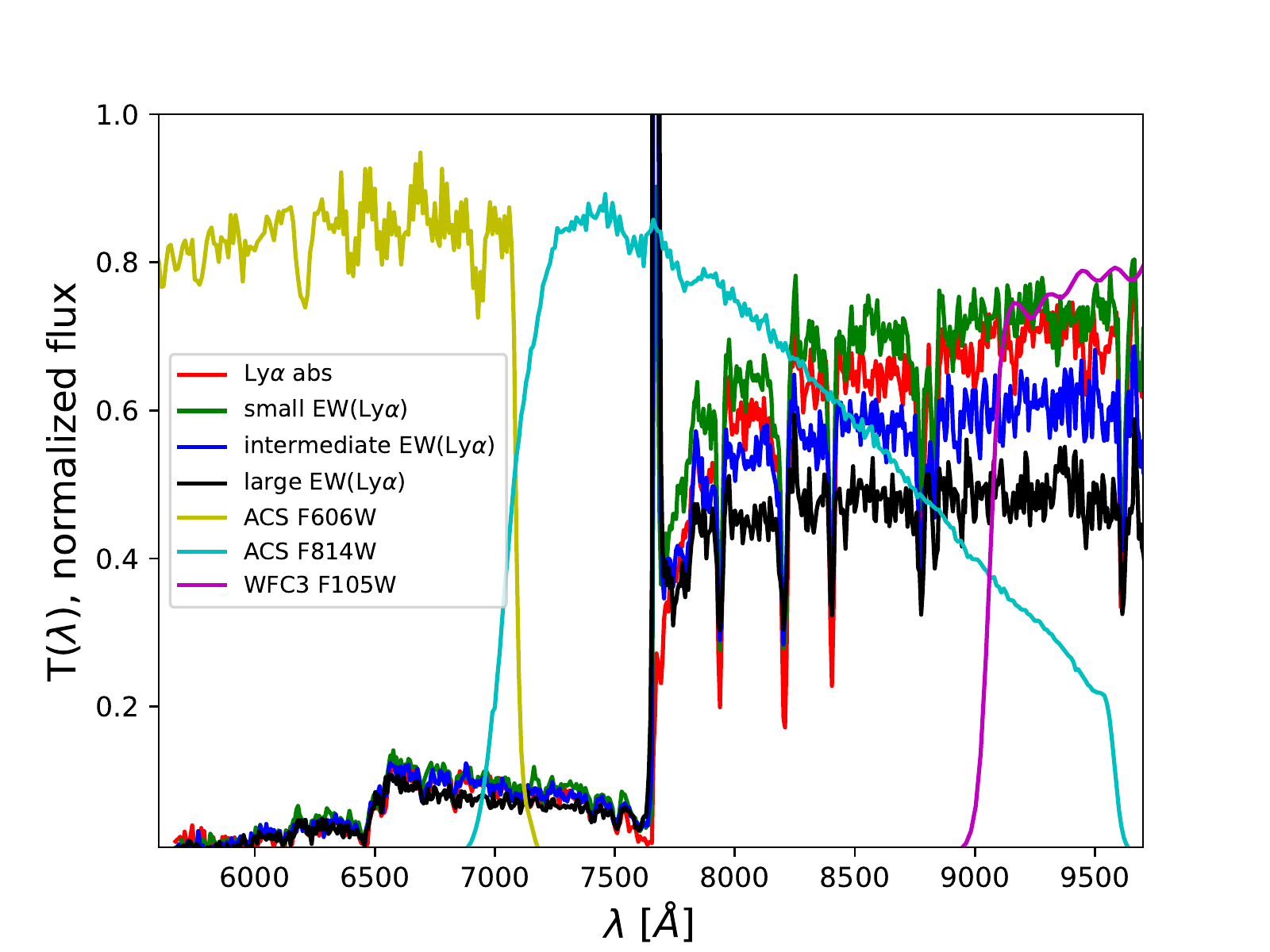}
\caption{$HST$ filter transmission curves ($ACS$ F606W in yellow, $ACS$ F814W in cyan, $WFC3$ F105W in magenta) and representative spectra of star-forming galaxies at $z\sim5.3$ \citep[Ly$\alpha$ absorber in red, Ly$\alpha$ emitter with a small, intermediate, and large EW(Ly$\alpha$) in green, blue, black, respectively from][]{Shapley:2003}. 
We corrected all the representative spectra for the intergalactic-medium absorption at $z=5.3$ by assuming the formalism in \citet{Madau:1995}, before using them.
}
\label{HSTfilters}%
\end{figure*}

We also investigated a method to identify a $z\sim5$ star-forming galaxy from the $HST$ F606W, F814W, F105W, F125W, and F160W images. 
%
As we can see in Fig. \ref{HSTfilters}, such a source is characterized by a sharp decrement between the F606W and F814W filters. We considered representative spectra of star-forming galaxies with Ly$\alpha$ in absorption and in emission \citep{Shapley:2003} and we convolved them with the $HST$ filter transmission curves. We found that a star-forming galaxy at $z\sim5.3$ is characterized by a V$_{606}$--I$_{814} $ color of about 3, given the non detection in F606W, and a flat spectrum redder than the F814W filter. 
We ran Source Extractor in dual mode, with the F105W image as a detection and the other HST-filter images as measurement images.
We chose a detection minarea equal to the PSF, a detection threshold of 2, and an optimal extraction aperture given the image PSF. 
Also, we asked for a signal-to-noise ratio of 3 in the detection flux. We found one source with color consistent with a $z\sim5.3$, that is also contained in the COSMOS2015 catalog with $z_{phot}=5.2$ (COSMOS2015\_842471). However, the MUSE spectrum does not show any significant emission line at the wavelength of Ly$\alpha$.

The best photometric filter to identify the members of the AzTEC-3 protocluster would be centered at $\sim$7600 {\AA} (the wavelength of Ly$\alpha$ at $z\sim5.3$). 
\citet{Sobral2018} studied the galaxies detected in the I767 filter and compiled a sample of Ly$\alpha$ emitters at $z\simeq5.3$. However, all the sources in their sample are located outside our MUSE coverage. 

Along the same line, we extracted the MUSE spectra of the galaxies discovered by \citet{Capak2011} in the AzTEC-3 overdensity. The ones showing Ly$\alpha$ emission at $z=5.3$ in our MUSE datacube are LBG1447524 and LBG1447526 (LBG-1 and LBG-3). 

\subsection{Other catalogs from the literature}
\label{other}

We also inspected other catalogs from the literature covering the AzTEC-3 area, the catalog of CO detections from \citet{Pavesi2018} and the radio-continuum detection catalog from \citet{Smolcic2017} and \citet{VanderVlugt2021}. 
We required non detection in the F606W image and we found only one source with Ly$\alpha$ at $z\sim5.3$ in both catalogs, corresponding to the AzTEC-3 galaxy.  In the Pavesi's catalog there is a source at $z=5.3$ (in addition to AzTEC-3 ) without Ly$\alpha$ emission from our MUSE observation that is shown in Fig. \ref{3Ddistr} (see next section).


\subsection{Space distribution of all the candidates at $z\simeq5.3$}
\label{location}

\begin{figure*}
 \centering
\includegraphics[width=10cm]{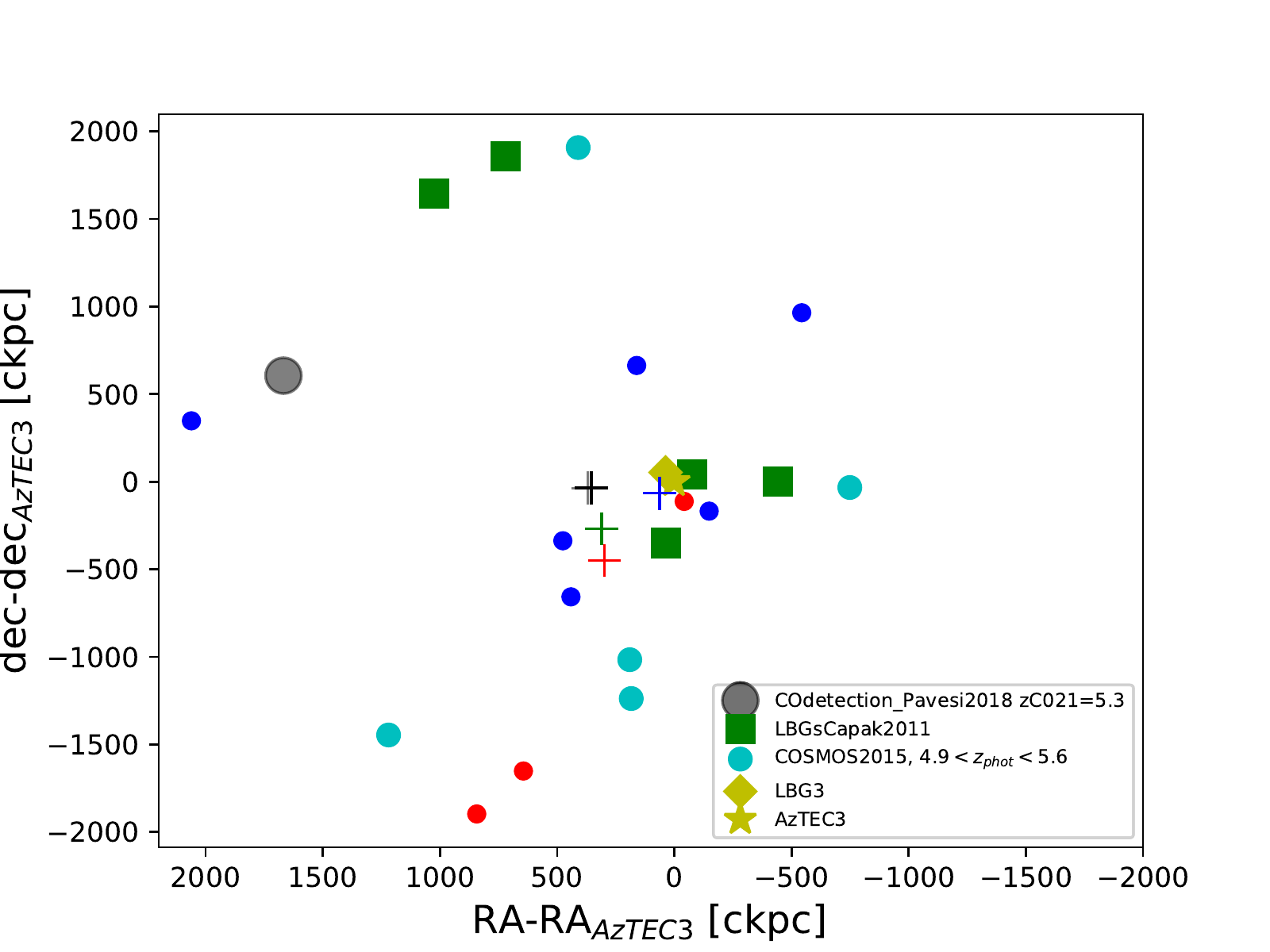} 
\caption{Location of the Ly$\alpha$ emitting sources detected in the AzTEC-3 environment. The ones with Ly$\alpha$ central wavelength larger (red) and smaller (blue) than that of the AzTEC-3 system are shown as small circles. 
The AzTEC-3 submillimeter and the LBG-3 galaxies are shown as yellow star and yellow diamond. Sources from the literature with photometric redshifts consistent with $z\sim5.3$ are also shown, five COSMOS2015 sources with $4.9<z_{phot}<5.6$ (cyan big circles), Lyman break galaxies listed in \citet{Capak2011}, but without Ly$\alpha$ emission in our MUSE datacube (green squares), one CO detection from \citet{Pavesi2018} at $z=5.3$ 
 (gray circle). 
The black cross is the center of coordinates of all the sources shown in the figure and the gray cross the center obtained without considering AzTEC-3. The blue is the center of coordinates only of the sources with stellar mass larger than $10^9$ M$_{\odot}$ and the red cross is the weighted mass barycenter of the Ly$\alpha$ emitting sources detected in the field. The green cross is the weighted mass barycenter of the sources with a stellar mass estimation but without considering AzTEC-3.
}
\label{3Ddistr}%
\end{figure*}

In Fig. \ref{3Ddistr}, we show the location in the RA--dec plane of the Ly$\alpha$ emitting sources detected in the MUSE datacube and of the sources from the literature with photometric redshift consistent with being at $z\simeq5.3$. 
To verify if the group of sources around AzTEC-3 constitutes an overdensity of galaxies and confirms the presence of a protocluster, we compare them with the catalog of Ly$\alpha$ emitting sources detected with MUSE in the Hubble Ultra Deep Field \citep[][mosaic datacube]{Inami2017}.  For a minimum Ly$\alpha$ luminosity of $4\times10^{41}$ erg sec$^{-1}$, 
there are 77 galaxies at $4.9< z<5.6$ (redshift bin of 0.7) in the mosaic area of $3'x3'$ (30800 sources per deg$^{2}$), which means 2200 sources per deg$^{2}$ in the redshift range corresponding to the peak of our Ly$\alpha$ line detections ($5.28<z<5.33$, redshift bin of 0.05). At the same limiting Ly$\alpha$ luminosity, we detected eight emitters in the Main sample, taking AzTEC-3 and LBG-3 as two separated galaxies, and 11 if we also consider the Supplemental sample. Therefore, we found $9\pm2$ emitters in our field (=$4 \times 10^{-4}$ deg$^2$), that can be translated into $9\pm3$ emitters if we take also into account the 40\% chance of fake detections due to the fidelity-cut criterion in the uncertainty budget or $22500\pm7500$ sources per deg$^{2}$. This implies an overdensity of $10\pm3$. 
We, also, considered all the sources at $4.9<z_{phot}<5.6$ in the COSMOS2015 catalog located outside the area covered by our MUSE observations and found a density of $1862\pm300$ galaxies per deg$^2$. The overdensity of Ly$\alpha$ emitters in the MUSE area is $12\pm5$ times the density of all the sources in the COSMOS2015 catalog, with and without Ly$\alpha$ emission.
These estimations confirm the presence of an overdense region around AzTEC-3. 

For comparison, 
the SSA22 protocluster \citep{Steidel2000}, which also contains a few SMGs, is an overdensity of about six at $z\sim3.1$.  
The overdensity of Ly$\alpha$ emitters discovered by \citet{Shi2019} contains about four times more galaxies than the field.
In the Hubble Deep Field, \citet{Walter2012, Calvi2021} found an overdensity around the submillimeter galaxy HDF850.1 at $z = 5.183$. They presented 23 spectroscopically confirmed star-forming galaxies, including Ly$\alpha$ emitters, at $z=5.2$ 
with a surface density of at least a factor of two higher than the field. None of the sources, in addition to HDF850.1, are detected at submillimeter wavelengths.
\citet{Pavesi2018b} discovered an overdensity at $z\sim5.7$ similar to that around AzTEC-3. At the center of the overdensity there is a dusty starburst, probably in the phase of an on-going merger. A star-forming galaxy as massive as LBG-1 is located at 13$''$ away from the starburst. An overdensity of Ly$\alpha$ emitters was discovered in the same area with an overdensity parameter of about ten. This and our area of observations are therefore ones of the regions at $z>5$, characterized by the highest overdensities in the COSMOS field. 

Other overdensities of Ly$\alpha$ emitters have been discovered in the literature at $4<z<6$ around radio galaxies or around quasars expected to trace dense regions of the Universe.  
For instance, \citet{Venemans2002} discovered that the region around the luminous radio galaxy TN J1338-1942 at $z=4.1$ is 15 times more dense than the field; \citet{Venemans2004} found a density up to six times higher than in the field around the radio galaxy TN J0924-2201 at $z = 5.2$; \citet{Zheng2006} found an overdensity on the order of six around the radio-loud quasar SDSS J0836+0054 at $z=5.8$. 

Assuming that all the galaxies shown in Fig. \ref{3Ddistr} belong to the protocluster, we can see that they follow the location of our Ly$\alpha$ emitting sources. 
The center of coordinates of the most massive sources with stellar mass larger than 10$^{9}$ M$_{\odot}$ and the weighted mass barycenter of the sources with a stellar mass estimation, even without considering AzTEC-3, are close to the submillimeter galaxy on the side of its Ly$\alpha$ emission peak. This implies that the region of the protocluster where the gravitational potential is the strongest is close to the SMG, which in turn is very close to the center of the protocluster.

\section{Spectroscopic properties from the MUSE spectra}

\subsection{AzTEC-3}
\label{AzTEC3MUSE}

\begin{figure*}
 \centering
\includegraphics[width=15cm]{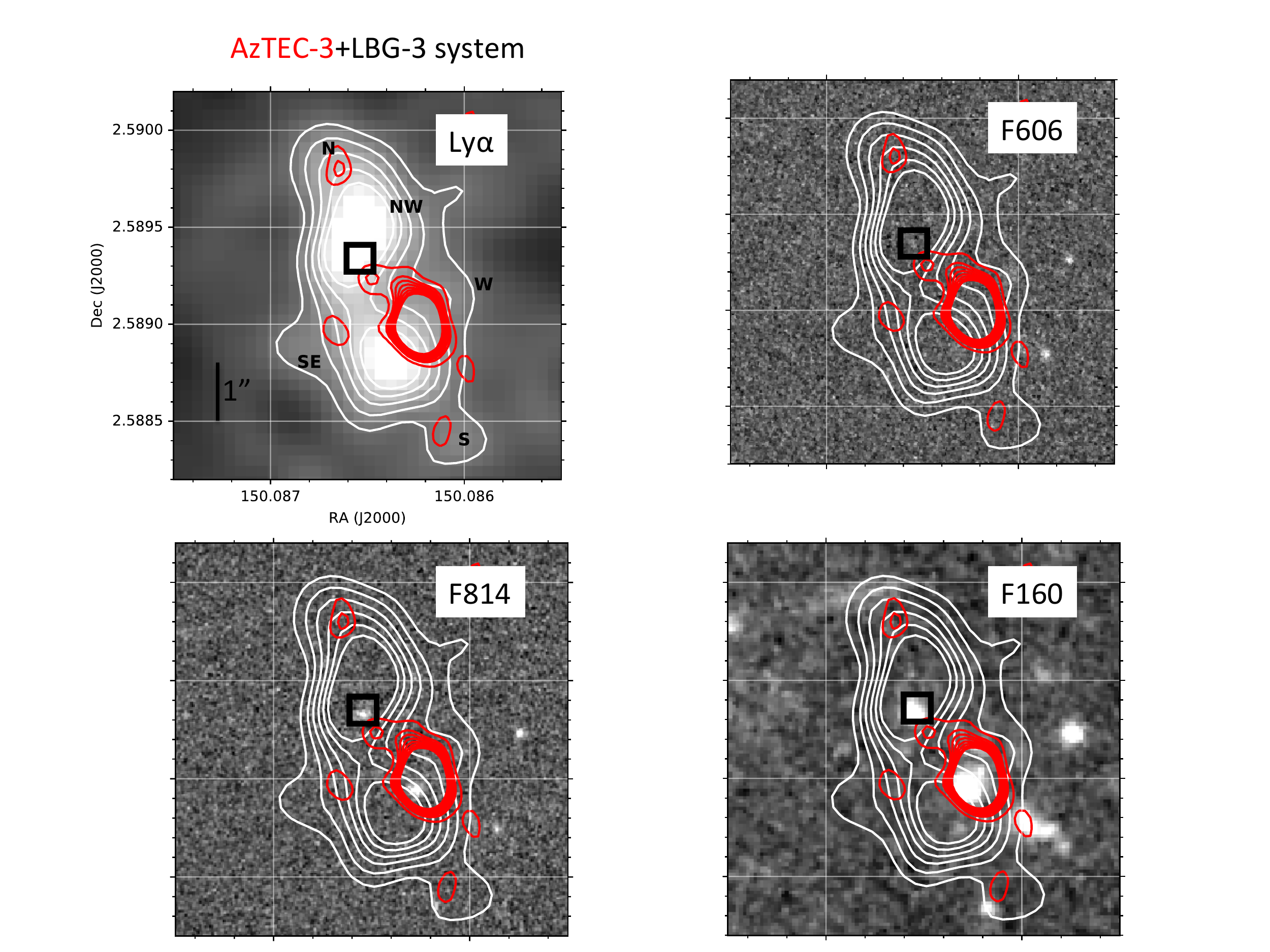} 
\caption{
$Upper ~left:$ Signal-to-noise optimally extracted narrow-band image \citep{Borisova2016} of the continuum-subtracted Ly$\alpha$ emission toward the AzTEC-3+LBG-3 system, rescaled to surface brightness units, and smoothed using a 2-pixel Gaussian kernel. 
The narrow band is built considering the 13 wavelength channels of the detection. The 5 white contours reveal the 1.3 to 3.8 erg sec$^{-1}$ cm$^{-2}$ arcsec$^{-2}$ levels, which correspond to 3-10$\sigma$ in this image.
The black letters indicate the directions we refer to in the text (N=north, NW=northwest, S=south, SE=southeast, W=west of the SMG). The small black segment indicates the 1" scale.
We also show the continuum-subtracted Ly$\alpha$ surface brightness contours overplotted to the $HST$ ACS F606W ($upper ~right$), F814W ($lower ~left$),  
and F160W ($bottom ~right$) images. The 0.5$''$-side square indicates the center of the F160W position of LBG-3. The PSF of our MUSE data is $0.7''$, corresponding to about 4 kpc or 27 ckpc at $z=5.3$.
The eight red contours correspond to the [C${\sc II}$] emission (from 3$\sigma$ to 10$\sigma$ where 1$\sigma=2.38  \times 10^{-4}$ Jy beam$^{-1}$) 
at 301.671-302.142 GHz, corresponding to -230 to 230 km sec$^{-1}$ from the peak emission. The synthesized beam of the [C${\sc II}$]  ALMA observations is $0.63''\times 0.56''$ as reported in \citet{Riechers2014}.
The 3$\sigma$ contour extends toward LBG-3.}
\label{AzTEC3optimalNBonallHST}%
\end{figure*}

In Fig. \ref{AzTEC3optimalNBonallHST}, we show the continuum subtracted Ly$\alpha$ surface brightness toward AzTEC-3 corresponding to 
7658--7670 {\AA}. 
The original narrow band image and the continuum subtracted one are produced within CubEx optimizing the signal-to-noise ratio, as described in \citet{Borisova2016}.
The two main peaks of the emission roughly correspond to the positions of AzTEC-3 and the LBG-3. 
The emissions toward AzTEC-3 and LBG-3 are blended at the 5$\sigma$ level. The entire emission occupies an area of $4.4''\times6.3''$ (about $27\times38$ kpc$^2$ and $170\times240$ ckpc$^2$).

As expected for a source at $z\sim5$, the sources are not visible in the F606W filter and the Ly$\alpha$ wavelength is contained in the F814W filter. 
The peak of the Ly$\alpha$ emission toward the SMG is offset toward the southeast with respect to the star-formation knots seen in the $HST$ images and traced by the [C${\sc II}$] detection. This may be related to the presence of dust on the main knot of star formation of the SMG (see next section). 
The LBG peak is offset by 0.5$''$ (20 ckpc) with respect to the source seen in the $HST$ images and elongated to the north. Since the Ly$\alpha$ emission is very sensitive to the even small presence of dust, this is in agreement with the idea that on the main star-formation knot of LBG-3 some of the Ly$\alpha$ photons could be absorbed by dust (see next Sect.). 

%
Even if the Ly$\alpha$ emissions toward the SMG and the LBG seem connected, 
 the spatial shape of the highest S/N contours indicates that the peak Ly$\alpha$ emission of the SMG is quite symmetrical. However, that of the LBG seems to be elongated and tilted toward the north (N position in the figure) and extends up to 2$''$ (almost 80 ckpc). Also, the Ly$\alpha$ emission of the SMG shows extended components toward various directions. 
It is worth noting that a low S/N emission from the SMG toward the direction of LBG-3 was also observed in the [C${\sc II}$] observations, despite the main [C${\sc II}$] flux was found to be concentrated in a compact position on top of the main knot of star formation, as described in \citet{Riechers2014}.  

\begin{figure*}
 \centering
\includegraphics[width=14cm]{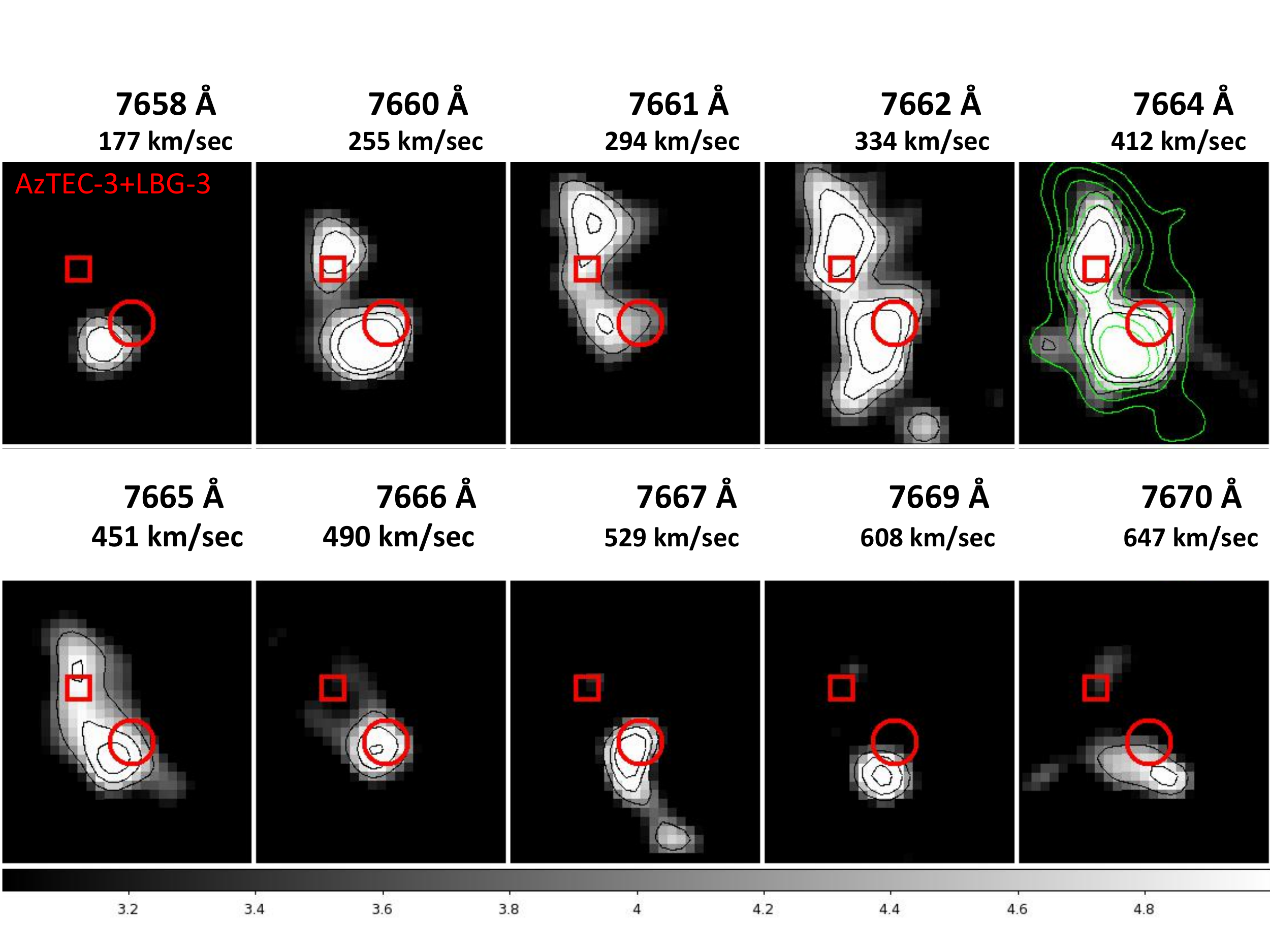} 
\caption{
Ly$\alpha$ channel maps toward the AzTEC-3+LBG-3 system. Each panel corresponds to the indicated wavelength 
and velocities with respect to the systemic redshift inferred from the [C${\sc II}$] emission line. The white contours represent the isophotes at S/N = 4, 5, and 6. The red square corresponds to the F160W position of LBG-3 and the red circle to the location of the peak of the [C${\sc II}$] emission of AzTEC-3. The green contours in the fifth panel represent the contours of the continuum-subtracted Ly$\alpha$ surface brightness. The horizontal gray-color bar shows a signal-to-noise ratio from 3 to 5. We refer to Fig. \ref{AzTEC3optimalNBonallHST} for the 1" scale.}
\label{AzTEC3channels}%
\end{figure*}

In Fig. \ref{AzTEC3channels}, we show the Ly$\alpha$ channel maps of the AzTEC-3+LBG-3 system. Each channel is separated by approximately 1.5 {\AA}. 
The figure shows that the Ly$\alpha$ emissions toward the two galaxies peak at similar wavelengths, 7662 {\AA} for LBG-3 and 7664 {\AA} for AzTEC-3, respectively. 
The Ly$\alpha$ emission associated to the SMG is visible in all the channels at 4$\sigma$, while that probably associated to LBG-3 is significant at 7660--7665 {\AA}.
At 7665 {\AA}, the emission is mainly concentrated on top of the SMG, but shows a kind of a $bridge$ between SMG and LBG, which could represent a region of interaction between the two close-by galaxies (see Sect. 6). The Ly$\alpha$ emission of the bridge is observed from 7660 to 7666 {\AA}, with a maximum at 7664 {\AA}. It occupies a wide range in Ly$\alpha$ wavelengths, that could correspond to a region of a large range of gas velocities, together with large HI column densities. 
These gas characteristics could be the result of the interactions in the AzTEC-3+LBG-3 system. 

At the position of the main star-forming knot of the SMG and of the peak of the Ly$\alpha$ emission associated to the SMG, the Ly$\alpha$ emission extends from 200 to 600 km sec$^{-1}$. The velocity 
could be associated to a star-formation driven outflow of neutral gas, departing from the SMG and mixing with the interacting gas in the bridge between SMG and LBG. This interaction could produce some of the extended tail seen also in the Ly$\alpha$ 1D spectrum. The blue shift of low-ionization absorption lines, directly tracing the kinematics of the HI gas, together with the radiative-transfer modeling of the Ly$\alpha$ emission (Sect. 6), sensitive to kinematics and amount of HI gas, could provide support of this hypothesis. 

 At 7662 {\AA}, the Ly$\alpha$ emission appears elongated to the north of LBG-3 (N position in Fig. \ref{AzTEC3optimalNBonallHST}) and to the southeast of AzTEC-3 (SE position in Fig. \ref{AzTEC3optimalNBonallHST}). 
Above LBG-3, the Ly$\alpha$ emission is compact in wavelength. 
The Ly$\alpha$ emission is compact in wavelength (around 300 km sec$^{-1}$) also in the region below the peak of the Ly$\alpha$ emission associated to the SMG. In this position, the gas kinematics could be 
associated with neutral gas flows between the SMG and the mosaic\_199 source (Table \ref{NBcandidates}).

Since Ly$\alpha$ photons can be scattered by HI gas and so reveal its presence, the elongated shape of the Ly$\alpha$ emission to the north of LBG-3 (N position in Fig. \ref{AzTEC3optimalNBonallHST}), in the bridge between SMG and LBG, and to the southeast of the SMG (SE position in Fig. \ref{AzTEC3optimalNBonallHST}) could be the result of the interaction between AzTEC-3, LBG-3, and possibly mosaic\_199. As shown in the simulations described in \citet{Yajima2013}, at the first passage of a merging event between star-forming galaxies, the Ly$\alpha$ emission coming from the progenitors can be intense due to the triggered star-formation event and its shape could also follow the gas distorted during the interaction in the outer regions.

In Fig. \ref{integratedspectrumzoom}, we can see the integrated 1D spectrum of the AzTEC-3+LBG-3 system in the Ly$\alpha$ 
wavelength ranges, 
free of strong sky line residuals.
Neither metal nor AGN-diagnostic lines are detected in the spectrum. 
As expected from Fig. \ref{AzTEC3channels}, the main peak of the Ly$\alpha$ line is redshifted with respect to the systemic redshift by about 400 km sec$^{-1}$. The entire line including its red tail occupies a velocity range up to 900 km sec$^{-1}$. As shown in \citet[][]{Verhamme2006}, the shift of the Ly$\alpha$ red peak can be 2 or 3 times the velocity of the outflowing gas depending on the HI column density conditions. 
\citet{Riechers2014} measured that the central velocity of the OH163$\mu$m doublet from the ALMA observations was blueshifted of about 100 km sec$^{-1}$ with respect to the [C${\sc II}$] line and this could be associated to an outflow, maybe produced by the AzTEC-3 starburst.

The integrated flux of the Ly$\alpha$ emission of the entire system is $(16.92\pm0.37) \times 10^{-18}$ sec$^{-1}$ cm$^{-2}$. 
By estimating a continuum on the order of $3 \times 10^{-19}$ erg sec$^{-1}$ cm$^{-2}$ {\AA}$^{-1}$ on the red side of Ly$\alpha$ by linear fit, we calculate a rest-frame EW(Ly$\alpha$) = 13$\pm$2  {\AA}. The FWHM(Ly$\alpha$) of the main peak excluding the long tail is 220 km sec$^{-1}$. 

\begin{figure*}
 \centering
\includegraphics[width=9cm]{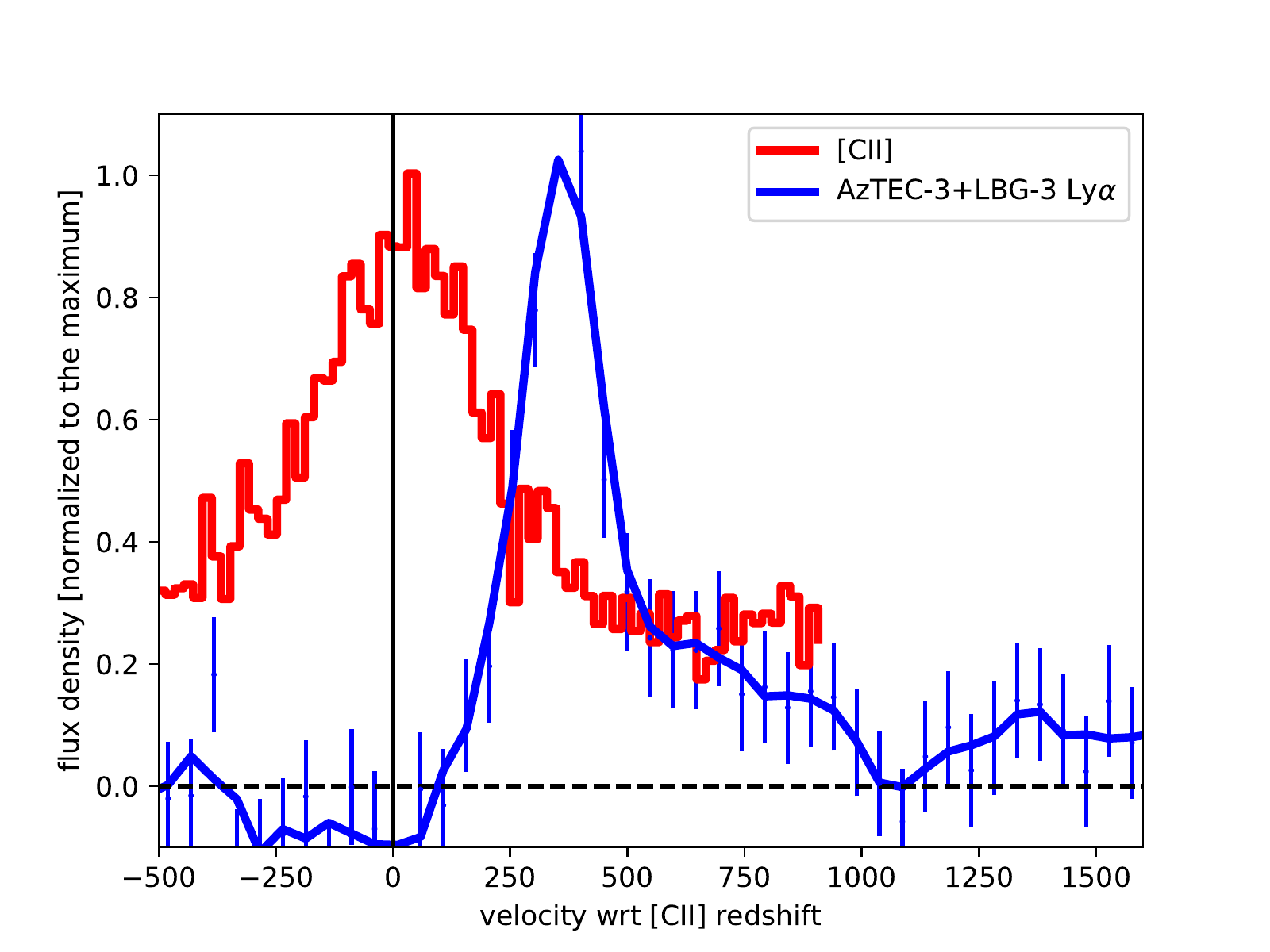}
\includegraphics[width=9cm]{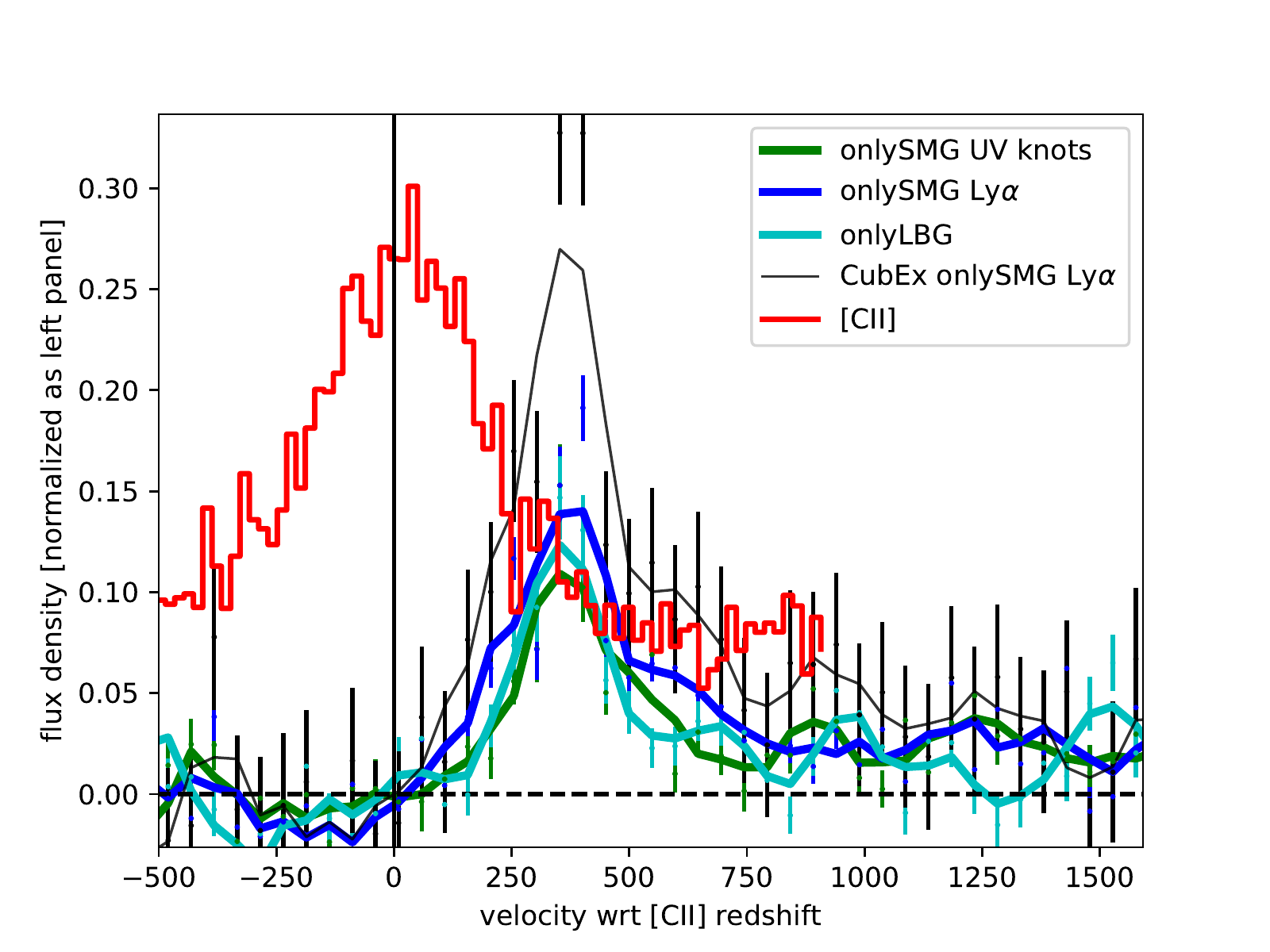}
\caption{$Left ~panel:$ Ly$\alpha$ (blue) and [C${\sc II}$] (red) profiles of the AzTEC-3+LBG-3 system in velocity space and normalized to the maximum.  The [C${\sc II}$] spectrum is taken from \citet{Riechers2014} and it is associated to AzTEC-3 (see Fig. \ref{AzTEC3optimalNBonallHST}). $Right ~panel:$ normalized [C${\sc II}$] profile (red), Ly$\alpha$ profile of the emission extracted at the position of the main Ly$\alpha$ peak associated to the SMG (blue and black curves), extracted from the position of the rest-frame UV knots of AzTEC-3 as seen in the F160W image (green), and extracted at the position of the main Ly$\alpha$ peak associated to LBG-3 (cyan). The extraction is done in an aperture of 0.7$''$ radius for the blue, green, and cyan profiles and increasing the S/N of the connected pixels in CubEx for the black profile. For simplicity, we call the spectra as onlySMG Ly$\alpha$, onlySMG UV knots, and onlyLBG. The normalization of these spectra is performed with the same factor used to normalize the spectrum of the AzTEC-3+LBG-3 system in the left panel to show the different intensities. The normalization of the [C${\sc II}$] profile is also performed accordingly.}
\label{AzTEC3CIIMGLBGLya}%
\end{figure*}

In Fig. \ref{AzTEC3CIIMGLBGLya}, we show the spectroscopic comparison between [C${\sc II}$] and Ly$\alpha$ emission in velocity space. 
 It is worth remembering that the Ly$\alpha$ emission contains contributions from AzTEC-3 and also from LBG-3, while the [C${\sc II}$] emission is only concentrated at the location of the SMG as seen in the $HST$ images. The zero velocity is given by the [C${\sc II}$] systemic redshift. 
Both AzTEC-3 and LBG-3 contribute to the main peak and to the extended tail (at more than 500 km sec$^{-1}$) of the integrated Ly$\alpha$ emission, even if the tail is mostly dominated by the SMG. 

We separated the 1D Ly$\alpha$ spectrum of the SMG and of LBG-3 in two ways, with CubEx and extracting a fixed-aperture spectrum from the MUSE datacube.
To make CubEx detect two separated (not blended) sources, we increased the S/N of the connected pixels from 3 to 5, the spatial and spectral S/N from 3 to 7, and decreased the minimum number of voxels from 30 to 5 in the detection parameters. This way CubEx is able to detect the brightest Ly$\alpha$ emission associated only to the SMG and that associated only to the LBG-3. Also, 
we extracted MUSE spectra within a 0.7$''$-radius aperture and obtained the spectra at the position of the brightest Ly$\alpha$ regions only coming from the SMG (onlySMG Ly$\alpha$ spectrum) and only from the LBG (onlyLBG spectrum).  
The profile of the Ly$\alpha$ emission just coming from the SMG presents a main peak which is fainter and broader than the main peak of the emission coming from the entire AzTEC-3+LBG-3 system (left panel of Fig. \ref{AzTEC3CIIMGLBGLya}). 
The Ly$\alpha$ profile, obtained from an aperture including all the rest-frame UV knots of the SMG (onlySMG UV knots spectrum) shows a sharper slope on the red side. However, the overall shapes of the Ly$\alpha$ emissions are comparable. 
Even if we focus on the brightest part of the Ly$\alpha$ emission only associated to the SMG (as the [C${\sc II}$] emission is), we see that the Ly$\alpha$ main peak is redshifted with respect to the [C${\sc II}$] by the same amount as the entire AzTEC-3+LBG-3 system. 

It is worth noting that 
\citet{Riechers2014} estimated a star-formation rate per unit of area of 530 M$_{\odot}$ yr$^{-1}$ kpc$^2$ \citep[see also][for a discussion]{Riechers2020}, which is larger than the limit of 0.1 M$_{\odot}$ yr$^{-1}$ kpc$^{-2}$, satisfied by local starbursts and high-$z$ Lyman Break galaxies to sustain starburst-driven outflows \citep{Heckman2001}. 
Starburst-driven outflows could favor the escape of Ly$\alpha$ photons even from dusty interstellar media \citep[e.g.,][]{Kunth1998, Verhamme2006}, as that of the SMG, and they could play an important role in producing the observed Ly$\alpha$ surface brightness from AzTEC-3. 
At 34 ckpc (0.9$''$) from the star-forming region traced by the [C${\sc II}$] emission peak toward the southeast of the SMG (SE position in Fig. \ref{AzTEC3optimalNBonallHST}), we measure a Ly$\alpha$ surface brightness of $\sim4.4 \times 10^{-18}$ erg sec$^{-1}$ cm$^{-2}$ arcsec$^{-2}$, indicating a L(Ly$\alpha)\sim$ SB(Ly$\alpha) \times 4 \pi 0.9^2 \times 4\pi$D$_{L}^2(z=5.3) \sim 1.3\times 10^{43}$ erg sec$^{-1}$. 

It is interesting to investigate what could be the main mechanism of the production of the Ly$\alpha$ photons that extend up to at least 30 ckpc. The molecular gas mass of $5.7\times 10^{10}$ M$_{\odot}$ \citep{Riechers2020} and the SFR$_{FIR}$ of 1100 M$_{\odot}$ yr$^{-1}$ imply that the starburst could be maintained for a time scale on the order of 50 Myr, as an upper limit assuming that no significant gas mass is lost due to the outflow itself. An outflow of a constant velocity of 800 km sec$^{-1}$ would reach a maximum distance of 40 kpc in a time scale of 50 Myr and could have already reached part of that distance in a shorter time scale.  
This outflow could channel the escape of ionizing radiation, ionized, and neutral gas.

Moreover, the Ly$\alpha$ luminosity of $1.3\times 10^{43}$ erg sec$^{-1}$ at a distance of 34 ckpc could be explained in terms of recombination of atoms ionized by a starburst of a thousand solar masses per year. In fact, following equation 7 in \citet{Cantalupo2017}, the luminosity could be translated into a ionization luminosity of $\sim1 \times 10^{54}$ photons sec$^{-1}$ consistent with that of a starburst of 1000 M$_{\odot}$ yr$^{-1}$ \citep[e.g.,][]{Leitherer1999}. Also, following equation 5 from the same paper, it can be shown that, for a volume of 4/3$\pi$(34ckpc)$^3$, 
a T=$10^{4}$ K, and a filling factor of the interstellar medium clouds of $10^{-4}$ \citep[e.g.,][]{Cantalupo2014}, a cloud density of $\sim3$ atoms cm$^{-3}$ would imply the observed L(Ly$\alpha$) at 34 ckpc. At this density and temperature, recombination is a plausible scenario to explain the observed L(Ly$\alpha$) and surface brightness. 

The luminosity at 90 ckpc would imply a lower cloud density of less than 1 atom cm$^{-3}$ that difficultly explains a recombination mechanism from the starburst. However, at this larger distance the HI scattering (free of dust absorption) could play a role in the case of N${\sc HI}$ $>10^{21}$ atoms cm$^{-2}$. The radiative-transfer-model fit of the 1D integrated Ly$\alpha$ profile of the SMG and of the AzTEC-3+LBG-3 system can give insight on the N${\sc HI}$ quantity. The gas at the peak of the Ly$\alpha$ emission and at 90 ckpc from the star-forming region could be the result of the interaction between the SMG and the mosaic\_199 source (see below).

\subsection{LBG-1}
\label{LBG1MUSE}


Among the star-forming galaxies with Ly$\alpha$ in emission detected in the MUSE datacube, we focus here on LBG-1, because it is the only galaxy, apart from AzTEC-3, for which we have a systemic redshift, inferred from the [C${\sc II}$] emission line \citep{Riechers2014}. In the next section, we consider two additional Ly$\alpha$ detections with photometric redshifts from the COSMOS2015 catalog (Fig. \ref{NBcandidates}). 
 
\begin{figure*}
 \centering
\includegraphics[width=14cm]{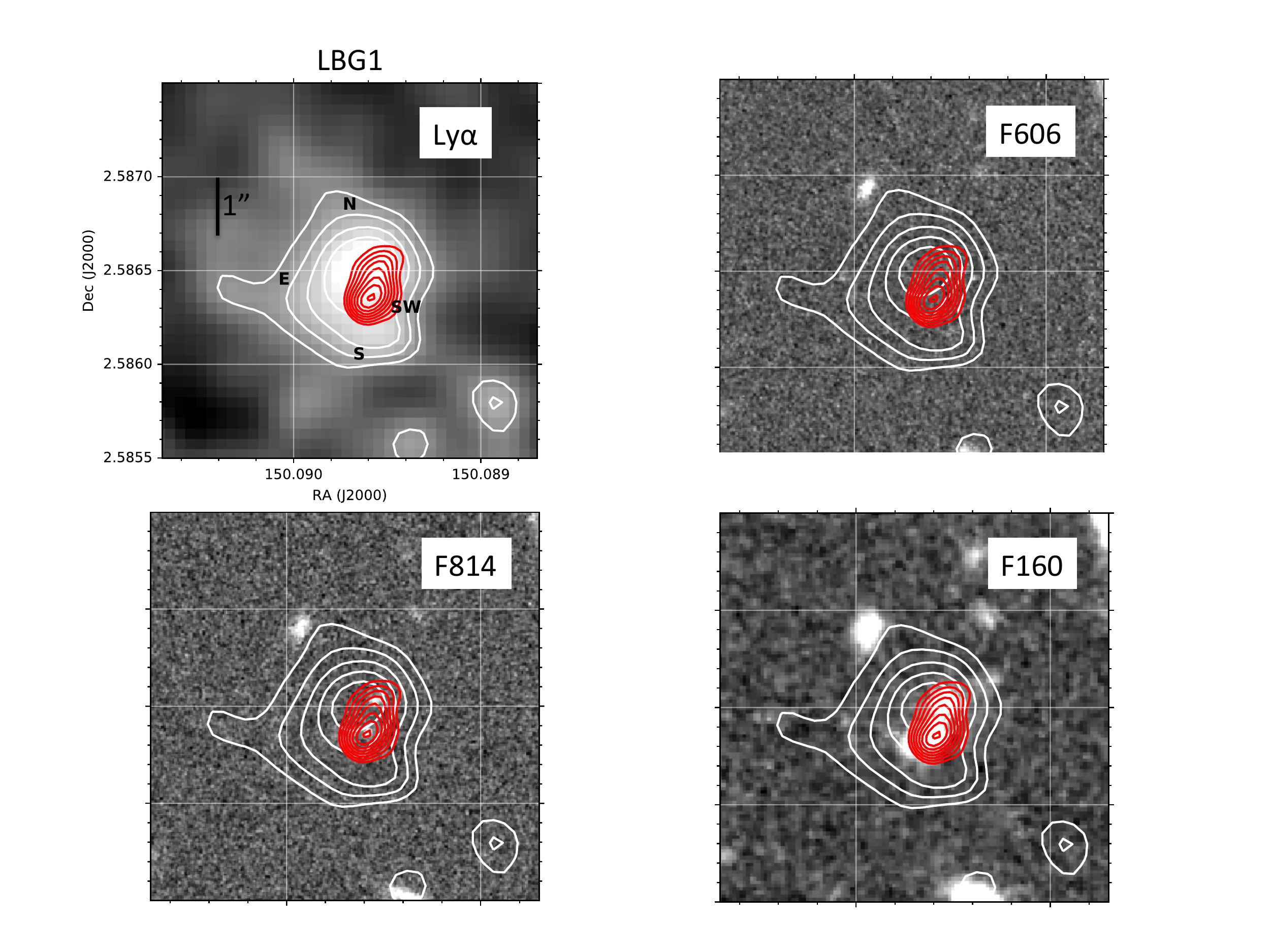} 
\caption{$Upper ~left$: Signal-to-noise optimally extracted narrow-band image \citep{Borisova2016} of the continuum-subtracted Ly$\alpha$ emission, rescaled to surface brightness units, and smoothed using a 2-pixel Gaussian kernel of LBG-1. The five white contours reveal the 0.8 to $2\times 10^{-18}$ erg sec$^{-1}$ cm$^{-2}$ arcsec$^{-2}$ levels, which correspond to 3 and 8$\sigma$ in this image. The letters indicate reference positions described in the text. 
The small black segment indicates the 1" scale.
We also show the Ly$\alpha$ surface brightness contours overplotted to the F606W ($upper ~right$), F814W ($lower ~left$) ACS, 
and F160W WFC3 ($lower ~right$) $HST$ images. 
The PSF of our MUSE data is $0.7''$, corresponding to about 4 kpc or 27 ckpc at $z=5.3$.
The eight red contours correspond to the 3-10$\sigma$ (1$\sigma=2.38 \times 10^{-4}$ Jy beam$^{-1}$) of the [C${\sc II}$] emission at 301.847-302.093 GHz, corresponding to -120 to 120 km sec$^{-1}$ from the peak emission.
The synthesized beam size of the ALMA [C${\sc II}$] observations is $0.63''\times0.56''$ as reported in \citet{Riechers2014}.
%
%
}
\label{LBG1optimalNBonallHST}%
\end{figure*}

\begin{figure*}
 \centering
\includegraphics[width=16cm]{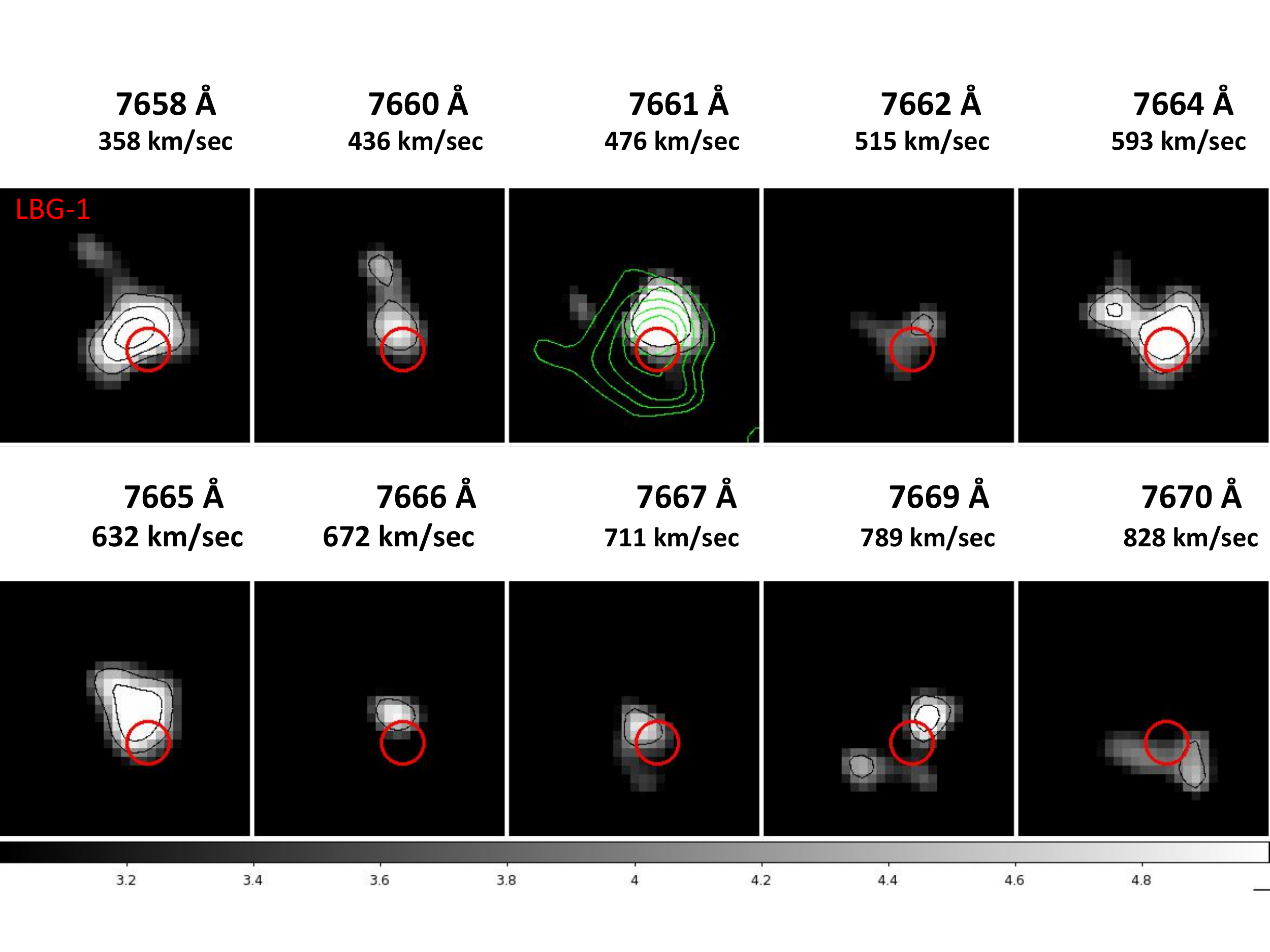} 
\caption{
Ly$\alpha$ channel map of LBG-1. Each panel corresponds to the indicated wavelength 
and velocities with respect to the systemic redshift inferred from the [C${\sc II}$] emission line. 
The white contours represent the isophotes at S/N = 4, 5, and 6. 
The red circle shows the F160W position of LBG-1 as in Fig. 1, 
drawn to drive the eye. 
The green contours in the third panel represent the contours of the continuum-subtracted Ly$\alpha$ surface brightness. The horizontal gray-color bar shows a signal-to-noise ratio from 3 to 5. We refer to Fig. \ref{LBG1optimalNBonallHST} for the 1" scale.
}
\label{LBG1channels}%
\end{figure*}

In Fig. \ref{LBG1optimalNBonallHST}, we show the emission detected by CubEx at a wavelength around 7661 {\AA}, corresponding to Ly$\alpha$ toward LBG-1. 
This galaxy was studied in \citet{Capak2011, Riechers2014, Pavesi2019}. Its known physical properties are reported in Table \ref{finallistMUSEdetectionsat53}.
Unlike the UV which clearly shows at least three components, the Ly$\alpha$ emission is smooth and encompasses the three UV components.
The integrated L(Ly$\alpha$) is equal to $(15.9\pm0.7) \times 10^{41}$ erg sec$^{-1}$  (($5.5\pm0.3) \times 10^{42}$ erg sec$^{-1}$ if corrected by dust extinction, following \citet{Calzetti2000} and A$_V=0.2$) and rest-frame EW(Ly$\alpha$)= $2.0\pm0.2$ {\AA}, by assuming the continuum magnitude (z\_p=23.71) from the COSMOS2015 database (the continuum is very noisy and affected by sky line residuals redwards of the Ly$\alpha$ wavelength in the MUSE spectrum).
The size of the entire Ly$\alpha$ emission is $3.2''\times3.2''$ (about $20\times20$ kpc$^2$ or $125\times125$ ckpc$^2$).
As shown in \citet{Riechers2014}, the [C${\sc II}$] emission at 301.974 GHz encompasses the three knots with a lower signal-to-noise tail toward the southwest (SW position in Fig. \ref{LBG1optimalNBonallHST}) as well. 
At MUSE resolution, we see that significant Ly$\alpha$ flux comes from the three components of LBG-1, maybe indicating that a low dust extinction is allowing Ly$\alpha$ photons to escape also close to the regions of star formation.
In Fig. \ref{LBG1channels}, we present the Ly$\alpha$ channel maps toward LBG-1.
The figure shows that the Ly$\alpha$ emission is concentrated on the northern region of LBG-1 
and that only 
at the longest wavelengths the emission comes only from the south (S and SW positions in Fig. \ref{LBG1optimalNBonallHST}). This could indicate that toward the south, LBG-1 is characterized by larger HI column densities and higher outflow velocities. 

\begin{figure*}
 \centering
\includegraphics[width=19cm]{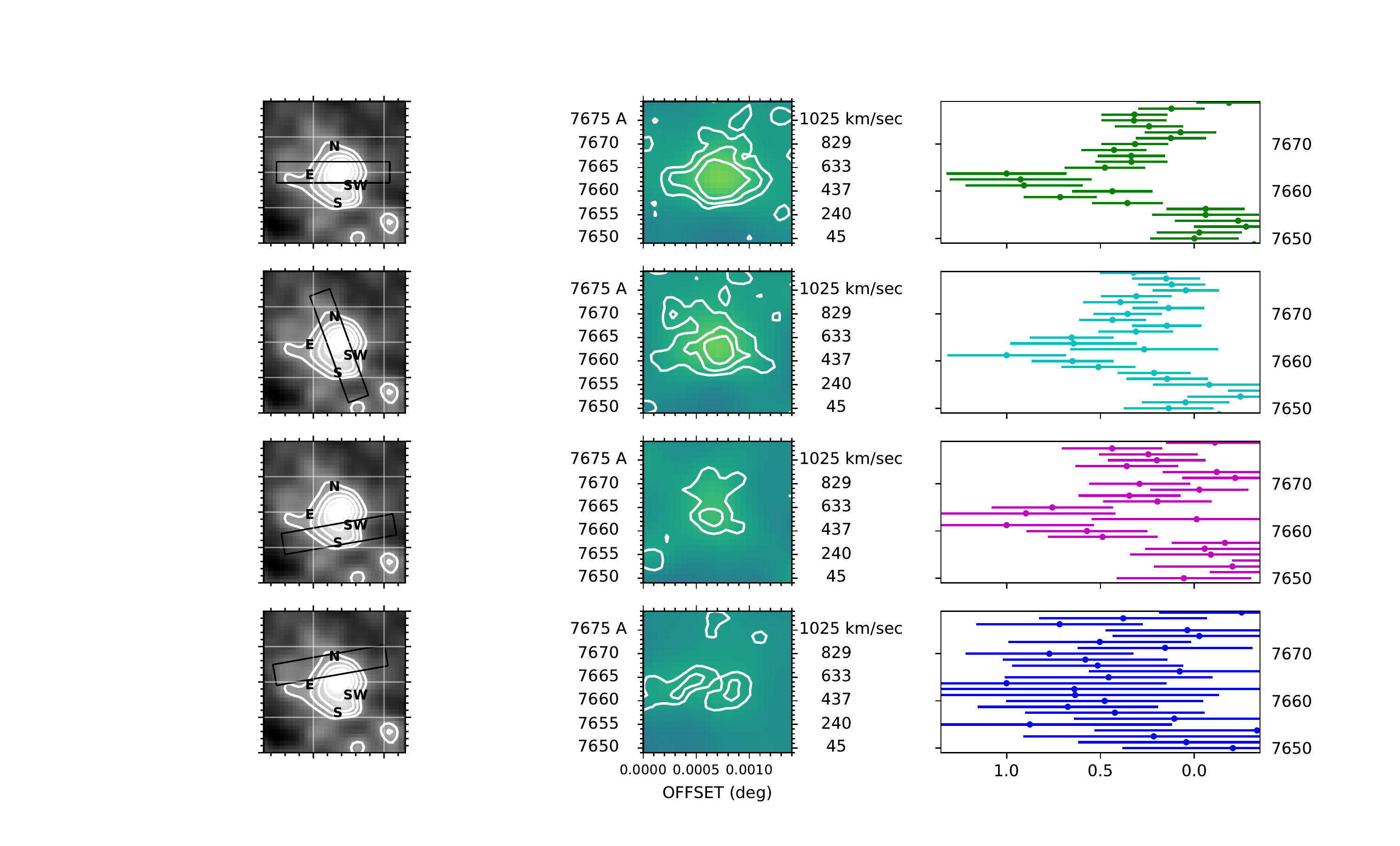} 
\caption{$Left ~panels:$ Pseudo slits oriented horizontally on the Ly$\alpha$ emission peak (first row), tilted by 110 degrees on the Ly$\alpha$ emission peak (second row), tilted by ten degrees below the Ly$\alpha$ emission peak (third row), and tilted by ten degrees above the Ly$\alpha$ emission peak (fourth row). $Middle ~panels:$ Position-velocity diagrams obtaining slicing the continuum-subtracted MUSE datacube with a pseudo slit oriented as shown in the $left ~panels$. 
In the left y-axis, we indicate the wavelength and in the right y-axis the velocity with respect to the systemic redshift inferred from the [C${\sc II}$] emission line, the x-axis corresponds to the position along the pseudo slit in degrees (from the east to the west, except in the second row where it is from the south to the north). The white contours correspond to the 1, 2, 3$\sigma$ level. 
$Right ~panels:$ 1D spectra corresponding to the position-velocity diagrams of the $middle ~panels$. Wavelength is on the y-axis in the same range as in the $middle ~panels$ and the flux normalized to the maximum is on the x-axis. The color coding of these spectra is like in Fig. \ref{LBG1LyaCII}.
}
\label{LBG1PV}%
\end{figure*}

To increase the S/N and investigate the origin of the Ly$\alpha$ emission coming from the different directions of LBG-1, we slice the MUSE cube along the pseudo slits shown in the left side of Fig. \ref{LBG1PV}, horizontal on the Ly$\alpha$ emission peak (first row), tilted by 110 degrees on the Ly$\alpha$ emission peak (second row), tilted by ten degrees below the Ly$\alpha$ emission peak (third row), and tilted by ten degrees above the Ly$\alpha$ emission peak (fourth row). 
These slit orientations are chosen to isolate the Ly$\alpha$ emission coming from the star-forming knots of LBG-1 to that coming from above and below of them.
We can see that at the position of the main Ly$\alpha$ peak, the emission extends from 7658 to 7672 {\AA}. The highest velocity regions come from the center toward the southwest (S and SW position in Fig. \ref{LBG1optimalNBonallHST}).
Toward the extreme southwest (SW position in Fig. \ref{LBG1optimalNBonallHST}), the Ly$\alpha$ emission reaches 7672 {\AA}, corresponding to 400 km sec$^{-1}$ with respect to the main peak wavelength. This value can be related to the gas kinematics, in a wrap of the material of the three knots and/or to high HI column densities (see Sect. 6). The kinematics of the gas could be related to the interplay of the gas in the merger of the three knots, as suggested by \citet[][]{Riechers2014}, the high column density could suggest that the merger is "wet" and/or there is a large gas reservoir supplying the star formation in the entire region.
%
By slicing above the main peak, the Ly$\alpha$ emission is confined in wavelength.

\begin{figure*}
 \centering
\includegraphics[width=7.8cm]{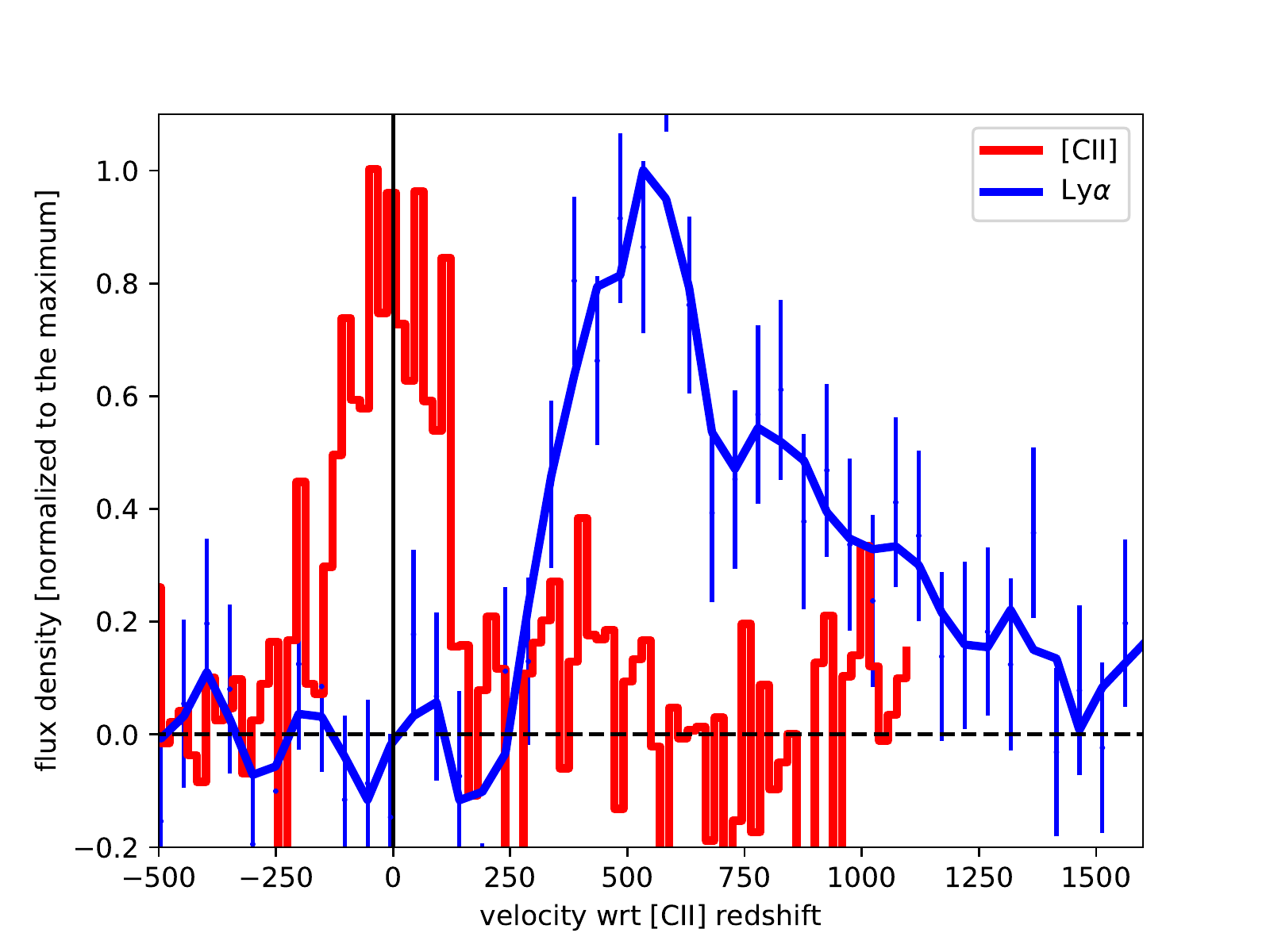}
\includegraphics[width=10.3cm]{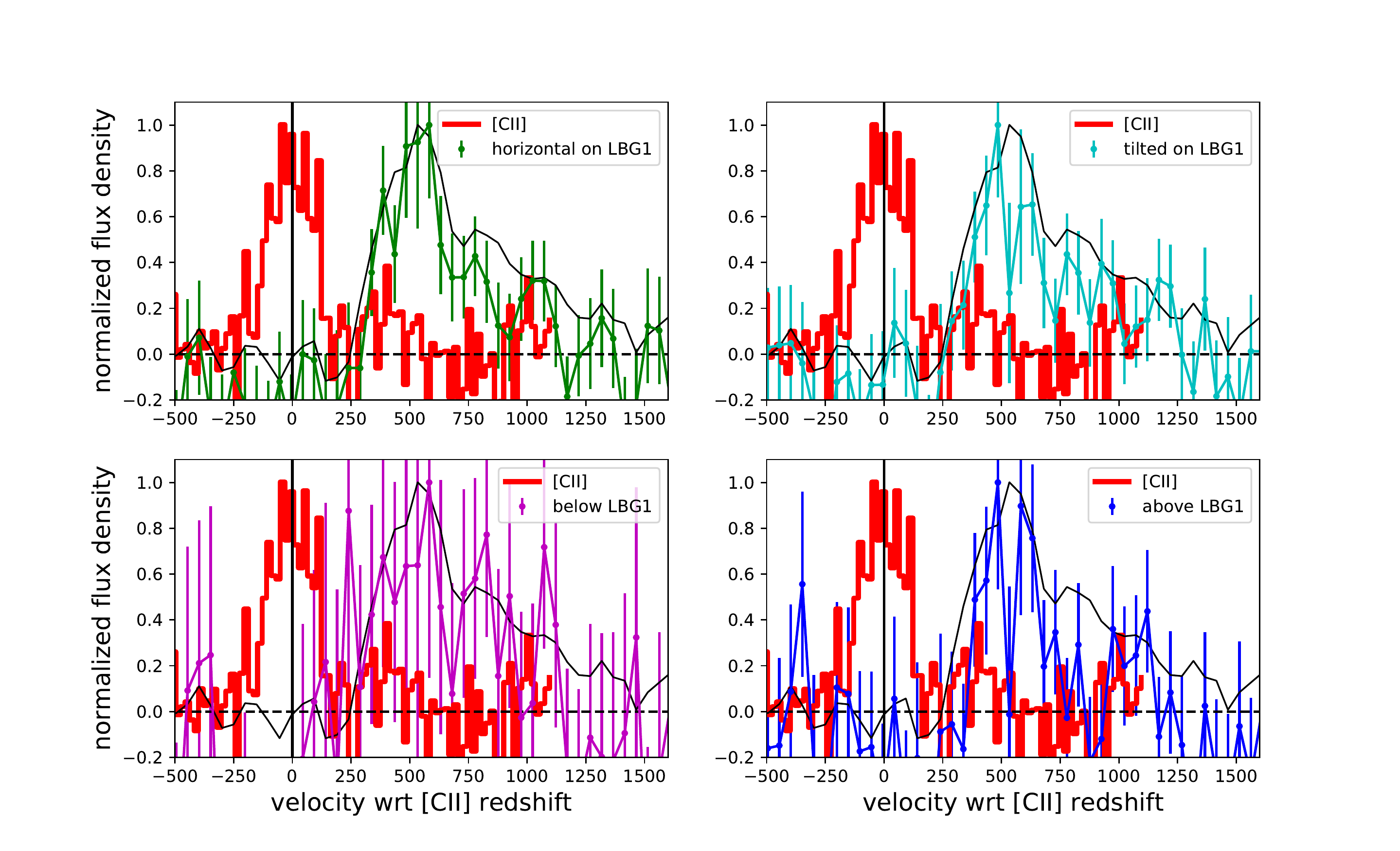}
\caption{$Left ~panel:$ 1D profiles of the Ly$\alpha$ (blue) and  [C${\sc II}$] (red) emissions toward LBG-1. The [C${\sc II}$] profile is taken from \citet{Riechers2014} and corresponds to the region outlined in Fig. \ref{LBG1optimalNBonallHST} (red contours).  $Right ~panel:$  [C${\sc II}$] profile in red as in the left panel and Ly$\alpha$ profile obtained slicing the Ly$\alpha$ emission horizontally on top of LBG-1 (green), tilted by 110 degrees on top of LBG-1 (cyan), tilted by ten degrees below LBG-1 (magenta), and tilted by ten degrees above LBG-1 (blue).  The color coding of these spectra is like in Fig. \ref{LBG1PV}. The two former spectra are the ones with higher signal to noise as can be seen by the smaller error bars. The thin black curve corresponds to the spectrum of the integrated Ly$\alpha$ emission as the blue curve of the left panel.}
\label{LBG1LyaCII}%
\end{figure*}

%
In Fig. \ref{LBG1LyaCII}, we show the [C${\sc II}$] and Ly$\alpha$ profiles as a function of the velocity with respect to the [C${\sc II}$] systemic redshift. 
The integrated 1D spectrum of the Ly$\alpha$ emission line is composed by a narrower peak 
and a much broader component that reaches the continuum at 7680 {\AA}. 
The FWHM of Ly$\alpha$ 
is larger than that of [C${\sc II}$]. 
This is not unusual 
in high-redshift Ly$\alpha$ emitting galaxies \citep[e.g.,][]{Matthee2019} 
and it could be related to the variety of HI kinematics properties that can condition the formation and escape of Ly$\alpha$ photons.
Also, the tail of the Ly$\alpha$ emission could indicate that there is gas in a wide range of kinematic conditions and with different HI column densities around the three star-forming knots. It could be consistent with rotation among the three knots as well. In fact, the wavelength range of this tail corresponds to the feature visible in the second and third rows of Fig. \ref{LBG1PV} at large velocities and coming from the southwest. 
It is possible that this feature comes from the location of the [C${\sc II}$] emission and also from the region below the three knots. 
No metal lines are detected at enough signal to noise in the MUSE spectrum. 

In Fig. \ref{LBG1LyaCII}b, we show the [C${\sc II}$] profile in velocity space together with the Ly$\alpha$ emission extracted slicing the Ly$\alpha$ image with pseudo slits oriented as in Fig. \ref{LBG1PV}. These spectra are very noisy, but we can see that 
the highest peak of the integrated Ly$\alpha$ emission receives a contribution from the three knots, while the tail may not be originated from the northern region of LBG-1 (N position in Fig. \ref{LBG1optimalNBonallHST}). This could indicate the presence of a gas with higher velocity to the south and may suggest a 
merger of the three components. Alternatively, the gas could be characterized by larger N${\sc HI}$. By studying the Ly$\alpha$ profile in terms of radiative transfer models could inform about the two possibilities (see next section). 

\section{Ly$\alpha$ radiative transfer model}
\label{model}

We use the most updated version of the FLaREON radiative transfer model \citep{Gurung-Lopez2018} to quantify the properties of the interstellar, circum-galactic, and intergalactic (IGM) medium that can condition the shape and intensity of the Ly$\alpha$ emission in the AzTEC-3 protocluster. FLaREON was based on LyaRT \citep{Orsi2012}, a radiative transfer Monte Carlo code of Ly$\alpha$ emission, and predicted the Ly$\alpha$ line profile escaping a galaxy through different outflow configurations. Several outflow geometries were implemented, such as thin shell of HI gas \citep[see also][]{Verhamme2006, Gronke2016}, and a galactic wind \citep[Fig. 1 in][]{Gurung-Lopez2019}. 
In the new version of FLaREON, called zELDA, \citet{Gurung-Lopez2021b} developed a new thin shell model in which the intrinsic galaxy spectrum contains a continuum in addition to a (Gaussian) Ly$\alpha$ emission line. 
The variable set delivered by zELDA includes the systemic redshift, z$_{in}$, the outflow expansion velocity, V$_{exp}$, the HI column density, N${\sc HI}$, and the dust optical depth, $\tau$. 
Inflows are predicted such as the outflows, but with negative expansion velocities.

To be able to perform a quantitative comparison between our observed Ly$\alpha$ profiles and the ones predicted by zELDA, a Gaussian kernel representing the MUSE resolution is applied to the predicted profiles and the systemic redshift is provided. Also, the observed-spectrum sampling and signal to noise are taken care of in the observed frame \citep[see Sect. 6.1 of][for details]{Gurung-Lopez2021}. The quantitative comparison involves obtaining the model that produces the best fit of the observed spectrum. The best fit is obtained with a Markov Chain Monte Carlo (MCMC) approach. In each step of the chain, a line profile corresponding to the variables mentioned above is computed and compared with the observed spectrum. The MCMC takes into account the uncertainties in the observed spectrum. After the MCMC run, we consider the model corresponding to the 50th percentile of the probability distribution function of the variable as the best-fit profile and the 16th (84th) percentile as the model corresponding to the lower (upper) limit of each variable.

As shown in \citet{Gurung-Lopez2021b}, there is a grid of parameters where the MCMC chains can look for the best combination of parameters to reproduce the observed spectrum. Expansion velocities from 0 to 1000 km sec$^{-1}$ and HI column densities from 10$^{17}$ to 10$^{21.5}$ atoms cm$^{-2}$ are explored (their Sect. 2). However, the starting point of each chain is set by performing an initial optimization, which is an attempt to narrow the parameter space (their Sect. 4). 
The uncertainties of the best-fit zELDA parameters depend on the resolution, sampling, and signal to noise of the observed spectra (their Sect. 4). The zELDA performance is tested for a signal to noise larger than 5 at the wavelength of the main Ly$\alpha$ peak. The dust optical depth is poorly constrained even in the best signal-to-noise scenario, while the uncertainty on V$_{exp}$ can increase by a factor of 25\% when the signal to noise changes from ten to 5 and the uncertainty on N${\sc HI}$ is almost four times larger. It means that by fitting only the Ly$\alpha$ profile with zELDA, we are not able to provide a significant estimation of the dust absorption, but we can aim at a reliable estimation of V$_{exp}$ and N${\sc HI}$ when the signal to noise of the observed Ly$\alpha$ spectrum is larger than ten.

According to the analysis in \citet{Gurung-Lopez2021b}, we can expect an anticorrelation between Ly$\alpha$ luminosity and N${\sc HI}$. This is explained considering that a gas with higher N${\sc HI}$ would produce a larger number of scattering events and so a longer escaping path that can be associated to a larger dust attenuation and so a lower escape fraction of Ly$\alpha$ photons than in the case of lower N${\sc HI}$. However, the scattering events could also produce larger Ly$\alpha$ nebulae depending on the viewing angles.

We performed the zELDA fits of the spectra of the AzTEC-3+LBG-3 system (mosaic\_1513), of LBG-1 (mosaic\_1496), of the fixed-size aperture spectra extracted on top of the SMG Ly$\alpha$ emission peak (onlySMG Ly$\alpha$), on top of the UV knots of the SMG (onlySMG UV knots), on top of LBG-3 (onlyLBG), on the bridge between SMG and LBG (bridgeSMGLBG), extracted from pseudo-slits oriented horizontally on LBG-1, tilted, tilted below, and tilted above LBG-1 (see Fig. \ref{LBG1PV} and Fig. \ref{LBG1LyaCII}). Also, we performed the fit for the sources mosaic\_1548 and mosaic\_1520 for which we have a signal-to-noise ratio larger than ten and an estimation of the physical parameters and photometric redshift from the  
COSMOS2015 catalog. 
We report the best fit values of the most reliable parameters, V$_{exp}$ and N${\sc HI}$, in Table \ref{zELDAbestfit}, together with the $\chi^2$ of the best fit. We also report the ranges of the parameters contained within the 16th and 84th percentiles to show the extent of the parameter space.

\subsection{Radiative transfer modeling of the AzTEC-3+LBG-3 system}

\begin{figure*}
 \centering
\includegraphics[width=8cm]{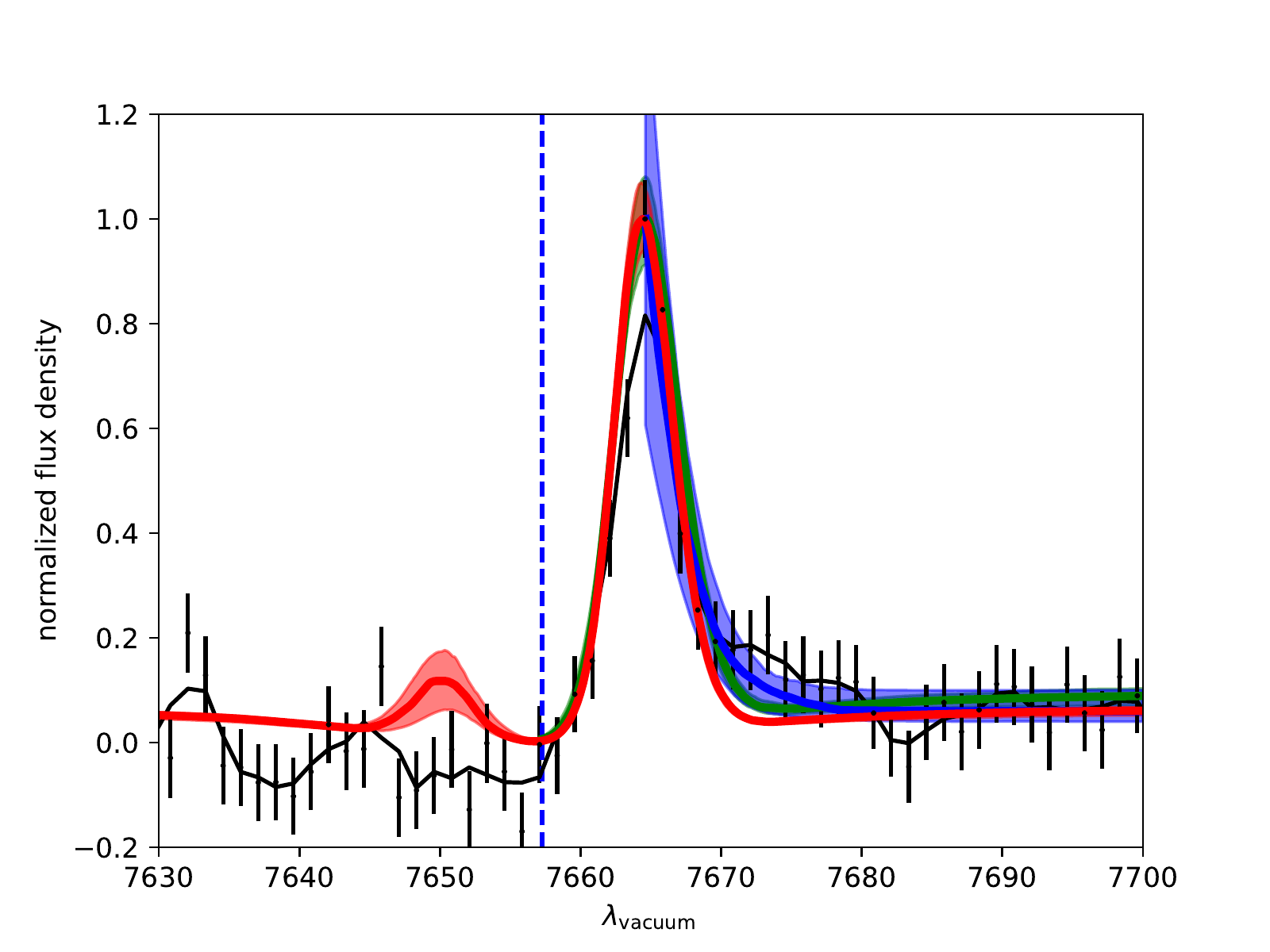} 
\includegraphics[width=9cm]{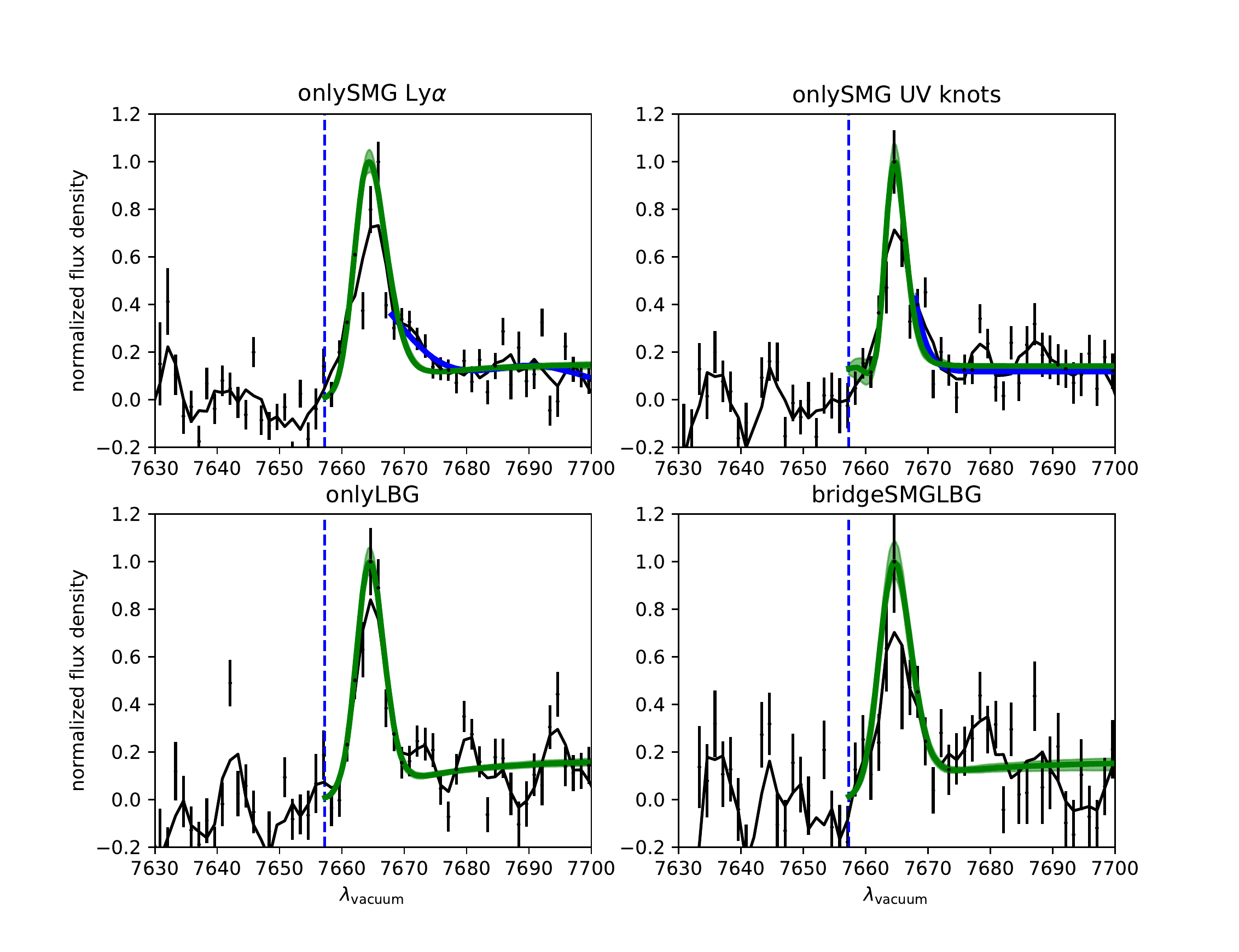}
\caption{
$Left ~panel:$ Observed-frame spectrum at the Ly$\alpha$ wavelength of the AzTEC-3+LBG-3 system in the vacuum. The observed spectrum is shown as data points with error bars and the smoothed spectrum as a black curve like in Fig. \ref{AzTEC3CIIMGLBGLya}. The best-fit zELDA model that corresponds to the 50th percentile is shown in green and the green shaded area contains the models within the 16th and 84th percentiles obtained fitting the spectrum at wavelength larger than the systemic redshift. In blue, we show the models obtained fitting the spectrum only at the wavelength larger than the maximum. The red curve corresponds to the best fit in the wavelength range 7630-7700 {\AA} for the AzTEC-3+LBG-3 spectrum corrected for the IGM absorption, by using the prescription of \citet{Madau:1995}.  
$Right ~panel:$ Best-fit zELDA models for the spectra extracted in a 0.7$''$ aperture located on the position of the main Ly$\alpha$ emission associated to AzTEC-3 ($upper ~left$), the position of the UV knots of the SMG ($upper ~right$), the position of the main Ly$\alpha$ emission associated to LBG-3 ($lower ~left$), and the bridge between SMG and LBG ($lower ~right$). Black dots with error bars are the observed-frame spectra as shown in Fig. \ref{AzTEC3CIIMGLBGLya}, the black lines are the smoothed spectra, the green (blue) curves and shaded areas are the best-fit models and the models within the 16th and 84th percentile of the model parameter space obtained fitting the spectra at wavelength larger than the systemic redshift (larger than 7669 {\AA}). Vertical blue dashed lines indicate the Ly$\alpha$ redshift given by the AzTEC-3 [C${\sc II}$]  detection. The zELDA fits are performed fixing the redshift to the systemic inferred by the [C${\sc II}$] emission peak.
}
\label{zELDAfitSMGLBG}%
\end{figure*}

In Fig. \ref{zELDAfitSMGLBG}a, we show the models that best fit the observed spectrum of the AzTEC-3+LBG-3 system. The observed spectrum is shown in the vacuum framework as the zELDA models. 
By fixing the systemic redshift to that provided by the [C${\sc II}$] emission line, we estimated the MCMC model that best fits the data at wavelengths larger than the Ly$\alpha$ systemic wavelength (green curve in the figure). In fact, the zELDA models do not take into account the IGM absorption, but at $z\simeq5.3$ the IGM could condition the spectrum at wavelengths bluer that Ly$\alpha$. 
Also, we tried to correct the bluer side of the spectrum for the IGM effect by using the average prescription by \citet{Madau:1995} and run an MCMC chain of the corrected spectrum of the AzTEC-3+LBG-3 system (red curve in the figure). The best fit zELDA models in these two cases provided similar combinations of parameters. 
However, since the IGM could affect the blue side of the Ly$\alpha$ emission line all the way through the maximum of its red peak, we tried also a zELDA fit only of the reddest part of the Ly$\alpha$ emission line, at $\lambda_{vacuum}>7664 {\AA}$ (blue curve in the figure). 
The spectrum of the AzTEC-3+LBG-3 system is composed by one main peak and an extended tail at $\lambda_{vacuum}>7668 {\AA}$. By masking the wavelengths bluer than the maximum of the Ly$\alpha$ peak, zELDA finds the best compromise between the red peak and the extended tail of the line as the best fit, which corresponds to models with a wide range of HI column densities (see Table \ref{zELDAbestfit}) and V$_{exp}$ up to 90 km sec$^{-1}$.

In general, zELDA mainly accounts for the main peak of the Ly$\alpha$ emission line, which is dominant in intensity with respect to the extended tail. To identify the regions that most contribute to the extended tail at $\lambda_{vacuum}>7668 {\AA}$, we ran zELDA fits of the onlySMG Ly$\alpha$, onlySMG UV knots, onlyLBG, and bridgeSMGLBG spectra (Fig. \ref{zELDAfitSMGLBG}b). A main red peak and an extended tail are seen in the onlySMG spectra and we found a zELDA best fit for the main peak and one for the extended tail for them. 
For the onlySMG Ly$\alpha$ spectrum, the Ly$\alpha$ profile is consistent with zELDA models characterized by V$_{exp} \leq 50$ km sec$^{-1}$ and N${\sc HI}$ on the order of 10$^{20}$ atoms cm$^{-2}$ for the main red peak and by V$_{exp} \sim 800$ km sec$^{-1}$ and N${\sc HI}$ on the order of 3 $\times$ 10$^{20}$ atoms cm$^{-2}$ for the extended tail. 
The N${\sc HI}$ value could be responsible for the scattering of Ly$\alpha$ photons up to 90 ckpc as mentioned in Sec. \ref{AzTEC3MUSE}. On the contrary, in the region of the UV knots of star formation, the AzTEC-3 Ly$\alpha$ profile is consistent with zELDA models with V$_{exp} < 100$ km sec$^{-1}$ and N${\sc HI}$ $< 10^{19}$ atoms cm$^{-2}$. 

The best fit zELDA models of the bridgeSMGLBG and onlyLBG spectra are both characterized by V$_{exp} \sim 20$ km sec$^{-1}$ and N${\sc HI}$ on the order of 10$^{20}$ atoms cm$^{-2}$. 
Therefore, the HI column density could be responsible for the escape of Ly$\alpha$ photons in the distorted region above LBG-3 (N position in Fig. \ref{AzTEC3optimalNBonallHST}) and in the region of interaction between AzTEC-3 and LBG-3. In this region of interaction, the gas could be turned into stars, produce star formation under favorable conditions, and allow the formation of new Ly$\alpha$ photons as well.
In Fig. \ref{cartoon}a, we present a cartoon to better visualize the combination of parameters inferred by zELDA for the AzTEC-3+LBG-3 system.

\begin{table*}
\caption{Best fit parameters of the zELDA models}             
\label{zELDAbestfit}      
\centering                          
\scalebox{0.9}{
\begin{tabular}{c| c| c| c}        
\hline              
spectrum & V$_{exp}$ & N${\sc HI}$ &  $\chi^2/dof$ \\    
 & km sec$^{-1}$ & $10^{19}$ atoms cm$^{-2}$  & \\ 
\hline                        
AzTEC-3+LBG-3$^1$ & 21[15-28] & 19[18-22]  & 0.92\\
AzTEC-3+LBG-3$^2$ & 20[3-93] & 1.3[0.1-7.8]  & 1.65\\
AzTEC-3+LBG-3$^3$ & 25[20-30] & 19[18-20]  &  0.48\\
onlySMG Ly$\alpha ^a$ & 42[34-56] & 16[14-18] & 1.14\\
onlySMG Ly$\alpha ^b$ & 785[754-825] & 32[22-46]&  2.24 \\
 onlySMG UVknots$^a$ & 60[40-127] & 0.3[0.1-0.8]  &  0.89\\
onlySMG UVknots$^b$ & 13[1-71] & 0.1[0.02-2]  &  1.39\\
onlyLBG3$^a$ & 15[6-23] & 21[18-28] & 0.93\\
bridgeSMGLBG$^a$ & 15[7-25] & 19[17-28] &  0.63 \\
\hline 
LBG-1$^a$ &  12[4-30] &  68[53-102] & 1.09  \\
LBG-1$^b$ & 83[25-293] & 24[2-92]   & 1.07\\
onLBG1horizontally$^a$  & 10[2-28] & 70[48-107] & 0.54\\
onLBG1tilted$^a$  & 14[4-30] & 60[43-94] & 0.47\\
belowLBG1$^a$  & 9[1-58] &2.2[0.1-37] & 0.38\\
aboveLBG1$^a$ & 12[2-28] & 69[45-103]& 0.41\\
\hline                                   
mosaic\_1548$^a$ & 30[20-46]&0.1[0.02-0.3]($z=5.304$) & 0.54\\
mosaic\_1548$^{inflow}$ & 10[4-30] &0.8[0.1-1.8]($z=5.308$)  & 0.46\\
mosaic\_1520$^a$ &  26[6-61] & 0.8[0.1-8]($z=5.298$)& 0.36\\
\hline
\hline
\end{tabular}
}
\tablefoot{
Expansion velocity, V$_{exp}$ and HI column density, N${\sc HI}$ 
of the zELDA models that correspond to the best fits of the spectra in the first column. The last column provides an estimation of the goodness of the fit in terms of reduced $\chi^2$. 
The numbers in the second and third columns correspond to the 50th percentile, while the two numbers in parentheses are the variables corresponding to the 16th and 84th percentiles of the parameter space of the V$_{exp}$ and the N${\sc HI}$ values of the models.
AzTEC-3+LBG-3$^1$ corresponds to the fit performed at wavelength larger than the systemic redshift (green curve in Fig. \ref{zELDAfitSMGLBG}a); AzTEC-3+LBG-3$^2$ at wavelength larger than the maximum of the main peak (blue curve in Fig. \ref{zELDAfitSMGLBG}a); AzTEC-3+LBG-3$^3$ corresponds to fit performed on the spectrum corrected for IGM absorption based on the prescription of \citet{Madau:1995} (red curve in Fig. \ref{zELDAfitSMGLBG}a). $^a$ corresponds to the fit of the main peak. $^b$ corresponds to the fit of the extended tail. $^{inflow}$ proposed best fit solution with inflow. The fit of the last three models was performed leaving the redshift as a free parameter. The best fit systemic redshift in indicated in parenthesis.}
\end{table*}

\subsection{Radiative transfer modeling of LBG-1}

\begin{figure*}
 \centering
\includegraphics[width=8cm]{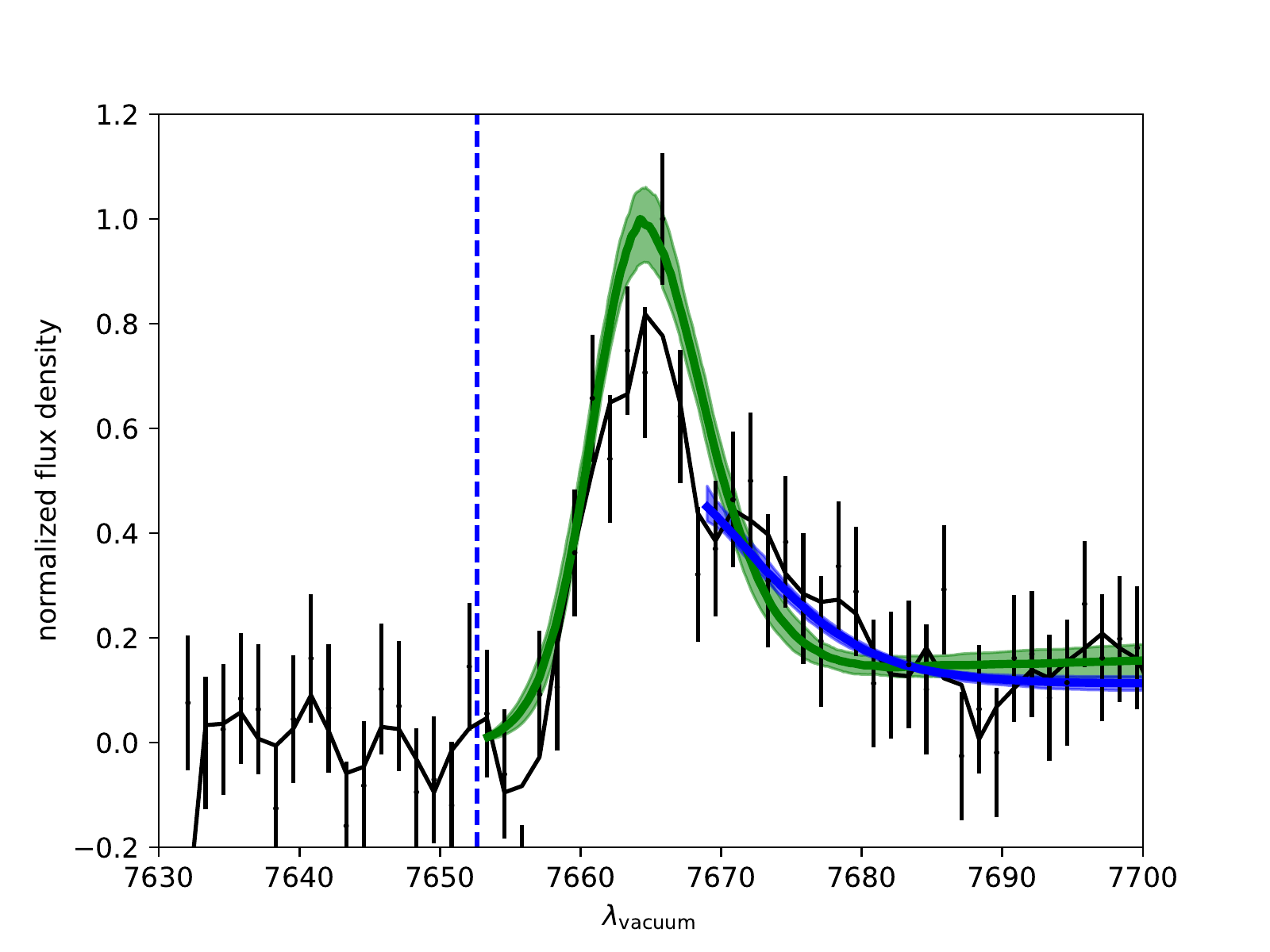}
\includegraphics[width=9cm]{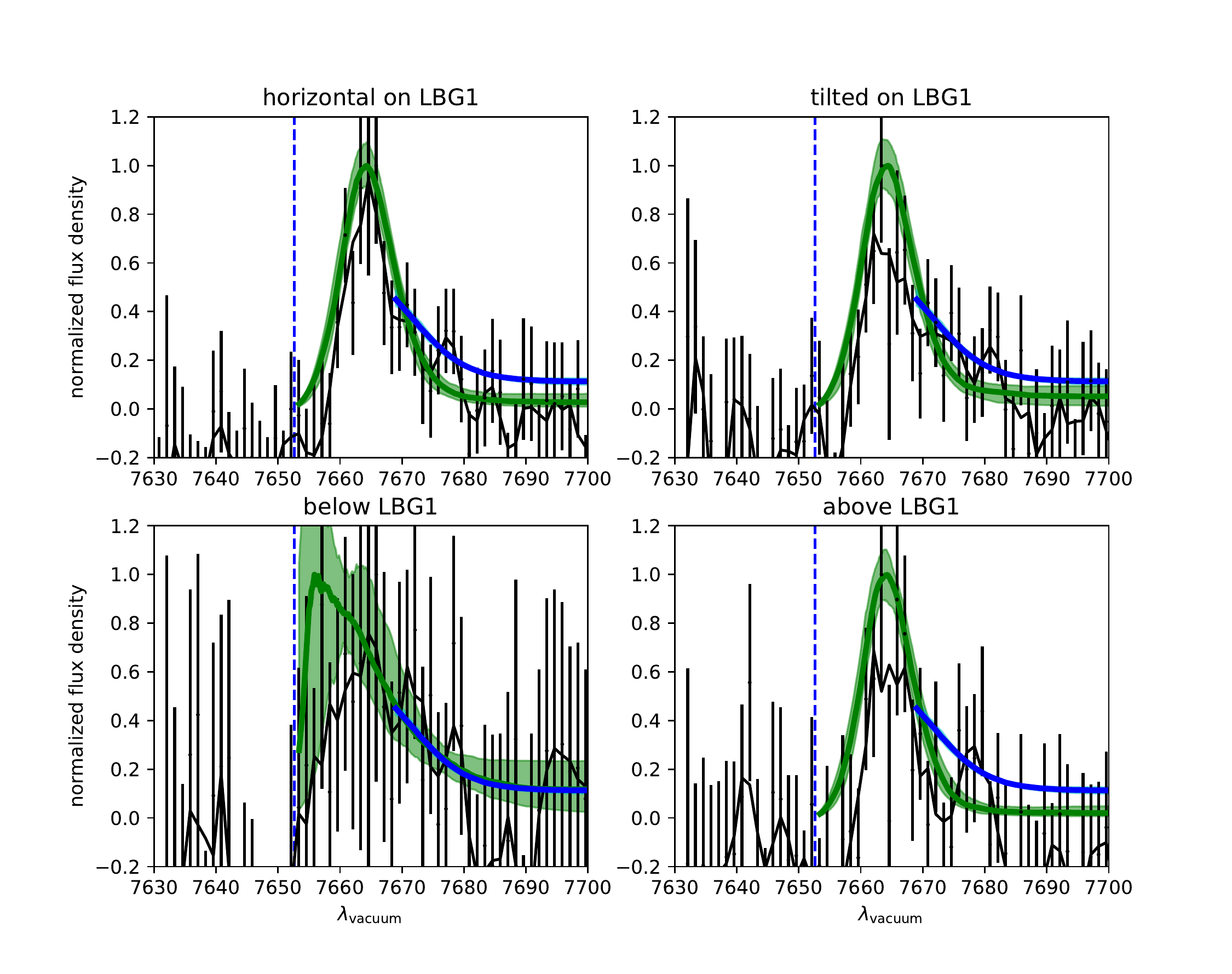}
\caption{$Left ~panel:$ Observed-frame spectrum at the Ly$\alpha$ wavelength of LBG-1. The observed spectrum in vacuum wavelengths is shown as black datapoints with error bar, the black curve is the smoothed spectrum. The green curve is the best-fit zELDA model that corresponds to percentile 50 and the green shaded area contains the models within percentiles 16 to 84. The blue curve corresponds to a zELDA fit at wavelengths larger than 7669 {\AA}, obtained with the scope of finding a zELDA model for the extended tail at wavelengths larger than the main red peak. $Right ~panels:$ zELDA fits of the spectra obtained collapsing the LBG-1 spectrum as in Fig. \ref{LBG1LyaCII}. In green, we show the best fit and the models within the 16th and 84th percentile obtained for wavelengths larger than the systemic redshift. 
The blue curves correspond to the best fit of the extended tail for the spectrum in the left panel. Vertical blue dashed lines indicate the Ly$\alpha$ redshift given by the LBG-1 [C${\sc II}$] detection, we fixed while performing the fits. 
}
\label{zELDAfitLBG1}%
\end{figure*}

In Fig. \ref{zELDAfitLBG1}a, we show the results of modeling the Ly$\alpha$ profile of LBG-1. We performed the fit at wavelengths larger than the systemic redshift and at $\lambda_{vacuum}>7669 {\AA}$ to account only for the tail. 
In the LBG-1 spectrum, the main red peak has intensity more comparable to the extended tail than in the case of the AzTEC-3+LBG-3 spectrum, making the best fit zELDA models broader on average to account for the tail. The best fit of the main peak is consistent with a model with V$_{exp}< 30$ km sec$^{-1}$ and N${\sc HI}=5-10 \times 10^{20}$ atoms cm$^{-2}$. The extended tail is consistent with models with a large range of expansion velocities, up to 300 km sec$^{-1}$, and HI column density up to $9\times 10^{20}$ atoms cm$^{-2}$. 

\begin{figure*}
 \centering
\includegraphics[width=14cm]{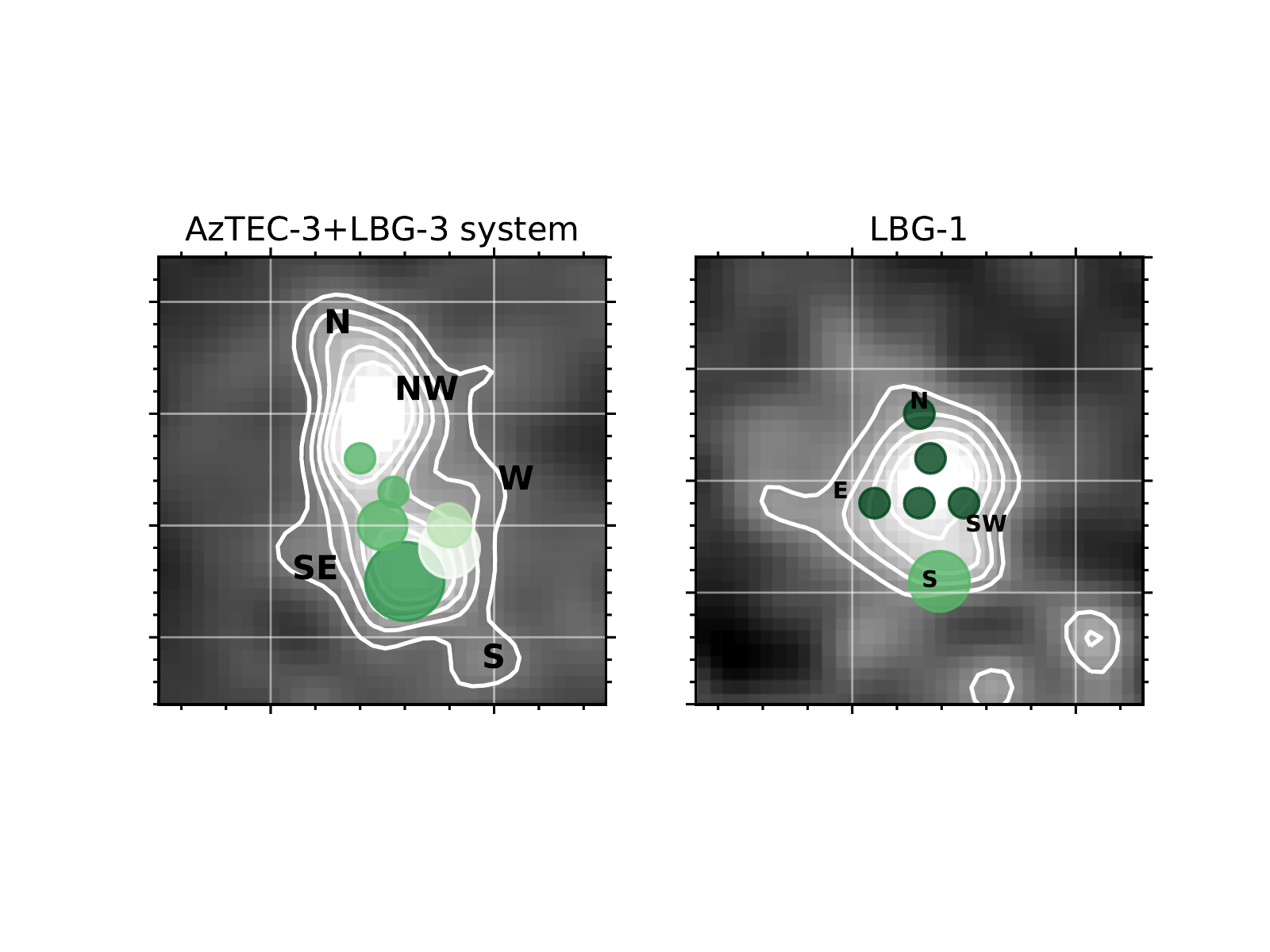}
\caption{Qualitative representation of the combination of the zELDA parameters that best fit the Ly$\alpha$ emission of the AzTEC-3+LBG-3 system ($left ~panel$) and of the LBG-1 region ($right ~panel$). The size of the dots in the cartoon is proportional to V$_{exp}$ and the color scale corresponds to the N${\sc HI}$ value in a way that fainter color indicates lower N${\sc HI}$. The smallest and darkest dots show V$_{exp}< 30$ km sec$^{-1}$ and N${\sc HI} \sim 10^{21}$ atoms cm$^{-2}$. The largest dot corresponds to V$_{exp}$ up to 800 km sec$^{-1}$, the faintest color to N${\sc HI} \sim 10^{19}$ atoms cm$^{-2}$.
}
\label{cartoon}%
\end{figure*}

To investigate which regions of space and combination of parameters better reproduce the extended tail, we performed zELDA fits of the profiles obtained slicing the Ly$\alpha$ emission like in Fig. \ref{LBG1PV} and we show the results in Fig. \ref{zELDAfitLBG1}b. Given the noisy spectra, we obtained a large uncertainty in the best fit models and their parameters. However, we can see that the best fit models of the slices containing Ly$\alpha$ emission coming from the N and E positions (included in the horizontal, tilted pseudo slit located on top of LBG-1, and the one above LBG-1, see Fig. \ref{LBG1optimalNBonallHST}) are consistent with HI column density up to 10$^{21}$ atoms cm$^{-2}$ and V$_{exp}< 30$ km sec$^{-1}$. The best fit is consistent with models of lower N${\sc HI}$ and larger V$_{exp}$ in the other position. This could support the idea of random distortion of the gas due to the merger of the three components toward the south of LBG-1.
The combination of parameters inferred by zELDA for the LBG-1 region are shown in Fig. \ref{cartoon}b.

\subsection{Radiative transfer modeling of the Ly$\alpha$ emitting sources without systemic redshift}

\begin{figure*}
 \centering
\includegraphics[width=7cm]{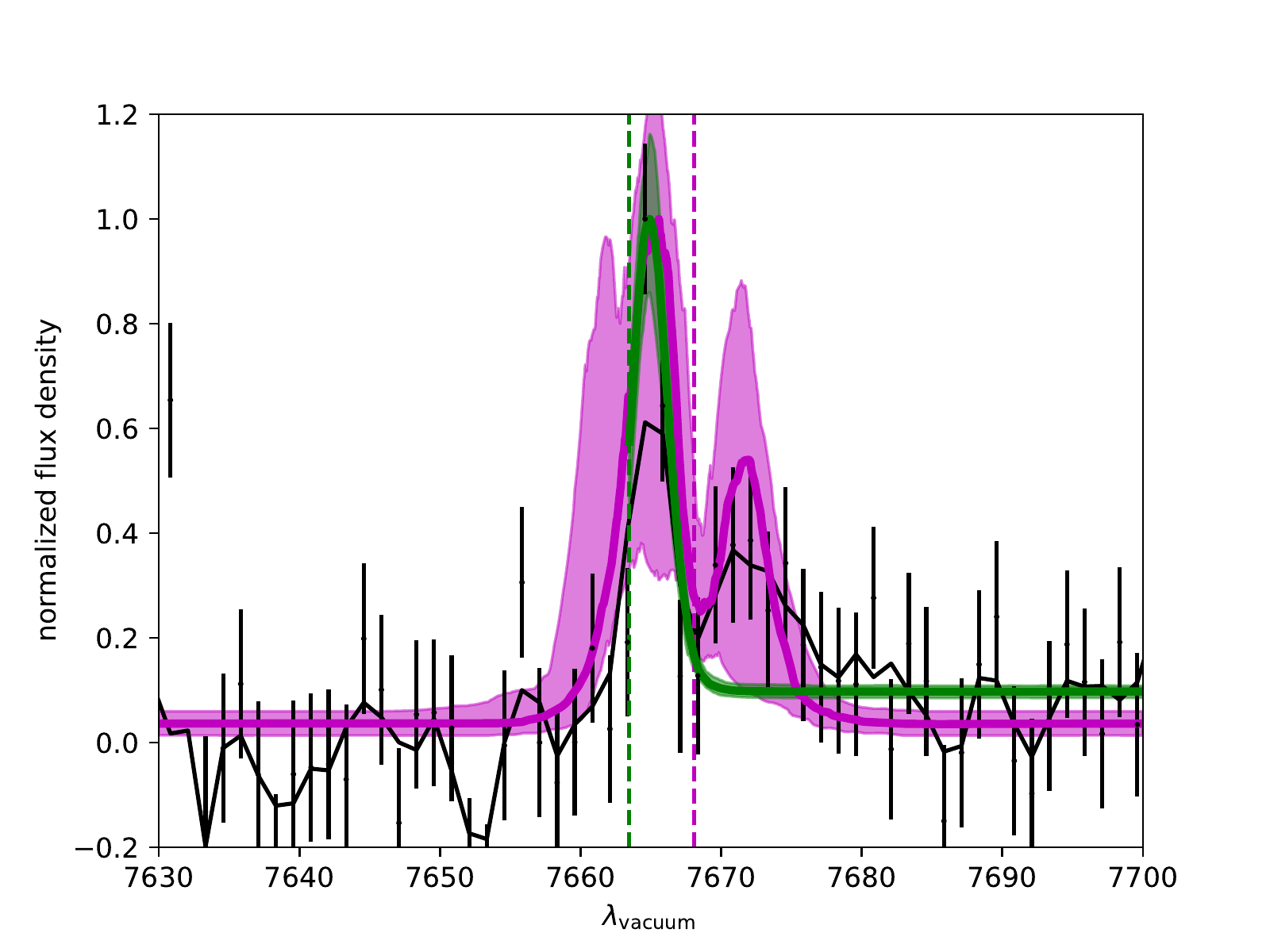}
\includegraphics[width=7cm]{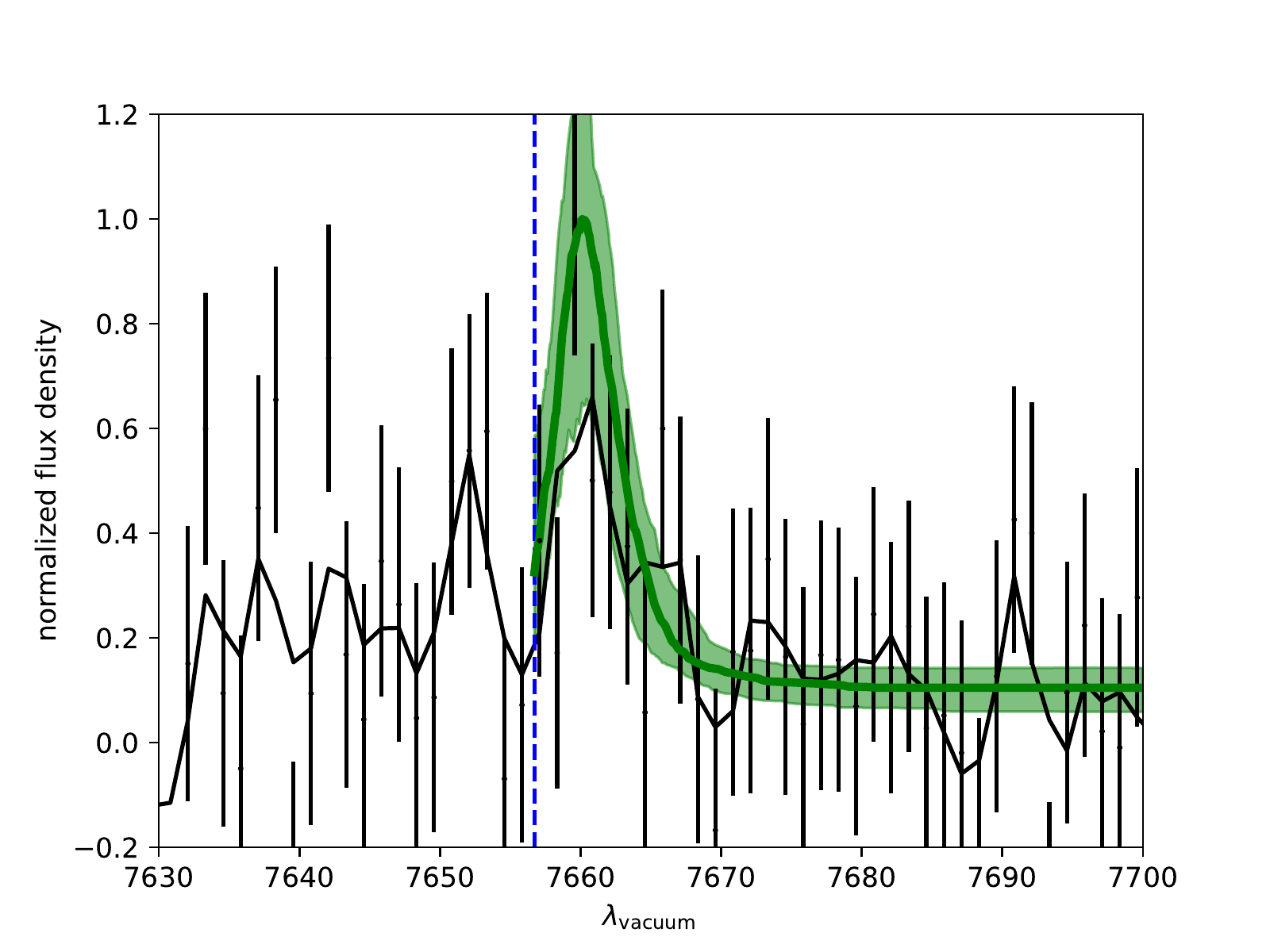}
\caption{Observed-frame spectrum at the Ly$\alpha$ wavelength of mosaic\_1548 ($left$), mosaic\_1520 ($right$). 
 The observed spectrum in vacuum wavelengths is shown as black data points with error bars, the black curve is the smoothed spectrum. The green curve is the best-fit zELDA model of a redshifted Ly$\alpha$ main peak that corresponds to percentile 50 and the green shaded area contains the models within percentiles 16 to 84. The vertical blue dashed lines indicate the Ly$\alpha$ redshift given by the 50th percentile model. For the left hand side spectrum, we also show in magenta a zELDA fit of the two visible peaks at 7665 and 7672 {\AA}. The systemic wavelength corresponding to this fit is at 7668.1 {\AA} and the model implies a blue peak higher than the red peak as in the case of an inflow (see Table 2 for the best fit parameters).
}
\label{zELDAfitothers}%
\end{figure*}

We ran zELDA leaving the redshift as a free parameters for the Ly$\alpha$ emitting sources in the AzTEC-3 environment without systemic redshift (Fig. \ref{zELDAfitothers}). We focus on the sources with a counterpart in the COSMOS2015 catalog and with a photometric redshift consistent with $z\sim 5.3$. The best fit parameters are shown in Table \ref{zELDAbestfit}, where we also report the best fit systemic redshift value. As expected, the best fit redshift is consistent with $z\sim5.3$. The best fit zELDA models are consistent with galaxies with HI column densities up to 10$^{20}$ atoms cm$^{-2}$ and expansion velocities on the order of a few tens of km sec$^{-1}$. This combination of parameters is consistent with the fact that these sources are characterized by low masses and moderate star-formation rates (see Table 2). In particular, the mosaic\_1520 source is characterized by a star-formation rate of more than 100 M$_\odot$ yr$^{-1}$ and it is plausible that this rate is accompanied by an outflow with a velocity up to 60 km sec$^{-1}$. 
Since the CubEx detection map shows a S/N larger than 5 from 7662 to 7675 {\AA}, we propose an alternative zELDA solution for the spectrum of the mosaic\_1548 source. In this solution, we found a combination of parameters that account for two peaks, a higher blue one at 7665 {\AA} and a lower red one at 7672 {\AA}. The best fit zELDA model suggests the presence of an inflow of a few tens of km sec$^{-1}$ and a gas with a low column density up to $2\times 10^{19}$ atoms cm$^{-2}$. If confirmed, this solution indicates the presence of an inflow at the outskirt of the protocluster. However, the presence of an inflow, visible as a Ly$\alpha$ blue peak at $z\sim5.3$, would imply the presence of an ionized bubble in the vicinity of the region of the production of Ly$\alpha$ photons and that those photons survive the IGM absorption from $z=5$ to $z=0$.

\section{Discussion}
\label{discussion}
In the $1.4\times1.4$ arcmin$^2$ region around the submillimeter galaxy AzTEC-3, we found ten Ly$\alpha$ emitting sources, including one extended emission toward AzTEC-3 itself and one toward the Lyman break galaxy, LBG-1. Significant [C${\sc II}$] detections were also observed at the positions of AzTEC-3 and LBG-1 \citep{Riechers2014, Riechers2020, Pavesi2019} and a significant CO molecular gas detection was detected at the positions of AzTEC-3 \citep{Riechers2010}. In the following section, we investigate the relation between Ly$\alpha$, [C${\sc II}$], and CO emissions and we discuss the properties of the AzTEC-3 protocluster we can derive from them. 

\subsection{Dust and gas content revealed by Ly$\alpha$, [C${\sc II}$], and CO detections in the AzTEC-3 environment} 

As described in \citet{Riechers2010, Riechers2014, Riechers2020}, AzTEC-3 shows detections of the CO transitions 
from the Very Large Array, the NOrthern Extended Milllimeter Array, and the Plateau de Bure interferometer,
 and also rest-frame 1mm continuum, [C${\sc II}$], and OH163$\mu$m doublet emissions from an ALMA campaign and the COLDz survey.

The CO peak emission was observed to be offset from the $HST$ ACS F814W image of the SMG by about 0.5$''$ (3 pkpc). 
This offset could be partially due to dust obscuration in the rest-frame UV image \citep{Riechers2010}. The dust obscuration could also explain the difference between the UV- and FIR-derived star-formation rates of AzTEC-3 \citep{Capak2011}, suggesting that the regions of most intense star formation are highly dust obscured. In fact, a dust mass on the order of $3\times 10^{8}$ M$_{\odot}$ was estimated from the SED fit of the optical-through-IR spectral energy distribution of the SMG \citep{Riechers2014, Riechers2020}.
Our MUSE observation shows a shift between the $HST$ position of the SMG and the Ly$\alpha$ peak emission, which intensifies toward the southeast and the northeast. 
Ly$\alpha$ is not detected to the west. 
This could indicate that the dust on the main knots of star formation prevents the escape of Ly$\alpha$ photons on one side. 

The molecular gas detections were in favor of a scenario of two gas components, a diffuse (1.3$''$ = 50 ckpc), low-excitation component and a denser (1$''$ = 40ckpc), high-excitation component, where the diffuse gas has properties similar to those of normal high-$z$ galaxies while the dense gas components has properties more similar to those of high-$z$ FIR luminous quasars. However, as described in \citet{Riechers2010}, the size of the CO emission was observed to be overall quite compact, 2.3 pkpc.
Our MUSE observations show an even more diffuse HI gas, revealed by the Ly$\alpha$ emission, that extends up to 150 ckpc, embedding or blending with the HI gas of LBG-3. The zELDA modeling we performed on the Ly$\alpha$ profile informs about the presence of HI column densities up to $3\times 10^{20}$ atoms cm$^{-2}$ above LBG-3 (Sect. \ref{model}). 

The [C${\sc II}$] emission toward AzTEC-3 was also observed to be compact 
and its peak position to be consistent with that of CO \citep{Riechers2014}.
The fact that the CO and [C${\sc II}$] peak emissions are spatially consistent implies that [C${\sc II}$] is tracing the position of the star-forming regions and the location of the molecular gas clouds rather than other gas phase regions of the SMG. The star-forming regions of AzTEC-3 would then contain some metals, unlike LBG-3 that lacks [C${\sc II}$] detection probably due to its low metallicity. The Ly$\alpha$ emission is much more extended than the star-forming regions traced by [C${\sc II}$], probably revealing the shape and distribution of the HI gas in the system, elongated above LBG-3 and extending southern than the SMG. The elongated shape could be the result of the interaction between AzTEC-3 and LBG-3. 

At the compact position of the [C${\sc II}$] emission, 
F(Ly$\alpha)=(1.9\pm0.7) \times 10^{-18}$ erg sec$^{-1}$ cm$^{-2}$, 
rest-frame EW(Ly$\alpha)=6.7\pm1.6$ {\AA}, and L(Ly$\alpha)=(1.4\pm0.5) \times 10^{8}$ L$_{\odot}$, giving a ratio of L(Ly$\alpha$)/L([C${\sc II}$])= $(2.1\pm0.7) \times 10^{-2}$ \citep[see also][]{Riechers2020}.
Extracting the MUSE spectrum within a 0.7$''$-aperture on top of LBG-3, we estimate a 
F(Ly$\alpha)=(3.8\pm0.6) \times 10^{-18}$ erg sec$^{-1}$ cm$^{-2}$, a rest-frame EW(Ly$\alpha)=9.5\pm1.7$ {\AA}, and L(Ly$\alpha)=(2.8\pm0.4) \times 10^{8}$ L$_{\odot}$. At this position no significant [C${\sc II}$] emission was detected and L([C${\sc II}])<0.17\times 10^{9}$ L$_{\odot}$ \citep{Riechers2014}. So, on top of LBG-3, L(Ly$\alpha$)/L([C${\sc II}$])> $0.6$.   

\begin{figure*}
\centering
\includegraphics[width=19cm]{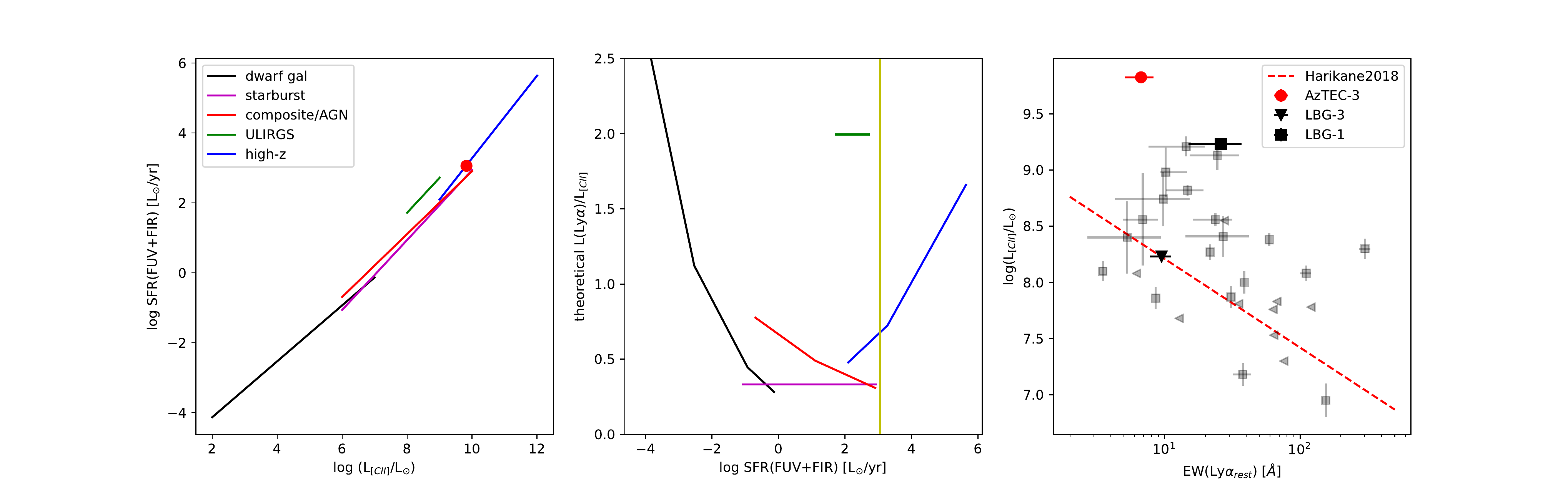}
\caption{$Left ~panel:$ total SFR vs [C${\sc II}$] luminosity from the empirical relations of \citet{DeLooze2014}. The red circle corresponds to the total SFR and L([C${\sc II}$]) of AzTEC-3. $Middle ~panel:$ theoretical value of L(Ly$\alpha$)/L([C${\sc II}$]) versus total SFR obtained assuming that SFR(Ly$\alpha$) is equal to the total SFR and SFR(Ly$\alpha$) is calculated with the calibration from \citet{Kennicutt:1998}. This calculation is intended to give an idea of the theoretical value of L(Ly$\alpha$)/L([C${\sc II}$]) at the location where the [C${\sc II}$] emission is concentrated. The vertical yellow line correspond to the total SFR of AzTEC-3 for all the possible values of L(Ly$\alpha$)/L([C${\sc II}$]). $Right ~panel:$ observed L(Ly$\alpha$)/L([C${\sc II}$]) versus rest-frame Ly$\alpha$ equivalent width as shown in \citet{Harikane2018}. The average relation for all the galaxies in that work is shown as a red dashed line. The gray symbols are their data points for the sources with a Ly$\alpha$ detection, excluding the upper limits. The red and black symbols correspond to AzTEC-3 (red circle), LBG-3 (black facedown triangle), and LBG-1 (black square).
}
\label{SFRLCII}%
\end{figure*}

We can now investigate the L(Ly$\alpha$)/L([C${\sc II}$]) ratio in terms of its relation to star formation. 
A Ly$\alpha$ luminosity lower than the theoretical value and so a lower L(Ly$\alpha$)/L([C${\sc II}$]) ratio can be usually explained by the presence of dust. 
We refer to the SFR versus L([C${\sc II}$]) empirical relation obtained by \citet[][]{DeLooze2014} to estimate the L(Ly$\alpha$)/L([C${\sc II}$]) expected value for different kind of high-redshift star forming galaxies. 
We consider their equations 3 to relate the SFR and the luminosity of a SFR-sensitive emission line. Also, we assume that SFR is the total UV plus FIR star-formation rate and that it is equal to the SFR(Ly$\alpha$) in the case all the emitted Ly$\alpha$ photons escape the galaxy (Fig. \ref{SFRLCII}). 
We take the SFR(Ly$\alpha$) expression from \citet{Kennicutt:1998}, for which SFR=10 M$_{\odot}$ yr$^{-1}$ implies an intrinsic L(Ly$\alpha)=10^{43}$ erg sec$^{-1}$. By assuming the L([C${\sc II}$]) vs SFR relation of local starburst and high-$z$ star-forming galaxies \citep[equations 7, 8, 11, 14, and 17 in][]{DeLooze2014}, L(Ly$\alpha$)/L([C${\sc II}$])$\sim0.3-0.7$ at a SFR(UV+FIR) of about 1200 M$_{\odot}$ yr$^{-1}$ (yellow vertical line in Fig. \ref{SFRLCII}). The L(Ly$\alpha$)/L([C${\sc II}$]) theoretical ratio could decrease in the case of subsolar metallicity. 
A dust extinction corresponding to A$_V=0.5$ could explain the L(Ly$\alpha$)/L([C${\sc II}$]) ratio calculated for the spectrum extracted on top of the SMG, while the theoretical value is comparable with the ratio we calculate for the spectrum extracted on top of LBG-3.
%
%

\citet{Harikane2018} showed an anticorrelation between L([C${\sc II}$]) and rest-frame EW(Ly$\alpha$) for galaxies at $5<z<7$ (see the right panel of Fig. \ref{SFRLCII}). Based on the EW(Ly$\alpha$) we calculate, either the L([C${\sc II}$]) is much stronger in AzTEC-3, implying that AzTEC-3 is a source more metal rich than the galaxies studied in that work, or Ly$\alpha$ is more suppressed due to the higher dust content in AzTEC-3 (their Fig 18, 19, 20) which is more likely the case given the nature of AzTEC-3. Also, a log(L([C${\sc II}$])/SFR$_{FIR}$)=6.8 is lower than the value they calculated for a galaxy with a rest-frame EW(Ly$\alpha$) lower than 30{\AA}. This is not surprising given the fact that the L([C${\sc II}$])/L$_{FIR}$ ratio suggested the presence of a stronger-than-average, not normal, far-UV radiation field \citep{Riechers2014, Riechers2020}.

\citet{Riechers2014} showed that the [C${\sc II}$] emission is compact over the entire velocity range, but also showed a blue low-significance wing toward LBG-3, that could correspond to either a tidal feature or a [C${\sc II}$] outflow. 
A blueshift of the OH163$\mu$m doublet also supported this hypothesis.
The presence of an outflow is plausible due to the intense AzTEC-3 star-formation rate. In fact, the SFR value and the size of the star-formation region as given by the [C${\sc II}$] emission determine a star-formation rate per unit of area larger than the limit expected for supporting starburt-driven outflows \citep{Heckman2001, Riechers2020}.
Modeling the Ly$\alpha$ profile, we found best fit models that are consistent with the presence of outflows from the UV star-forming regions of the SMG and the presence of high N${\sc HI}$ where the Ly$\alpha$ emission is brighter, where, therefore, HI scattering could be efficient.  
This suggests that the gas kinematics around AzTEC-3 are the result of its ongoing starburst, and may be further shaped by its possible interaction with LBG-3.

%
The only other galaxy in the AzTEC-3 neighborhood with a significant detection of [C${\sc II}$] is LBG-1. \citet{Riechers2014} found that the [C${\sc II}$] emission is spatially resolved and covers all the three optical knots.  
The infrared SED fitting gave a dust mass less than $9\times 10^{7}$ M$_{\odot}$, which implies a dust extinction much lower \citep{Capak2011} 
than in the case of AzTEC-3 and a $SFR_{FIR}<54$ M$_{\odot}$ yr$^{-1}$, which  
makes LBG-1 consistent with being a main sequence galaxy at $z=5$, composed of a young starburst and an underlying older population. 
At the compact position of the [C${\sc II}$] emission, 
F(Ly$\alpha)=(1.3\pm0.8) \times 10^{-18}$ erg sec$^{-1}$ cm$^{-2}$, rest-frame EW(Ly$\alpha)=26\pm11$ {\AA} 
(assuming a continuum flux density of $1\times 10^{-20}$ erg sec$^{-1}$ cm$^{-2}$ ${\AA}^{-1}$), and L(Ly$\alpha)=(1.0\pm0.6) \times 10^{8}$ L$_{\odot}$, giving a ratio of L(Ly$\alpha$)/L([C${\sc II}$])= $0.06\pm0.04$ 
\citep{Pavesi2019}, that is consistent with the theoretical value by assuming a dust reddening corresponding to $A_{V}=0.2$ (see Table 2) and extinction law from \citet{Calzetti2000}.

In addition to LBG-1, we detected eight other Ly$\alpha$ emitting galaxies 
in the AzTEC-3 field. They tend to be located in an elongated configuration 
(see Fig. \ref{NBcandidates}). 
Some other objects with photometric redshift consistent with that of the SMG, but without a Ly$\alpha$ emission in our MUSE data, are also distributed around AzTEC-3 
(Fig. \ref{3Ddistr}). 
They could lack of Ly$\alpha$ emission due to a unfavorable amount and distribution of dust toward the line of sight, but they could also be lacking the HI gas reservoir that would shine in Ly$\alpha$. It could have been consumed by star formation or by tidal stripping within the AzTEC-3 environment. Instead, the elongated distribution of the Ly$\alpha$ emitting galaxies could follow the elongated distribution of HI gas falling toward the SMG. 
More sensitive data would be needed to confirm this hypothesis. 

\subsection{Evidence of merger phenomena in the AzTEC-3 environment} 

There is evidence in the literature that supports the idea that compact SMGs could be the result of major mergers \citep[e.g.,][]{Tacconi2006, Swinbank2008,Engel2010, Simpson2014}.
In the protocluster of our study, 
the high molecular gas mass (roughly 5 times the stellar mass) of AzTEC-3 and its SFR$_{FIR}$ equal to 1100 M$_{\odot}$ yr$^{-1}$ \citep{Riechers2014} suggest the presence of a heavily obscured starburst, possibly triggered by a major merger. 
The richness of the gas reservoir in the AzTEC-3 system (both dense and diffuse molecular gas, and diffuse atomic gas revealed by Ly$\alpha$) supports the idea of a gas-rich merger. 

The FIR luminosity surface density of AzTEC-3 was estimated to be $5\times 10^{12}$ L$_{\odot}$ kpc$^2$ \citep{Riechers2014} and it is at the limit for a starburst to be supported by radiation pressure. 
The [C${\sc II}$] to FIR luminosity ratio (L([C${\sc II}$])/L$_{FIR}$) and the no-detection in X rays \citep{Riechers2020} suggested the presence of a strong radiation field, 
but gives no direct indication for the presence of an obscured AGN \citep{Riechers2014}.  The merger of the SMG progenitors could have triggered the starburst we currently see.
Also, the observed Ly$\alpha$ luminosity supports the idea that the Ly$\alpha$ photons are mainly produced by recombination of atoms excited by a starburst radiation field (see Sect. 6). 

Based on the kinematics of our observations of Ly$\alpha$ emission,  
a fast outflow 
can depart from the starburst and extend to the south. 
We also detected a bridge of high column density gas between AzTEC-3 and LBG-3 that 
can be the result of the merger event between AzTEC-3 and LBG-3. 
The elongated Ly$\alpha$ emission north of LBG-3 suggests tidal interaction with the SMG gravitational potential. 
The gas could have been stripped and could have assumed a tidal structure as a consequence of a first passage between LBG-3 and AzTEC-3. 
The structure seen in the region to the south of the SMG suggests interaction between the SMG and the mosaic\_199 source, a third phenomenon of interaction (Fig. \ref{AzTEC3channels}) around AzTEC-3. 

In addition to this, in the field there is LBG-1 that itself could be a merging system. The Ly$\alpha$ observations and modeling suggest the presence of a gas with expansion velocities of 50 up to 200 km sec$^{-1}$ in the southern region of LBG-1 that could follow the merging rotation and an N${\sc HI}$ overall on the order of $10^{21}$ atoms cm$^{-2}$ that could make a gas-rich merger.
We do not have enough resolution to provide a more detailed description of the LBG-1 merger.  
%
However, as proposed by \citet{Riechers2014}, this system could be in an early stage of merger. 
The final stage could drive to a more intense star-formation rate than the one currently measured, but probably to a much fainter Ly$\alpha$ emission absorbed by the dust produced in the star formation merger-driven event \citep{Yajima2013}.

Other systems in the literature at $z \sim 6$, such as CR7 \citep{Matthee2020} and Himiko \citep{Ouchi2013} share properties with LBG-1.
For both sources from the literature, the Ly$\alpha$ emissions were found to be elongated along the direction connecting the multiple components and the Ly$\alpha$ peaks were seen offset with respect to the brightest UV emissions. The offset could be related to a merger event \citep[see also][]{Jiang2013} 
and the elongation of the Ly$\alpha$ emission along the axis of the multiple components could trace the underlying gas distribution \citep{Matthee2020}, which seems distorted toward the southwest in the case of LBG-1.
\citet{Riechers2014} measured a L([C${\sc II}$])/L$_{FIR}$ ratio of 3 - 9 $\times 10^{-3}$, consistent with the value for normal star-forming galaxies \citep[see also][]{Pavesi2019}, and a L([C${\sc II}$])/L(CO) $ >4600$ indicative of a moderate radiation field. Also, the FWHM and the EW(Ly$\alpha$) of LBG-1 are consistent with the values of normal star-forming galaxies rather than AGN.


\subsection{Study of AGN activity in AzTEC-3}

Our MUSE observations do not provide strong constrains on the AGN activity in AzTEC-3. However, by comparing the Ly$\alpha$ emission we detected with observations in the literature that support AGN activity, we found differences that point more to a merger-driven starburst rather than to an influential AGN.

\citet{Borisova2016} and \citet{denbrok2020} showed a large sample of Ly$\alpha$ nebulae around typeI and typeII AGN at $z=3$, detected trough the CubEx software. Despite the surface brightness dimming, the radii of those nebulae 
were observed to be larger than 450 ckpc (more than 110 pkpc) and between 220 and 450 ckpc (50-100 pkpc and up to 300 pkpc with 480 pkpc for the Slug nebula), respectively, much larger than in our case. Also, the integrated Ly$\alpha$ luminosities are 1 order of magnitude brighter than the value of our entire system, even in the cases of obscured tori. 
Also, their 1D spectra show Ly$\alpha$ peaks that are redshifted more than 1000 km sec$^{-1}$. 
The FWHM and the EW of the Ly$\alpha$ emission line we measured, even for the spectrum extracted just on top of the SMG, are consistent with the values measured for star-forming galaxies rather than AGNs \citep{Henry2015, Matthee2020}. However, the luminosity values we measured can still be dominated by the effect of the dust of the SMG. 
In fact, correcting the Ly$\alpha$ flux by the extinction of $A_V=0.8$ estimated for the starburst, the L(Ly$\alpha$) emitted at the compact  position of the SMG becomes comparable to the faintest sources studied in \citet{Borisova2016}.

Emissions of Ly$\alpha$, comparable in size to that of the AzTEC-3+LBG-3 system, were found around two star-forming galaxies at $z=6.5$ and 6.6 (VR7 and CR7 Matthee et al. 2019 and 2020). However, their Ly$\alpha$ luminosities are more than 5 times brighter than in our system, implying that the SMG dust could allow the escape of fewer Ly$\alpha$ photons, even if the N${\sc HI}$ conditions in the CGM could be similar and could produce an equally efficient scattering. 
For CR7, the AGN contribution was disfavored given the UV to IR luminosity ratio \citep{Matthee2020}. In the case of AzTEC-3, \citet{Riechers2014} pointed out that a luminous AGN component was also disfavored by the non detection of the high excitation CO(16-15), even though not detecting this transition is not a hard constraint. 
\citet{Algera2021} found that, based on its radio emission, AzTEC-3 does not show a sign of AGN activity either and the MUSE spectrum does not present typical AGN features.
Therefore, the data so far are not in favor of a scenario in which an AGN is powering the emission of AzTEC-3, even if a very obscured AGN, also a possible result of the major merger event, could still be present.

There are examples in the literature where other mechanisms of the production of the radiation field were proposed and that showed different properties in the Ly$\alpha$ emission in comparison to the AzTEC-3 system.
\citet{Daddi2020} found evidence of gas accretion toward a massive system also containing an SMG at $z=2.91.$ The phenomenon of gas accretion was supported by a blue shifted component of the Ly$\alpha$ emission line. However, the Ly$\alpha$ emission they observed peaks in an empty region located at the center of the halo potential well and it extends over 1000 ckpc (300 pkpc), much larger than what we detect.

The SSA22 protocluster is an overdensity with more than one extended Ly$\alpha$ emitting source. 
The overdensity was detected as an excess of LBGs \citep{Steidel2000} and more than 200 Ly$\alpha$ emitters \citep{Hayashino2004}, more than 30 Ly$\alpha$ nebulae \citep{Matsuda2004}, and more than 50 SMGs \citep{Umehata2014} were discovered in the entire area. Both the Ly$\alpha$ emitters and the SMGs trace the densest peak of the protocluster. The extended Ly$\alpha$ sources include two giant Ly$\alpha$ blobs with sizes larger than 100 pkpc \citep{Steidel2000}. They were discovered to be located in the intersection of the three filaments, traced by the Ly$\alpha$ emitters, that characterize the protocluster. In agreement with the theory that galaxy formation at high redshift occurs along large-scale filamentary overdense regions, the intersection of filaments could evolve in low-redshift clusters and the giant Ly$\alpha$ blobs could be revealing cluster progenitors. The Ly$\alpha$ luminosity of the two giant Ly$\alpha$ blobs on the order of 10$^{44}$ erg sec$^{-1}$ (about 20 times brighter than the Ly$\alpha$ emission associated to the AzTEC-3+LBG-3 system) and the association of one of the two to an SMG \citep[LAB1][]{Chapman2001} suggested they may be progenitors of very massive galaxies near the center of a massive cluster \citep{Matsuda2005} at $z=0$. 
The star-formation rate estimated for the SMG associated to LAB1 is $\sim 1000$ M$_{\odot}$ yr$^{-1}$ and the CO emission line implies a large amount of molecular gas \citep{Chapman2004}, comparable to the values of AzTEC-3. However, cooling flow could be a possible scenario for the Ly$\alpha$ emission \citep{Chapman2001} also in SSAA2, due to the lack of an optical counterpart in LAB1, lack of both an optical and a submillimeter source in the other blob, and for the large and bright Ly$\alpha$ emission, unlike what we observe in the AzTEC-3 system.

\subsection{Estimation of the mass of the AzTEC-3 protocluster and its fate at lower redshift} 

We found that six of the Ly$\alpha$ emitting sources detected in the AzTEC-3 environment, including AzTEC-3, LBG-3, and LBG-1, have a counterpart in the COSMOS2015 catalog and so we have an estimation of their stellar masses (from $5 \times 10^{8}$ to $2\times 10^{10}$ M$_{\odot}$). The LBGs in the field studied in \citet{Capak2011} have stellar masses between $6.3\times 10^{7}$ and $2.5\times 10^{9}$ M$_{\odot}$. By summing all these mass values, we obtain a minimum value of the stellar mass in the protocluster of $(5.8\pm2.3) \times 10^{10}$ M$_{\odot}$.  
If we assume that the COSMOS2015 sources shown in Fig. \ref{3Ddistr} also belong to the AzTEC-3 protocluster, the minimum value of the stellar mass in the protocluster is $(8.1\pm2.4)\times 10^{10}$ M$_{\odot}$. 

An estimation of the dark-matter halo mass of the protocluster can be obtained by assuming the relation between stellar mass and halo mass for the clusters studied in \citet{vanderburg2014}. For a stellar mass of $5.8\times 10^{10}-8.1\times 10^{10}$ M$_{\odot}$, we can calculate a minimum dark-matter halo mass of $4\times 10^{11} -8\times 10^{11}$ M$_{\odot}$, which is consistent with the lower limit of dark-matter halo mass estimated in \citet{Capak2011}. It is worth noting that the protocluster projected area can be as big as a 2cMpc-radius circle \citep{Capak2011} larger than the area probed by our MUSE observations. However, the stellar mass could be concentrated within 1cMpc, since the center of mass of the protocluster is close to the location of AzTEC-3 and the two most massive sources of the field are about 600 ckpc apart. 
The relation from \citet{vanderburg2014} was obtained for clusters at $z=1$ and there could be evolution between the properties of $z\sim5$ protoclusters and $z=1$ clusters. 
\citet{Cen2013} provided an approximate equation to relate the observed Ly$\alpha$ luminosity of a Ly$\alpha$ nebula and the dark matter halo mass of its central galaxies. By following their equation 2 and 4, we can give an estimation of the dark matter halo mass in which the AzTEC-3+LBG-3 Ly$\alpha$ nebula is located. The total Ly$\alpha$ luminosity of the AzTEC-3+LBG-3 system is $(5 \pm 1) \times 10^{42}$ erg sec$^{-1}$. This luminosity implies a dark matter halo mass of ($1.84\pm0.04)\times 10^{12}$ M$_{\odot}$ and a size of the nebula on the order of 8 arcsec$^2$ rescaled to $z=5.3$, which is consistent with what we observe. 
Considering the simulation in \citet{Chiang2013}, a protocluster with a halo mass of $10^{12}$ M$_{\odot}$ could evolve into a cluster of $2\times 10^{13}$ M$_{\odot}$ by $z=2$ and into a Fornax-type cluster at $z=0$ with a typical mass of $2\times 10^{14}$ M$_{\odot}$.
 \citet{Chiang2013} also showed that $z>5$ protoclusters with $10^{12}$ M$_{\odot}$ halo masses span a region in the sky within an effective radius of about 5 cMpc. Therefore, we could be observing just the central, most massive part of the entire overdensity.

Submillimeter galaxies are thought to be high-$z$ progenitors of present-day ellipticals that could be located in the core of present-day galaxy clusters.  \citet{Swinbank2008} showed that SMGs at $z=2$ could reside in a similar mass halo as SMGs as $z>4$, but could be characterized by twice their stellar mass. To achieve that stellar mass, major mergers are a plausible explanation \citep{Swinbank2008,Engel2010,Simpson2014,Dudzeviciute2020,Stach2021} and we have evidence of them in the protocluster around AzTEC-3.
Unlike some other overdensites of galaxies at $z>4$ from the literature, the AzTEC-3 environment contains only one submillimeter galaxy. As an example, the SPT2349-56 protocluster contains 14 submillimeter galaxies at $z=4.3$ and it is expected to evolve into a very massive cluster of more than $10^{15}$ M$_{\odot}$ at $z=0$ \citep{Miller2018}. Also, the overdensity discovered by \citet{Oteo2018}, containing at least ten dusty star-forming galaxies at $z\simeq4$, is expected to evolve in a cluster as massive as the Coma cluster at $z=0$.

\section{Conclusions}
\label{summary}

In this work we have analyzed the MUSE data of the environment around the AzTEC-3 submillimeter galaxy. We have made use of the CubExtractor \citep{Cantalupo2019} software to reduce and analyze the data, as shown in Sects. 2 and 3. 
We found ten Ly$\alpha$ emitting sources, including an extended emission toward AzTEC-3 and LBG-3 with a total Ly$\alpha$ luminosity of $(5 \pm 1) \times 10^{42}$ erg sec$^{-1}$ and another extended emission embedding the three components of LBG-1. The sources appear distributed in an elongated configuration of about $70''$ in extent. 
Five of the sources have a counterpart in the COSMOS2015 catalog. 
The presence of the ten Ly$\alpha$ emitting sources around AzTEC-3 confirms that the region is an overdensity at $z\sim5.3$ \citep[][and Sect. 5]{Capak2011}. 
We calculated the center of mass of the protocluster which is close to the AzTEC-3+LBG-3 system, the most massive of the protocluster.

We studied the spectroscopic properties of the Ly$\alpha$ emission in the AzTEC-3 environment in detail and we compared with the zELDA radiative transfer model \citep{Gurung-Lopez2021b}.
We note that Ly$\alpha$ photons are emitted toward the southeast of the SMG, probably due to dust obscuration to the west. The Ly$\alpha$ emission of the SMG is blended with that of LBG-3 and shows a bridge between the two galaxies, which could reveal the gas distribution due to their interaction. The Ly$\alpha$ emission is elongated to the north of LBG-3 resembling a tidal feature due to the interaction with AzTEC-3. An elongated feature is also seen toward the south of the SMG that could be produced by the interaction with the mosaic\_199 source. 
The presence of an outflow departing from the SMG is supported by the size of the star-forming region, the star-formation rate estimated from the [C${\sc II}$], and FIR observations \citep{Riechers2014, Riechers2020}. It is plausible that this outflow cleared channels for the escape of Ly$\alpha$ photons and that the Ly$\alpha$ photons are produced by recombination of HI atoms based on the energetics of the Ly$\alpha$ emission \citep[][and Sect. 4]{Cantalupo2017}. The HI scattering can explain the Ly$\alpha$ luminosity we observe at more than 90 ckpc. Also, the modeling of the Ly$\alpha$ line did not require any contribution from an AGN.

Based on the resolution of our MUSE observation, we found that the Ly$\alpha$ emission comes from the three components of LBG-1. This could be related to the low dust content in this galaxy. However, the emission is not symmetrical and could indicate the interaction of the three components. The Ly$\alpha$ emission line profile shows a main peak and an extended tail which is consistent with models of galaxies characterized by gas in a wide range of velocities up to 200 km sec$^{-1}$. Overall the Ly$\alpha$ emission is consistent with models of galaxies characterized by N${\sc HI}$ on the order of $10^{21}$ atoms cm$^{-2}$ that could indicate that the interaction of the three components is a gas-rich merger. 
The emission is redshifted with respect to the [C${\sc II}$] systemic redshift and the Ly$\alpha$ emission line is broader than [C${\sc II}$] in the 1D spectrum. This is not unusual for high-$z$ Ly$\alpha$ emitters and could be related to the variety of gas kinematics and column densities that condition the escape of Ly$\alpha$ photons and instead do not affect the escape of [C${\sc II}$] photons. 
The radiative transfer fit of the Ly$\alpha$ of two other star-forming galaxies indicate that they are galaxies with HI column densities up to 10$^{20}$ atoms cm$^{-2}$ and HI expansion velocities on the order of a few tens of km sec$^{-1}${. 

Given the availability of CO and [C${\sc II}$] observations from previous campaigns \citep{Capak2011, Riechers2010, Riechers2014, Pavesi2018, Pavesi2019, Riechers2020} and the Ly$\alpha$ information from our MUSE dataset, we can discuss weather the environment can play a role in the stage of evolution of AzTEC-3. We have evidence of starburst-driven phenomena and interactions around AzTEC-3 that could support the idea that the gravitational potential of the SMG is accreting nearby galaxies (like LBG-3 and possibly mosaic\_199) and accumulating gas. Due to the gravitational interactions, the submillimeter galaxy could increase its mass significantly by $z=2$. With LBG-1, we could be assisting to another merging event in the AzTEC-3 protocluster, and the two galaxies could all merge by $z=0$ \citep{Riechers2020}. We estimated that the dark matter halo mass of the protocluster is on the order of $10^{12}$ M$_{\odot}$. This value is lower than that of other protoclusters studied in the literature that contain more than one submillimeter galaxy \citep[e.g.,][]{Miller2018}. However, a dark matter halo of $10^{12}$ M$_{\odot}$ could evolve into a cluster of $2\times10^{13}$ M$_{\odot}$ by $z=2$ and into a Fornax-type cluster at $z=0$ with a typical mass of $2\times10^{14}$ M$_{\odot}$ \citep{Chiang2013}. 


\begin{acknowledgements}
We thanks Franz Bauer, Lodovico Coccato, Yara Jaffe, Evelyne Johnston, Sam Kim, Michael Maseda, Ian Smail, Frederic Vogt  
for useful discussions.
      LG is grateful to ETH for the use of computer facilities and to the CAS-CONICYT, CAS18016, fellowship. 
SC gratefully acknowledges support from Swiss National Science Foundation grant PP00P2\_190092 and from the European Research Council (ERC) under the European Union’s Horizon 2020 research and innovation programme grant agreement No 864361.
HD acknowledges financial support from the Agencia Estatal de Investigaci\'on del Ministerio de Ciencia e Innovaci\'on (AEI-MCINN) under grant (La evoluci\'on de los c\'umulos de galaxias desde el amanecer hasta el mediod\'ia c\'osmico) with reference (PID2019-105776GB-I00/DOI:10.13039/501100011033) and from the ACIISI, Consejer\'ia de Econom\'ia, Conocimiento y Empleo del Gobierno de Canarias and the European Regional Development Fund (ERDF) under grant with reference PROID2020010107.
\end{acknowledgements}

\bibliographystyle{aa}   
\bibliography{biblio}        


s

\begin{appendix} 
\section{Protocluster members with Ly$\alpha$ in emission}
\label{Lyamembers}

We show here the surface brightness profiles of the protocluster candidates we have identified through their Ly$\alpha$ emission. Fig. \ref{mosaic1548} to Fig. \ref{horizontalISN3199} show the surface brightness of the detections in the form of optimized S/N narrow-band images for the sources listed in Table 2. 
\begin{figure*}
 \centering
\includegraphics[width=15cm]{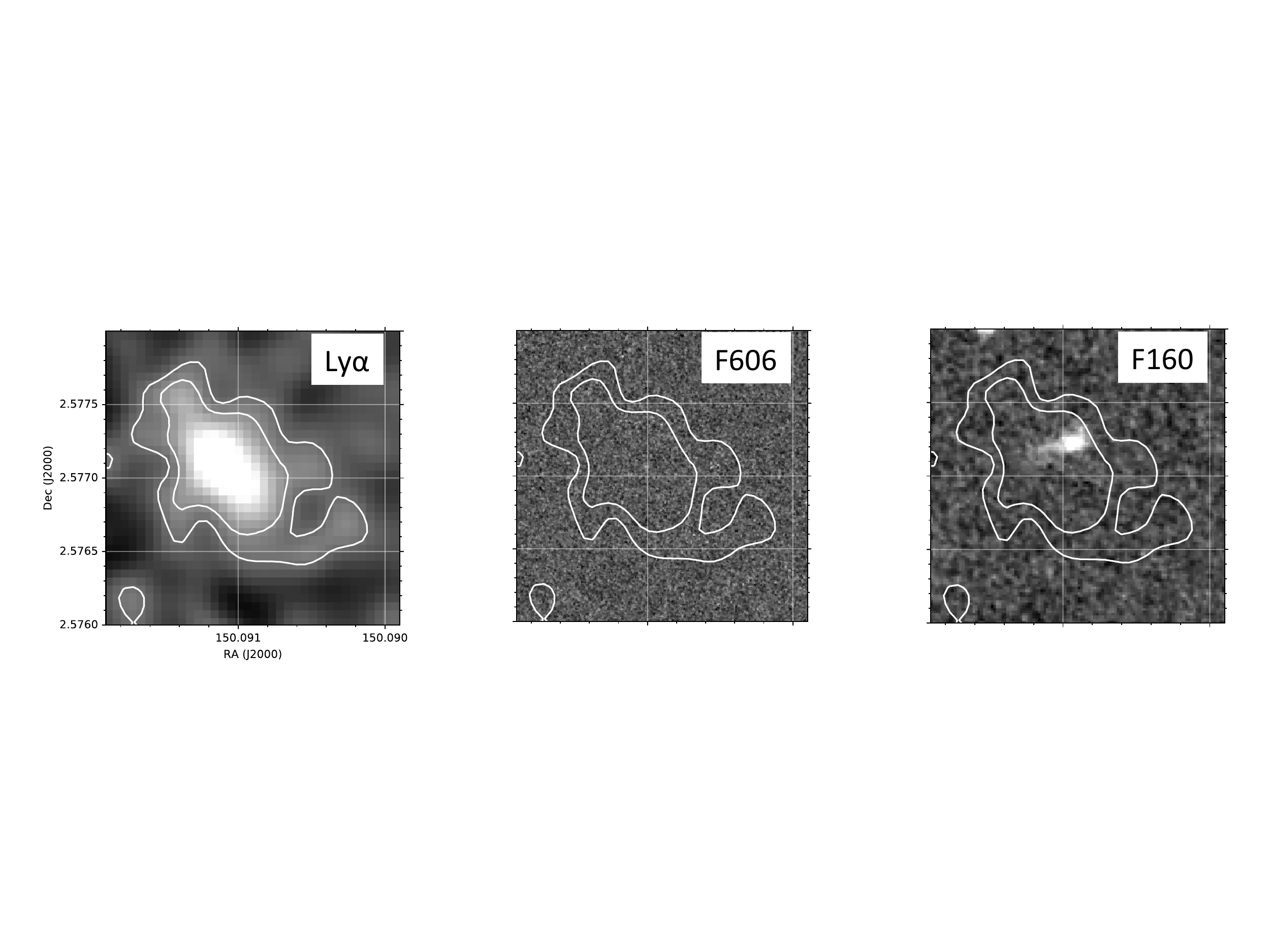} 
\caption{
Optimally extracted narrow-band images of the Ly$\alpha$ emission mosaic\_1548 ($left ~panel$) and contours in surface brightness units. The two white contours correspond to 3 and 4$\sigma$ above the background, based on the S/N map provided by CubEx. In this case they correspond to $0.7 \times 10^{-18}$ and $1.2\times 10^{-18}$ erg sec$^{-1}$ cm$^{-2}$ arcsec$^{-2}$. Stamps of the source in the $HST$ F606W and F160W images are shown in the $middle$ and $right$ panels, together with the narrow-band surface brightness contours.
The source within the contours of the right panel is COSMOS2015\_841844 and it has a 0.4$''$ and a 1$''$ axes in the F160W image.
The size of the pictures are adapted to contain exactly the entire emission.
}
\label{mosaic1548}%
\end{figure*}

\begin{figure*}
 \centering
\includegraphics[width=15cm]{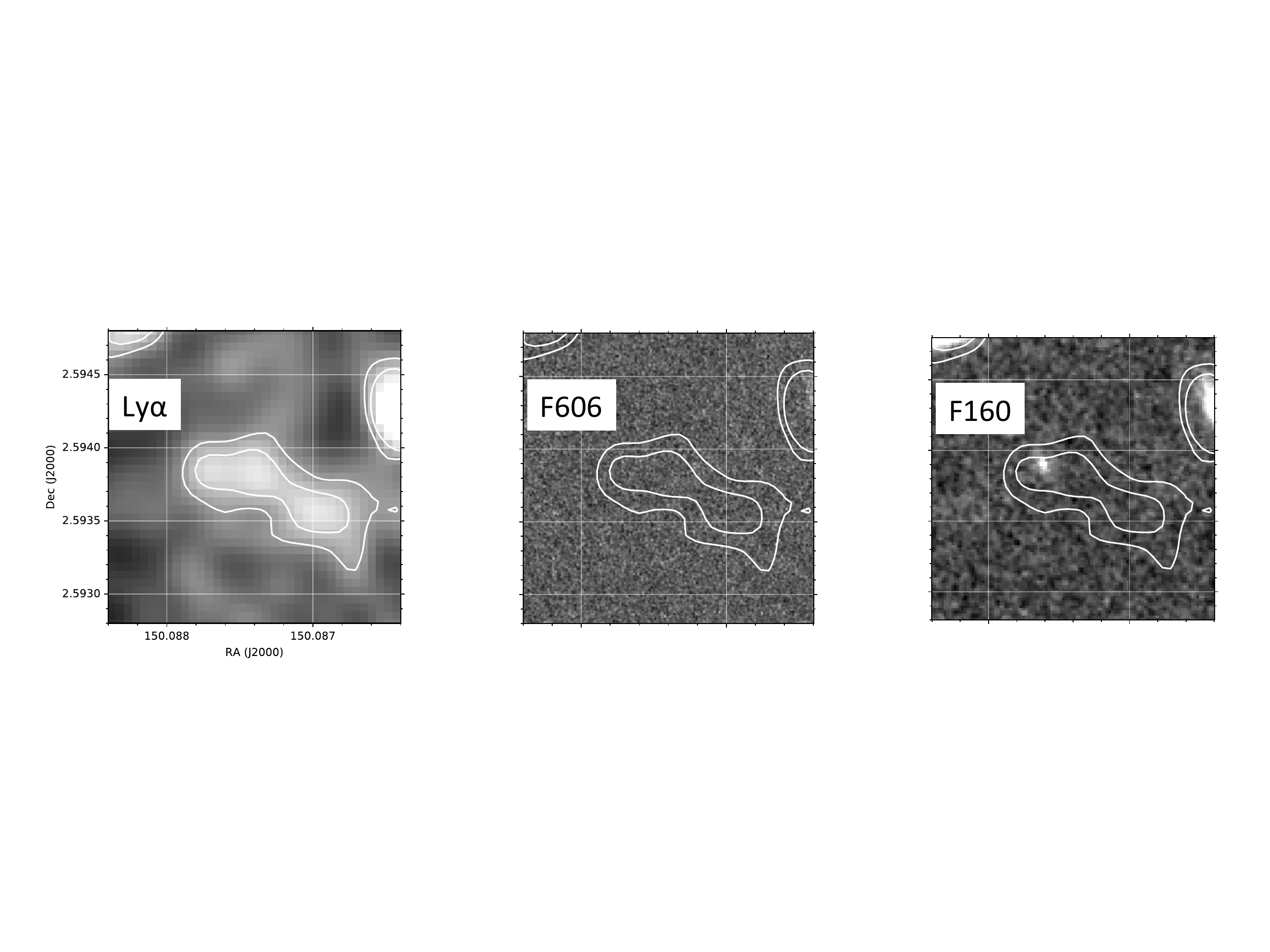} 
\caption{Same as Fig. \ref{mosaic1548}, but for the mosaic\_1520 source. The three white contours correspond to $0.5\times 10^{-18}$ and $0.8\times 10^{-18}$ erg sec$^{-1}$cm$^{-2}$ arcsec$^{-2}$. 
The source inside the contours of the right panel is COSMOS2015\_852474 and it has a 0.25$''$ radius in the F160W image.
}
\label{mosaic1520}%
\end{figure*}


\begin{figure*}
 \centering
\includegraphics[width=15cm]{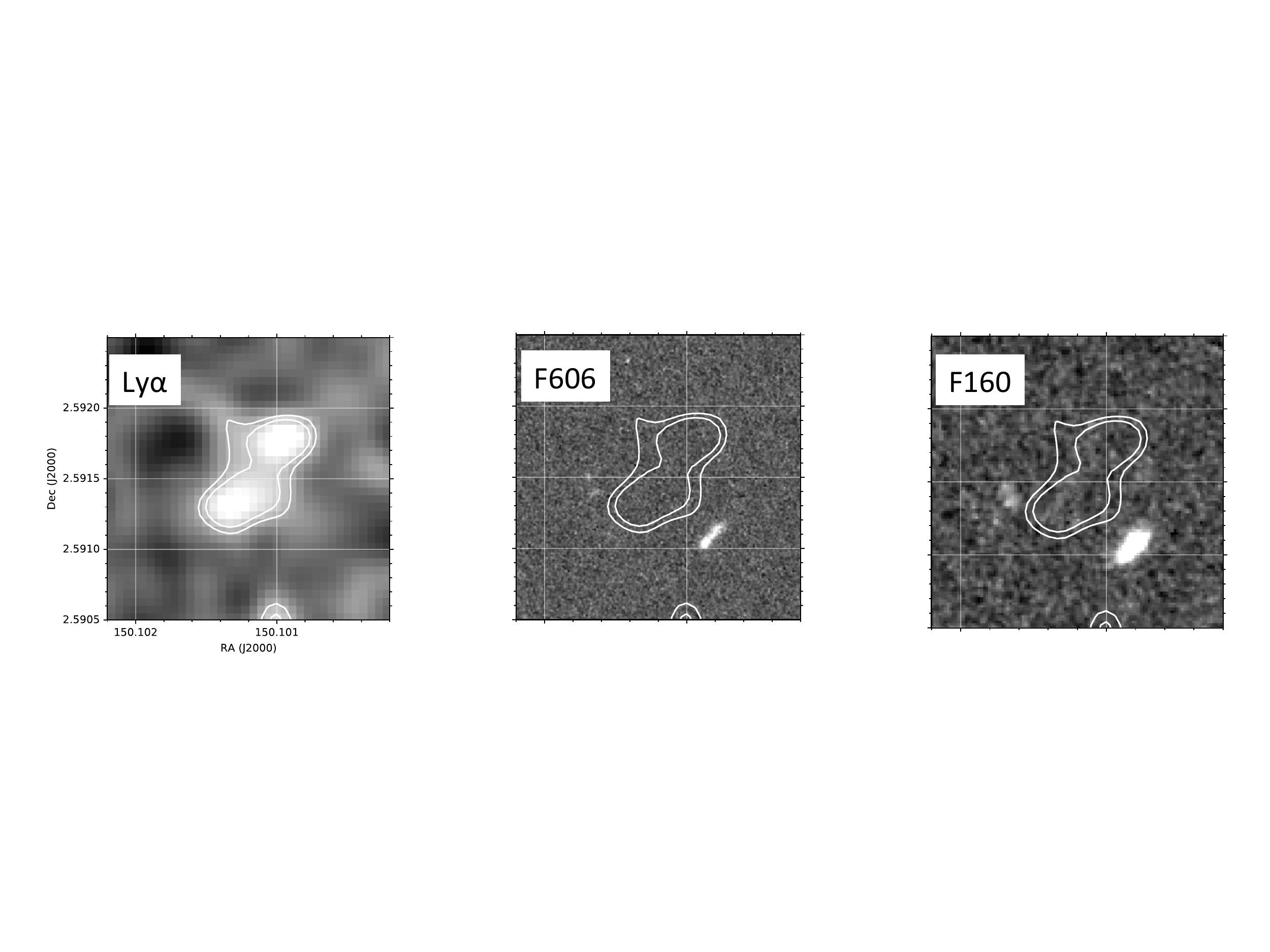} 
\caption{Same as Fig. \ref{mosaic1548}, but for the mosaic\_770 source. The two white contours correspond to $0.8\times 10^{-18}$ and $1.1\times 10^{-18}$ erg sec$^{-1}$ cm$^{-2}$ arcsec$^{-2}$.
}
\label{mosaic770}%
\end{figure*}

\begin{figure*}
 \centering
\includegraphics[width=15cm]{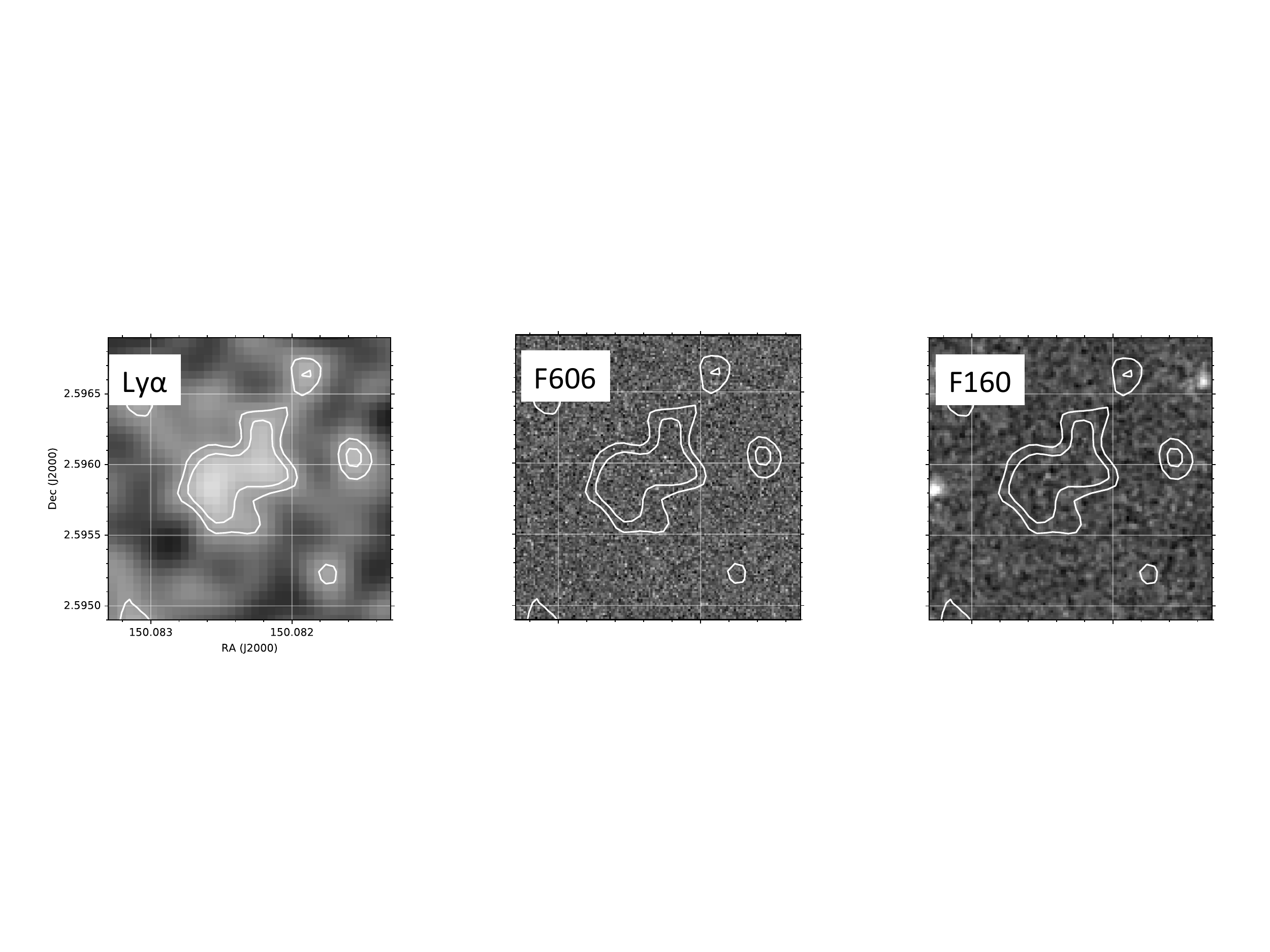} 
\caption{Same as Fig. \ref{mosaic1548}, but for the mosaic\_1035 source. The two white contours correspond to $0.6\times 10^{-18}$ and $0.8\times 10^{-18}$ erg sec$^{-1}$ cm$^{-2}$ arcsec$^{-2}$.
}
\label{mosaic1035}%
\end{figure*}

\begin{figure*}
 \centering
\includegraphics[width=15cm]{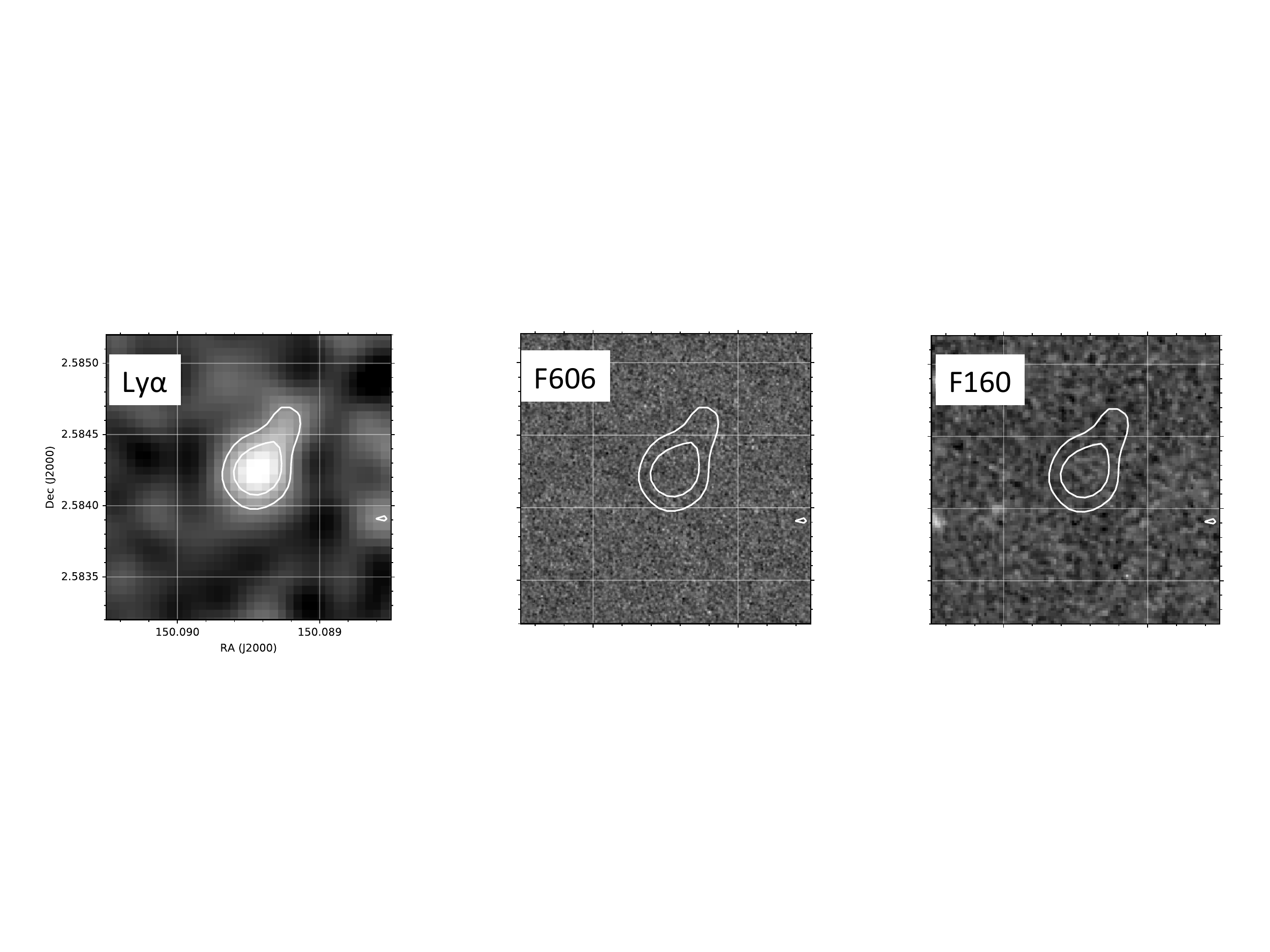} 
\caption{Same as Fig. \ref{mosaic1548} for the P3\_547 source. The two white contours correspond to $0.5\times 10^{-18}$ and $1.0\times 10^{-18}$ erg sec$^{-1}$ cm$^{-2}$ arcsec$^{-2}$.
}
\label{deep3547}%
\end{figure*}

\begin{figure*}
 \centering
\includegraphics[width=15cm]{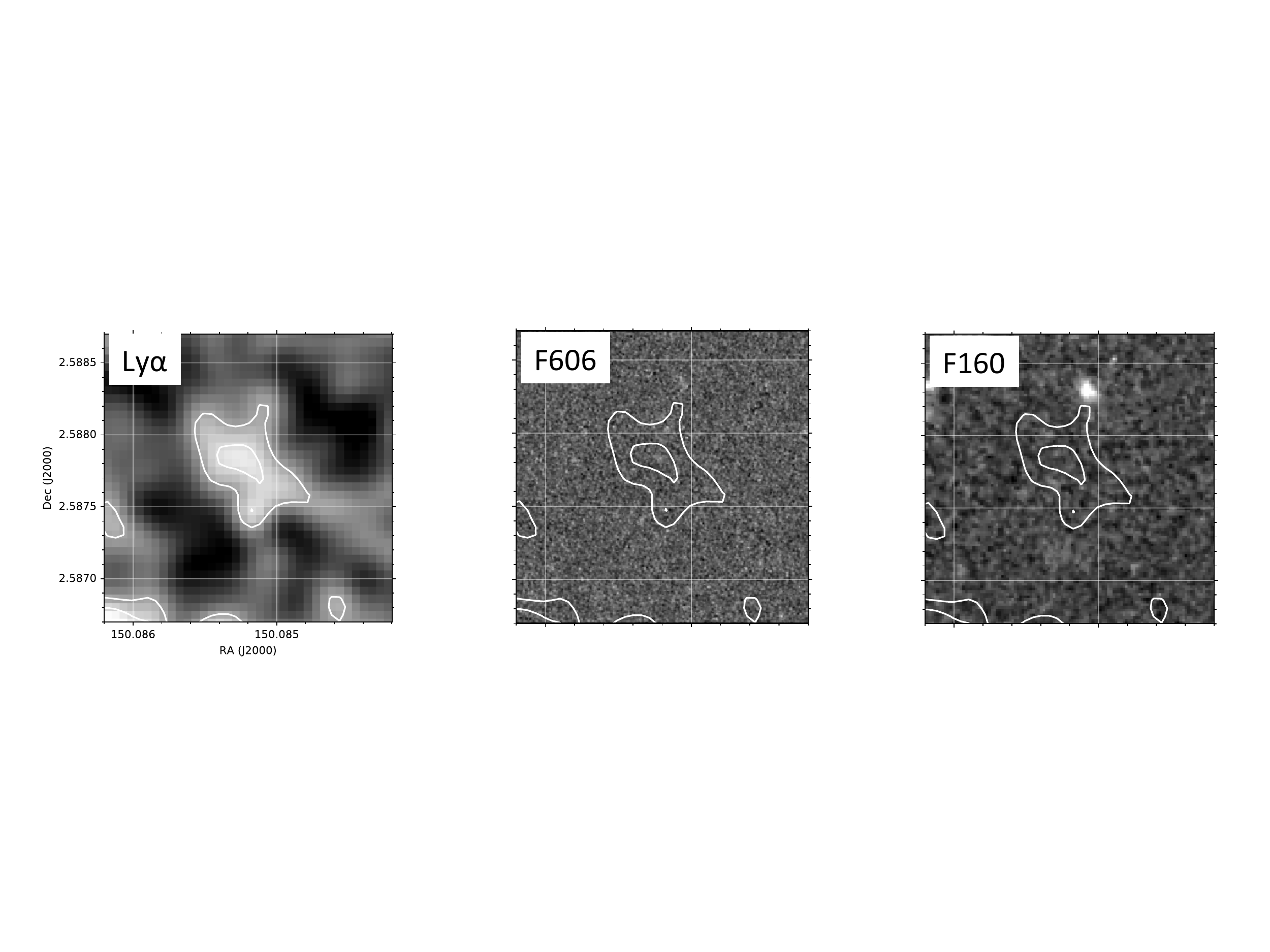} 
\caption{Same as Fig. \ref{mosaic1548} for the P2\_446 source. A source is visible in the $HST$ F160W image in the center of the Ly$\alpha$ emission, but does not have a counterpart in the COSMOS2015 catalog. The two white contours correspond to $0.5 \times 10^{-18}$ and $0.8\times 10^{-18}$ erg sec$^{-1}$ cm$^{-2}$ arcsec$^{-2}$.
}
\label{deep2446}%
\end{figure*}

\begin{figure*}
 \centering
\includegraphics[width=15cm]{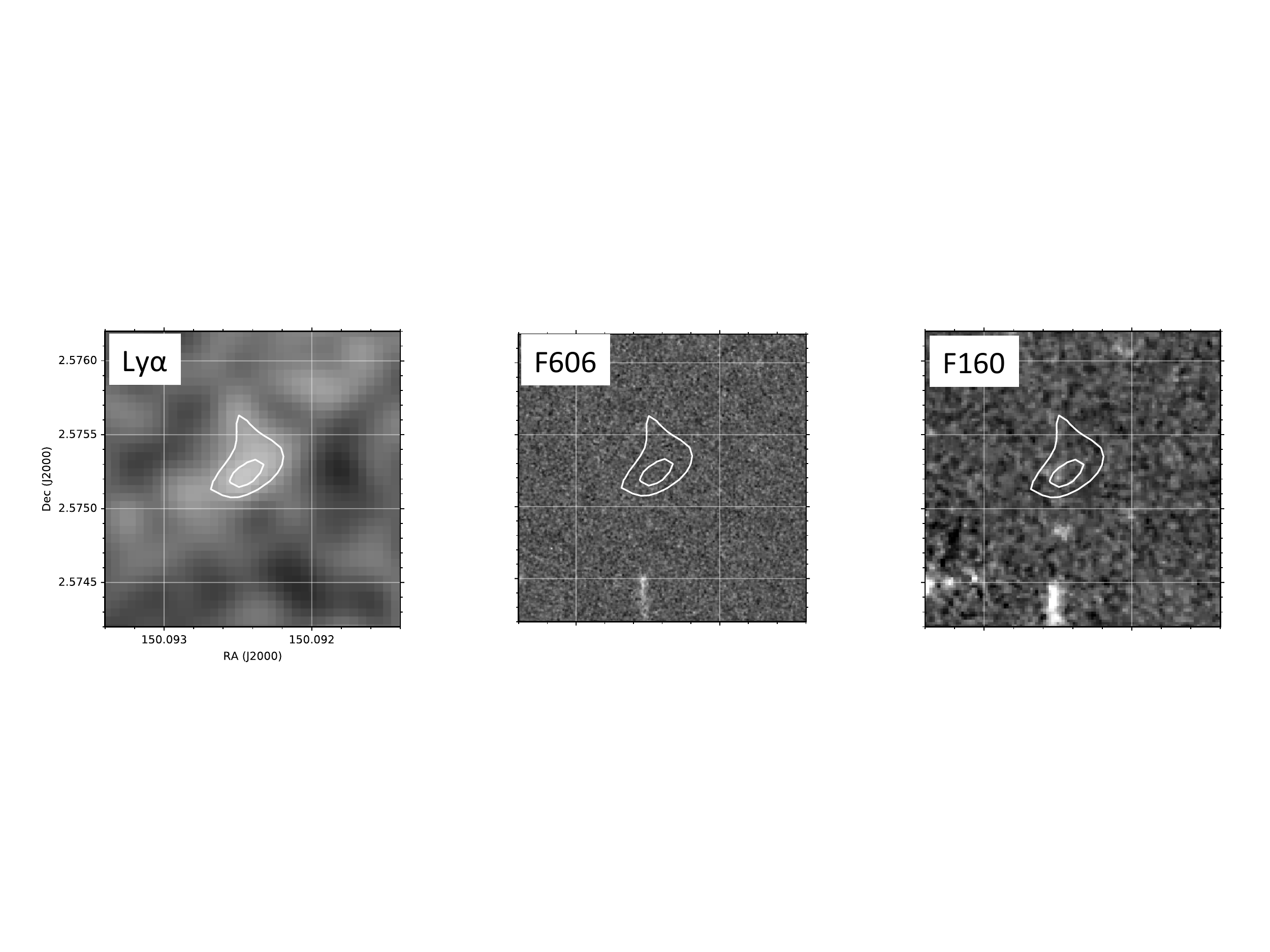} 
\caption{Same as Fig. \ref{mosaic1548}, but for the mosaic\_414 source. 
A source is visible in the $HST$ F160W image in the center of the Ly$\alpha$ emission, but does not have a counterpart in the COSMOS2015 catalog. The two white contours correspond to $0.5\times 10^{-18}$ and $0.7\times 10^{-18}$ erg sec$^{-1}$ cm$^{-2}$ arcsec$^{-2}$. 
}
\label{d3414}%
\end{figure*}

\begin{figure*}
 \centering
\includegraphics[width=15cm]{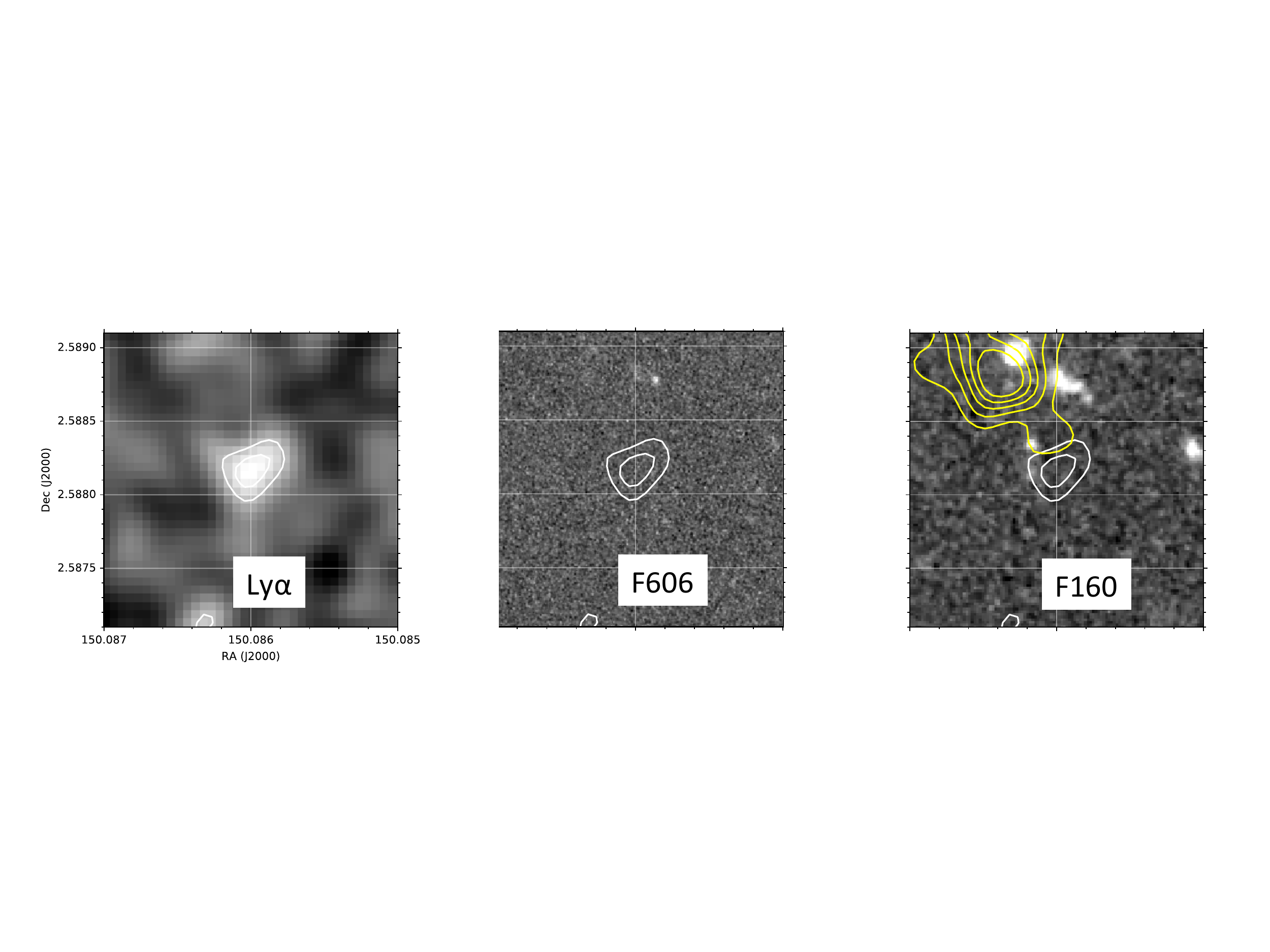} 
\caption{Same as Fig. \ref{mosaic1548}, but for the mosaic\_199 source. The two white contours correspond to $0.5\times 10^{-18}$ and $0.7\times 10^{-18}$ erg sec$^{-1}$ cm$^{-2}$ arcsec$^{-2}$.
A source is visible in the F160W image which is COSMOS2015\_848724. Some low SN emission could be blended with the AzTEC-3+LBG-3 system (yellow contours). 
}
\label{horizontalISN3199}%
\end{figure*}

\end{appendix}

\end{document}